\documentclass[prc,nofootinbib,showpacs,twocolumn]{revtex4-2}
\usepackage{graphicx}
\usepackage[usenames,dvipsnames]{color}
\usepackage{amsmath,amssymb,bbold}
\usepackage[normalem]{ulem}
\usepackage{tabularx}
\usepackage{array}
\usepackage{hyperref}

\graphicspath{{./figures/}}

\begin{document}

\title{Pulsar as a Weber detector of gravitational waves 
and a probe to its internal phase transitions}
\author{Partha Bagchi}
\email{parphy85@gmail.com}
\affiliation{School of Physical Sciences, National Institute of Science
Education and Research, Jatni, Odisha -752050, India}
\author{Oindrila Ganguly}
\email{oindrilacg@gmail.com}
\affiliation{The Institute of Mathematical Sciences, Chennai -600113, India}
\author{Biswanath Layek}
\email{layek@pilani.bits-pilani.ac.in}
\affiliation{Department of Physics, Birla Institute of Technology and Science, 
Pilani -333031, India}
\author{Anjishnu Sarkar}
\email{anjishnu@lnmiit.ac.in}
\affiliation{Physics Department, The LNM Institute of Information Technology, 
Jaipur -302031, India}
\author{Ajit M. Srivastava}
\email{ajit@iopb.res.in}
\affiliation{Institute of Physics, Bhubaneswar-751005, India}

\begin{abstract}
It is believed  that cores of neutron stars provide a natural laboratory
where exotic high baryon density phases of QCD may exist. In fact, the
theoretically well established {\it neutron
superfluid phase} is also believed to be found only inside neutron stars.
Focus on neutrons stars has tremendously intensified in recent
years with the direct detection of gravitational waves by LIGO/Virgo
from binary neutron star (BNS) merger events which has allowed 
the possibility of directly probing
the properties of the interior of a neutron star. A truly
remarkable phenomenon manifested by rapidly rotating neutron stars is in
their {\it avatar} as {\it Pulsars}. The accuracy of pulsar timing can reach
the level of one part in 10$^{15}$, comparable to that of atomic clocks.
Indeed, it was such a great accuracy which had allowed the first indirect
detection of gravitational waves from a BNS system. Such an incredible
accuracy of pulse timings points to a very interesting possibility. Any
deformation of the pulsar, even if it is extremely tiny, has the
potential of leaving its imprints on the pulses through introduction
of tiny perturbations in the entire moment of inertia (MI) tensor. While,
the diagonal components of perturbed MI tensor affect the pulse timings,
the off-diagonal components lead to wobbling of pulsar, directly affecting
the pulse profile. This opens up a  new window of opportunity for
exploring various phase transitions occurring inside a pulsar core, through
induced density fluctuations, which may be observable as perturbations
in the pulse timing as well as its profile. Such perturbations also naturally
induce a rapidly changing quadrupole moment of the star, thereby providing
a new source of gravitational wave emission. Another remarkable possibility
arises when we consider the effect of an external gravitational wave on
neutron star. With the possibility of detecting any minute changes in
its configuration through pulse observations, the neutron star has the
potential of performing as a Weber detector of gravitational wave.
This brief review will focus on these specific aspects of a pulsar.
Specifically, the focus will be on the type of physics which can be
probed by utilizing the effect of changes in the MI tensor of the pulsar on
pulse properties.
\end{abstract}

\maketitle


\section{Introduction}

Cosmos has always proved to be the ultimate laboratory where physical
systems may exist in extreme environments, even those which are beyond
the reach of any terrestrial experiments. Early hot and dense stages of 
the Universe is one such case where extremely high temperatures are
achieved. Fortunately, some of those stages, with temperatures reaching
10$^{12}$ K (few hundred MeV) are possible to be partially probed
in terrestrial experiments now in relativistic heavy-ion collision
experiments (RHICE) \cite{Csernai:1994xw,Wong:1995jf}. 
This possibility has put the physics of quark-gluon
plasma (QGP) phase of QCD matter at a central stage, the phase which is 
believed to exist in the Universe when it was few tens of microseconds old. 
Experimental observations at RHICE have already shown unexpected 
results, for example, a near-perfect fluid nature of QGP with a value
of shear viscosity to entropy ratio which is close to the proposed
lowest bound on this number \cite{Heinz:2013th}. With continued efforts in 
RHICE with varying collision energy, it has been possible to extensively 
investigate certain part of the phase diagram of quantum chromodynamics (QCD) 
which corresponds to relatively low baron number density. As the early 
Universe was filled with a matter with extremely low baryon number density, 
it allows one to claim that conditions like early Universe have been 
recreated in laboratory (at least for the strongly interacting matter part).

At the same time, theoretical investigations have revealed
the possibility of an entire new spectrum of phases of strongly
interacting matter which are expected to arise at very high baryon
chemical potential \cite{Rajagopal:2000wf}.  It is reasonably clear
now that much of this extremely rich part of the QCD phase diagram
may remain out of reach in these terrestrial experiments. These phases
are collectively referred to as the {\it color superconducting phases}
\cite{Alford:2007xm,Rajagopal:2001ngu}.  Attention is thus naturally directed
towards astrophysics where gravity assisted ultra high baryon density
objects routinely occur. Extreme conditions of high baryon density are
expected to be reached in supernova explosions, in neutrons stars, and
in matter undergoing collapse to a black hole. The densest such object,
which can be directly observed at present, is a neutron star. It is
speculated that cores of neutron stars provide a natural laboratory
where various exotic phases of QCD may occur \cite{Alford:2000sx}. Even
exotic forms of matter, stable only under extreme conditions of
density and pressure, may form in these objects, such as strange stars
\cite{Witten:1984rs,Farhi:1984qu,Alcock:1986hz,Horvath:1992wq,Weber:1994yx}.
Interestingly, the theoretically very well established
{\it neutron superfluid phase} has never been seen in
any terrestrial experiment.  At the same time it is expected to routinely
occur inside neutron stars \cite{Yang:1971oux}.  In fact, superfluid
vortices in such a phase provide the most convincing explanation of the
phenomena of pulsar glitches \cite{Prakash:2000jr}.

Even though a host of QCD phases are expected to arise in the cores of
neutron stars, possibly with much larger baryon densities than possible
in laboratory heavy-ion collision experiments, the evolutionary history
of neutron stars, from its formation stage to late times, strongly
restricts the range of thermodynamic properties of these phases. Apart
from a very brief duration at the time of its formation, when its
temperature can be as high as tens of MeV, a neutron star rapidly cools 
to a temperature of order KeVs in less than a year. Thus, for observations,
we only have access to high baryon density matter at very cold
temperatures. This is in complete contrast to the low energy
relativistic heavy-ion collisions where very high baryon density matter
can be directly probed at temperatures ranging from tens of MeVs to
hundreds of MeVs, thus allowing the possibility of observing temperature
driven phase transitions in these high baryon density phases. Of course,
heavy-ion collisions do not allow the possibility of observing these
high baryon density phases at KeV temperatures, because thermal
freezeout temperature  in these experiments is of order 100 MeV. Then,
there are properties of QCD matter which can be directly probed in these
heavy-ion collisions, such as measurement of viscosity from flow
observations, which can only be indirectly inferred from neutron stars.
With all these complementary aspects of QCD phases available in these
two systems, one hopes that these systems together allow a very vast
range of QCD phase diagram to become accessible to experimental
observations.

Neutron stars have long been investigated theoretically and experimentally,
especially with pulsar observations. Pulsars are rapidly rotating neutron 
stars with pulse timings which are observed on earth with incredible
accuracy, reaching one part in 10$^{15}$ for certain pulsars, comparable to 
that of atomic clocks.
This extreme accuracy of pulse timings had allowed the first indirect
detection of gravitational waves from a binary neutron star (BNS) 
system \cite{Hulse:1974eb,Taylor:1982zz}.
Neutron star physics has acquired a centre stage recently with the advent of
gravitational wave detectors. After the first direct detection of 
gravitational waves (GWs) by LIGO coming from a binary black hole merger
event \cite{LIGOScientific:2016aoc},
the stage was set for the detection of GWs 
from spiral-in of other compact dense objects. The first such event of 
binary neutron star merger was detected by LIGO/Virgo in 
2017 \cite{LIGOScientific:2017vwq}
and that opened the remarkable possibility of directly probing the properties 
of the interior of neutron stars.

Neutron stars thus acquire a unique status of providing a laboratory
for probing microphysics of exotic phases of QCD on one hand, while
providing a window to probe the physics of its interior using GW
detectors on earth in BNS merger events on the other.  This remarkable
story of neutron stars still allows for one more chapter, that relating
to its {\it avatar} as a pulsar with extreme accuracy of pulsar timing
observations.  Such an incredible accuracy of pulse timings points to
a very interesting possibility. Any deformation of the neutron star,
even if it is extremely tiny, has the potential of leaving its imprints
on the pulses through introduction of tiny perturbations in the entire
moment of inertia (MI) tensor. Clearly, it will directly
affect the pulse timing. However, a general deformation of NS will
change the entire MI tensor, including its off-diagonal components. The
diagonal components of perturbed MI tensor will affect the pulse timings,
at the same time, the perturbed off-diagonal components will induce
wobbling of pulsar. Wobbling of pulsar (on top of any previously
present) will directly affect the profile of pulses as observed
on earth.  Thus observations of changes in pulse timings, along with
any accompanying changes in the pulse profile will contain information
about details of minute changes in the configuration of NS, e.g. density
perturbations inside the NS, or its overall deformations. 

This opens
up a new window of opportunity for exploring various phase transitions
occurring inside a pulsar core, through induced density fluctuations,
which may be observable as perturbations in the pulse timing as well as
its profile  \cite{Bagchi:2015tna,Srivastava:2017itj,Bagchi:2021etv}.
Such perturbations also naturally induce a rapidly changing quadrupole
moment of the star, thereby providing a new source of gravitational
wave emission \cite{Bagchi:2015tna}.  Another remarkable possibility
arises when we consider the effect of an external gravitational wave
on neutron star. With the possibility of detecting any minute changes
in its configuration through pulse observations, the neutron star has
the potential of performing as a Weber detector of gravitational wave
\cite{Das:2018kvy}.  The possibility of such {\it Weber} detectors,
spread out in space, with their signals (carrying imprints of any GWs)
monitored at earth, has tremendous potential, especially in allowing
us to re-visit GW events whose signal may have passed through earth in
past \cite{Biswal:2019szu}.

In this brief review we will focus on this particular aspect of NS physics,
namely the range of phenomena which can be probed utilizing the effect of 
changes in the configuration of NS using high precision measurements
of pulses coming from a pulsar. We will begin in Section 
\ref{sec:qcdphase} with a discussion of salient features
of the QCD phase diagram, specially the high baryon density regime. We will 
briefly discuss theoretical expectations of different phases in this regime. 
We will also connect with the experimental situation and discuss what parts 
of QCD phase diagram can be probed by the present and future relativistic
heavy-ion collision experiments, and which regimes may remain out of reach
of these terrestrial experiments, leading us towards cosmos, in particular
to neutron stars. Section \ref{sec:neutron_stars} will then be devoted to 
basic physics of neutron stars, including the superfluid phase in its 
interior as well as the possibility of exotic QCD phases in the inner core. 
There are excellent reviews on this subject (as well as on the subject matter 
of Section \ref{sec:qcdphase}, i.e. QCD phase diagram). We will only 
recollect essential parts of these discussions for self-completeness of the 
discussion here. Thus, we will also briefly discuss how recent gravitational 
wave detections have allowed the probe of NS interior properties. 

Section \ref{sec:pulsars} will be devoted to pulsars 
recalling the extreme accuracy 
of pulsar timing observations. We will also recall here the first (indirect) 
detection of gravitational waves (GWs) by pulsar observations, as well as 
ongoing attempts of pulsar timing arrays for detection of ultra low frequency 
GWs.  In Section \ref{sec:observ} we will discuss various 
proposals from the literature for
possible observational signatures of various phases in NS interiors.
Among these, glitches take a prominent role as well established signals for
the existence of superfluid phase in NS interior. We will discuss here 
difficulty of this explanation in accounting for relatively recent observations
of anti-glitches. We will also discuss various proposals for detection of
the exotic color superconducting phases of QCD in NS interior. 
Here we will also point to a new possibility where possibly the highest
observable baryon density phases could arise, that is in the matter
undergoing collapse to a black hole. For example, for a stellar mass
black hole, the Schwarzschild radius is about one third of the typical
neutron star radius. Thus it is possible that baryon densities in matter
collapsing to black hole may become about 20 times larger than that in
a neutron star (depending on the density profile, also the density
contrast will be smaller for a realistic, more massive, collapse to 
black hole). Certainly, it will 
only be transient, lasting only for tiny fractions of seconds, but still 
may allow observational signatures of novel QCD phase transition which 
can occur with typical local time scale of QCD, i.e. fm/c.

In Section \ref{sec:phase_transition} we will discuss the 
implications of  phase transitions
occurring inside a pulsar on its pulses. The consequences
of phase transition (for example from nuclear matter to QGP)
occurring in the core of a neutron star in terms of its effect
on the  moment of inertia have been discussed in the literature
with its observational implications on the spin rate of the neutron
star \cite{Glendenning:1997fy,Heiselberg:1998vh}. These discussions
primarily focused on the change in the equation of state during the
phase transition, and hence the main implications related to changes in
the diagonal components of the moment of inertia tensor affecting the
spin rate of the neutron star.  However, phase transitions necessarily
produce density fluctuations (as long as relevant correlation 
lengths remain smaller than the system size), which perturb entire moment 
of inertia tensor, including its off-diagonal components.  This was discussed  
by some of us in \cite{Bagchi:2015tna,Srivastava:2017itj}, pointing out
that phase transitions induced density fluctuations modify the entire
moment of inertia tensor of the pulsar which affects
pulse timings, but also induces modulations of the pulse profile.
The detailed modification of the pulses carries the information of
statistical nature  of density fluctuations, and hence the precise nature
of phase transition occurring inside the NS interior. 

In Section \ref{sec:pulse_modification}
we will discuss a special case in detail when the density fluctuations
are modelled in terms of random components of MI tensor added to the
unperturbed diagonal MI tensor of the neutron star \cite{Bagchi:2021etv}.
We will see here the effect on pulse timings as well as the nature of
modulations expected in the pulse profile. We will discuss that even
for very tiny density fluctuations, even if changes in pulsar timings remain
extremely small, pulse profile modification may become relatively
large. This is because while pulse timing changes will be proportional
to typical density fluctuation magnitude $\epsilon$, the pulse profile
modification will be proportional to $\epsilon/\eta^2$ where $\eta$
is the NS deformation parameter. ($\eta$ is typically very small $
\sim 10^{-8} - 10^{-4}$. This observation will play an important role in
later discussion when we discuss possible detection of external GWs using
NS deformations.) We will also briefly discuss how the same technique
can be used to detect other perturbations occurring in neutron stars,
e.g. collision with an asteroid. In Section \ref{sec:grav_waves} we will 
discuss how phase transitions occurring inside
NS may provide a new {\it high frequency} source of GWs through density
fluctuation induced rapidly changing quadrupole moment. 

Section \ref{sec:weber_detector} will
change the direction of discussion towards the effects of an external
gravitational wave on the neutron star configuration. Clearly, expected
deformations in NS will be extremely tiny. However, we will discuss how
it may be possible to detect even such tiny deformations utilizing the
impressive accuracy of pulsar timing observations, and in particular
possible changes in the pulse profile from induced wobbling of pulsar
(recalling discussion from the results of Section 
\ref{sec:pulse_modification} that even if pulse
timing changes remain very small, pulse profile modulations may become
relatively large). We will therefore conclude in this section that pulsars
will effectively act as {\it remotely stationed} Weber detectors whose GW
perturbed signals may be observable on earth \cite{Das:2018kvy}.  A very
important part of discussion here will involve the so called {\it ringing}
of the pulsar which will allow folding of very large number of pulses,
thereby tremendously increasing the signal to noise ratio, exactly what is
done in a Weber detector. We will make a point here on the importance of
the quantity the {\it Quality factor} {\bf Q} of the NS matter which will
directly determine the strength of the ringing effect. While much focus
has been there in the literature in calculating the shear viscosity to
entropy ratio of the QCD matter (in the QGP phase as well as the hadronic
phase), there is no such discussion for the Quality factor {\bf Q}.

This possibility of pulsars acting as Weber detectors, spread in cosmos
will lead to a truly remarkable possibility of re-vising past GW events.
We will discuss this in Section \ref{sec:revist}, how some GW
event (collisions of black holes, neutron stars, supernova explosions
etc.) from far away, whose signal may have passed through earth in past,
may become observable on earth again via observations of certain pulsars
which also get affected by the same GW source, and transmit the
perturbed pulses which are then later detected on earth
\cite{Biswal:2019szu}. Knowing the location of GW source and various
pulsar coordinates, it is possible to predict at what time in future
such GW events may become observable again through specific pulsars.
Even the GWs from the earliest detected supernova SN185, observed in AD
185, may be observable through observations of perturbed pulses of
specific pulsars, (in this case, via pulsars J0900-3144 and pulsar
J1858-2216 with perturbed pulsar signal arrival date reaching earth
during 2016-2049). The final Section \ref{sec:conclusions} will
present conclusions and future directions where we will discuss various
limitations of these proposals, as well as what specific efforts are 
needed to make these proposed techniques more effective.

\section{QCD Phase diagram}
\label{sec:qcdphase}

This section will discuss  salient features of the QCD phase diagram, 
and possible experimental probes of different regimes of the phase diagram.
The earliest known part of the {\it QCD phase diagram} relates to
the liquid/gas transition of nucleonic matter. Indeed, it was the
liquid drop model of the nucleus which led Lise Meitner to propose
theory of fission \cite{1939Natur.143..471M,Sime:1998}. Further structure
in this phase diagram emerged from astrophysics, with the notion of 
neutron superfluid in the cores of neutron stars 
\cite{Yang:1971oux}.  
Up to this stage, it would be more appropriate to call it the phase 
diagram of nucleonic (hadronic) matter. With the discovery of
asymptotic freedom, Quantum Chromo Dynamics (QCD) was established as the
fundamental theory of strong interactions, with quarks and gluons
being the fundamental constituents, and gluons being the mediators
of the color force among these constituents. Hadrons were composed of
quarks and the nuclear (hadronic) interactions arise as residual 
interaction from this fundamental color interaction. With the color
gauge group $SU(3)$, the QCD Lagrangian is

\begin{equation}
L_{QCD} = -\frac{1}{4} F_{\mu\nu}^a F^{a\mu\nu} + \sum_\alpha
\overline{\psi}_\alpha \left(i \gamma^\mu D_\mu - m_\alpha\right) \psi_\alpha
\end{equation}
where $\alpha = u, d, c, s, t, b$ is the flavor index for quarks.

The covariant derivative is
\begin{equation}
D_\mu \psi_\alpha = \left(\partial_\mu - i g_s T^a A_\mu^a\right) \psi_\alpha
\end{equation}

$g_s$ is the strong interaction coupling constant. $\psi_\alpha$ is in
the 3-dimensional fundamental representation of color $SU(3)$, 
with the generators $T^a$ = $\lambda^a/2$, $a = 1,..8$. 
$A_\mu^a$ are the 8 gluon fields and $\lambda^a$ are 
the Gell-Mann matrices.  The field strength tensor is

\begin{equation}
F_{\mu v}^a = \partial_\mu A_\nu^a - \partial_\nu A_\mu^a + g_s f^{abc}
A_\mu^b A_\nu^c
\end{equation}
 
where $f^{abc}$ are the antisymmetric structure constants for the
Lie algebra of SU(3). As isolated quarks are not seen, confinement
is an essential part of color interaction. Isolated objects
can only be color singlets. Running of the strong interaction
coupling constant $g_s$ with momentum transfer $Q$ exhibits the 
famous {\it asymptotic freedom}. Writing $g_s^2/4\pi \equiv \alpha_s$
we have,
\begin{equation}
\alpha_s\left(Q^2\right) = \frac {4\pi}{\left(11-2n_F/3\right)
\,\ln\, Q^2/\Lambda^2}\,,
\end{equation}
where $\Lambda$ is the QCD scale typically taken to be of 
order $200$ MeV. Thus color interaction  becomes weaker with large momentum 
transfer, equivalently at short distances. On the other hand, at large 
distances, interaction becomes very strong, which is consistent with the 
notion of color confinement.

With asymptotic freedom, it is but natural to expect that the behavior
of strongly interacting matter (system of hadrons) at ultra high
temperatures, or ultra-high densities, may show qualitatively different
behavior compared to the standard confining hadronic phase.  One would
expect in such extreme regimes, typical interaction between quarks
will be at very large energies/short distances, hence will become much
weaker. In the limiting case quarks and gluons should become almost
free, hence the notion of an almost ideal gas of quarks and gluons,
the quark-gluon plasma (QGP) phase of strongly interacting matter.
Extreme limits of temperature are only found in the very early stages
of the Universe, while very large baryon densities are expected to
arise in cores of neutron stars. These two systems thus provide natural
laboratories for the QGP phase of the QCD phase diagram. However, the
QGP phase in the early Universe  is primarily probed theoretically, with
no reasonably direct observable signals expected for the present stage
of the Universe. (This is with the present understanding that for very
small net baryon densities, the quark-hadron transition is a cross-over.
Some time back when the possibility of a first order transition was
also there, there were tantalizing possibilities of forming quark
nuggets which could be candidates for dark matter \cite{Witten:1984rs}.)
Situation is different with neutron stars which are directly accessible
to observations. Observational data available for masses  and sizes
of neutron stars put strong constraints on the equation of state of
the matter inside the neutron star. Detailed information about phases
in the interior of a pulsar is provided by observations of pulses,
in particular from the pulsar glitches \cite{Anderson:1975zze}.
A remarkable new probe of the neutron star interior in terms of {\it
tidal deformability} became available from the direct detection of
gravitational waves by LIGO/Virgo coming from binary neutron star (BNS)
merger events \cite{LIGOScientific:2017vwq}.

Study of QGP phase, and exploration of QCD phase diagram in general,
has acquired new dimensions with the arrival of relativistic
heavy-ion collider experiments (RHICE). Heavy ions (e.g.  Lead,
Gold, etc.) are accelerated to ultra-relativistic speeds and made to
collide, thereby creating a transient ultra hot and dense system of
strongly interacting matter. For high enough center of mass energy,
the temperatures are expected to be high enough for the creation of
a thermalized system of quarks and gluons, the so called quark-gluon
plasma (QGP) phase of QCD. These experiments have allowed unprecedented
control over the properties of strongly interacting matter created,
reaching temperatures and densities which were only available so far
in cosmos. Controlled experiments have given wealth of knowledge about
the system, from thermodynamic properties to transport coefficients,
and even allow to study QCD matter in strong external electromagnetic
field. 

The initial goal of these experiments was to find the QGP phase,
hence the drive for larger and larger centre of mass energy. Indeed,
from Super Proton Synchrotron (SPS) at CERN to Relativistic Heavy-Ion
Collider (RHIC) at BNL, USA, and then to Large Hadron Collider (LHC)
at CERN, center of mass energy per nucleon-nucleon pair has increased
from few tens of GeV at SPS to 200 GeV at RHIC, and 5.36 TeV at LHC for
nucleus-nucleus collisions. The temperatures reached at such energies
have certainly been high enough to convincingly demonstrate creation
of the QGP phase. However, increasing collision energy at the same
time leads to lower baryon density for the produced system. This is
because of asymptotic freedom of QCD, larger momentum transfer leads to
weaker interaction, thereby reducing possibilities of baryon stopping
in the produced system. The QGP system produced at ultra high energies,
therefore, resembles more closely the QGP phase of the early Universe
which also had very small baryon number density.

\begin{figure}
\centering
\includegraphics[width=0.8\linewidth]{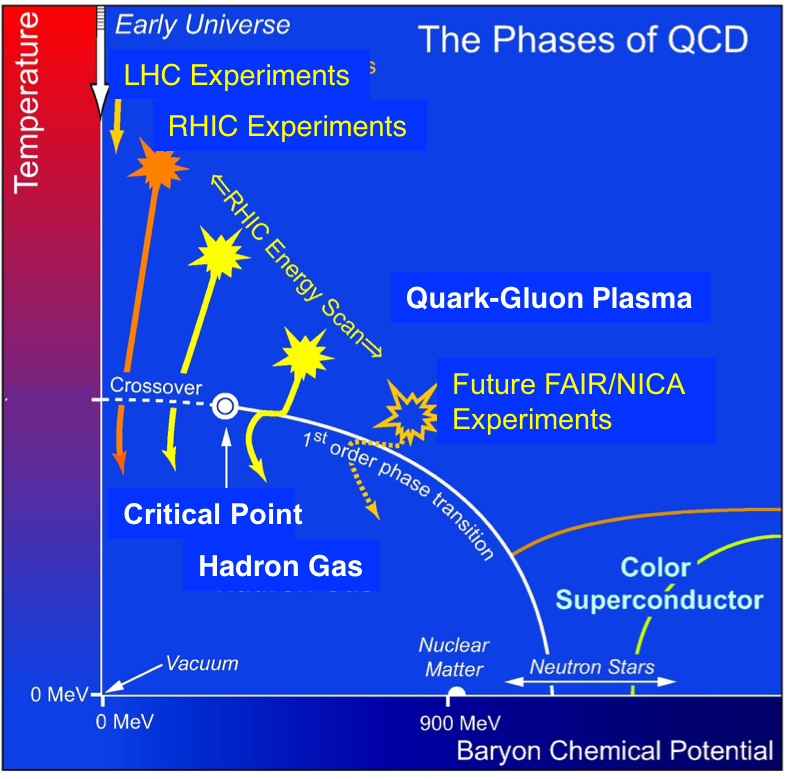}
\caption{QCD phase diagram in the $T,\mu$ plane,(from 
ref.\cite{Nayak:2020bjz})}
\label{fig:qcdphase}
\end{figure}

With these experiments at ultra high energy, creating a QGP system with
very small baryon densities, reaching a level of maturity, the attention
is now being focused on the other extreme condition, namely very high
baryon density regime of QCD matter. QCD matter is also expected to go
to QGP phase in this regime, but with different properties. While high
temperature low baryon density QGP teaches us about the early Universe,
the regime of very high baryon density QGP directly relates to the
interior of neutron stars. In fact, it is now realized that the QCD
phase diagram has extremely rich structure precisely in this ultra-high
baryon density regime. The situation is quite like the phase diagram of
water. While at high temperatures we have liquid and gas phases, at
ultra low temperatures and high densities there are numerous phases of ice
which appear. Possibilities of various phases of QCD are illustrated in
the QCD phase diagram in Fig. \ref{fig:qcdphase}. Here we show the phase
diagram in terms of two most important variables, temperature $T$ and
baryon chemical potential $\mu$ (representing baryon density). In
different situations, one can use other variables such as strangeness
chemical potential etc. For theoretical discussions it is also useful to
invoke another axis where quark masses can be varied. This allows for
the discussion of various possible phases, and critical points in the
phase diagram. We will confine our discussion to the standard 2-d phase
diagram as shown in Fig. \ref{fig:qcdphase}, in the $T,\mu$ plane.

The phase diagram in Fig. \ref{fig:qcdphase} shows several possible phases in 
different regions of temperature $T$ and baryon chemical potential $\mu_B$. 
The figure also shows where different regions of the phase diagram are 
expected to arise. It is useful to consider two different regimes in the 
entire phase diagram. One is for relatively low baryon density with baryon 
chemical potential values less than about 1 GeV, or ultra high
temperatures, and the other regime is for much higher values of $\mu_B$ 
and  relatively low temperatures.

\subsection{Low baryon density/ultra high temperature regime}

There are several reasons for separating this regime. First, there are only
two phases to discuss here. The hadron gas phase appears at low values of
$\mu_B$ and relatively low temperatures. Boundary of this phase is denoted
by the white curve, part solid, joined to  dashed white curve, the joining
point denoted as the {\it critical point}. The solid white curve denotes
line of first order transition, which ends at the critical end point
where the transition is second order (a continuous phase transition).
The dashed line denotes a cross-over transition, which is not a 
proper phase transition (in the sense that the partition function remains
analytic across this boundary).  The entire region of the phase
diagram outside this boundary, and above the solid yellow line is the
second phase, the quark-gluon phase. Importantly, all the present and future
planned heavy-ion collision experiments probe this part of the phase diagram
only. First system to discuss here will be the early Universe denoted by
the solid white arrow (almost) on the $\mu_B = 0$ line. The temperatures
reached in the Universe in its earliest stages can reach close to the 
Planck temperature ($\sim 10^{19}$ GeV), depending on the inflationary models.
The Universe, expanding and cooling from these earliest stages, reaches
the quark-hadron transition temperature (where the dashed line 
intersects the $T$ axis). 

Lattice calculations \cite{Borsanyi:2020fev, Endrodi:2011gv}
are under good control 
for $\mu_B = 0$ case, and give the value of quark-hadron transition 
temperature to be $T_c = 156 \pm 0.6 $ MeV. This value of temperature is 
expected to be reached in the early Universe when its age is few tens 
microseconds. Upto that time, since its birth, the Universe was filled with 
a plasma of quarks, gluons, as well 
as other elementary particles, e.g. leptons, photons. After this time, the 
Universe undergoes quark-hadron transition, (which is expected to be a smooth 
cross over from lattice calculations) when quarks and gluons get confined into
system of hadrons. As we mentioned, with smooth cross-over the earlier
much discussed quark nugget scenario \cite{Witten:1984rs} is no more 
possible.  Though, there have been alternate proposals for the formation
of quark nuggets which do not depend on the nature of quark-hadron
phase transition \cite{Ge:2017idw,Atreya:2014sca}.

The ultra-relativistic collisions at RHIC and LHC, with centre
of mass energies ranging from 200 GeV to 5.36 TeV (per nucleon-nucleon
collision) lead to a fireball with temperatures well above the
quark-hadron transition temperature with relatively low values of
$\mu_B$ ($\sim 50$ MeV for 200 GeV collision energy). As we mentioned,
increasing collision energy leads to smaller values of $\mu_B$ due to
asymptotic freedom of QCD. For these large values of collision energies,
confining forces become irrelevant, so basically the collision is
between quarks and gluons (partons) contained in each colliding nuclei. 
For very large collision energies, partons scatter little, basically
the colliding nuclei go through each other. Even with smaller scattering
at large energies, the energy available for secondary parton production
monotonically increases with collision energies, leading to higher
temperatures. But with smaller scattering, proton stopping is less, hence
net baryon number of the produced QGP system is monotonically decreasing
function of collision energy. (Strong interactions conserve baryon number,
so baryon density in the central fireball can only come from the stopping 
of initial valance quarks of the colliding nuclei.)

It is believed that at these collision energies, the expanding QGP
system undergoes quark-hadron transition through the dashed white line
in Fig. \ref{fig:qcdphase}, that is the cross-over line. This is with
several estimates of the location of the critical point. Though it has
not been possible to have a good control over lattice calculations for
non-zero $\mu_B$ (due to the so called {\it fermion sign problem}),
several techniques have been devised to extend lattice calculations for
non-zero $\mu_B$. Recent estimates suggest the location of the critical
point to be about ($T \lesssim  132 $ MeV, $\mu_B \gtrsim 485$ MeV)
\cite{Philipsen:2021qji}.  It should be mentioned that there are strong
theoretical arguments for the existence of a first order transition line
at non-zero values of $\mu_B$. These are typically based on various
effective field theory models. Combined with the solid knowledge of
cross-over transition at $\mu_B = 0$, it automatically follows that the
first order line has to end somewhere at a critical end point.

Physics near the critical point (in the critical regime) is
dominated by critical fluctuations showing universal properties.
Experimental evidence for such fluctuations will give deep insight
into this very important part of the QCD phase diagram. With this aim, 
the beam energy scan program was designed at RHIC, with collision
energies as low as 7.7 GeV (thereby creating the QGP system with
much higher values of $\mu_B$, hopefully evolving through the
critical regime).

One more region deserves mention in this part of the phase diagram,
denoted as {\it Nuclear matter} near $\mu_B = 900 $ MeV. There is a
short line of first order transition (not shown in Fig.
\ref{fig:qcdphase}) which corresponds to liquid-gas transition of
nuclear matter. What is more interesting is that at further larger
values of $\mu_B$ but still within the hadronic phase (so not crossing
solid white curve), there is the possibility of nucleonic superfluid
phase. It is precisely this superfluid phase, and associated superfluid
vortices which are supposed to play crucial role in neutrons stars,
especially in relation to the phenomenon of pulsar glitches. We will
have details discussion of this phase in the next section.

\subsection{Quarkyonic Matter Phase}

Arguments based on large $N_c$ (number of colors) expansion suggest
possibility of a novel phase of confined matter at baryon densities much 
higher than the QCD scale, where one would have expected deconfinement to 
set in.  Even though the matter here consists of mesons, baryons, and 
glueballs, typical interactions are at the quark level, being at very high 
energy scales \cite{McLerran:2007qj}. This hypothetical phase is termed as 
the {\it quarkyonic phase} and it may arise at moderately large baryon 
densities. (This phase is not shown in the QCD phase diagram in Fig.1).
In the 't Hooft limit of large $N_c$, with $g^2 N_c$ fixed ($g$ being
the gauge coupling), quark loops are suppressed (for fixed number of
flavors) being of order $N_c$, compared to gluon loops which will be
of order $N_c^2$. With quark loops suppressed, the color confinement
persists. Thus, the matter in this phase will consist of mesons, baryons and
glueballs, even at very high baryon density compared to the QCD scale. It
differs from the standard confining phase as, due to very high density,
typical interactions will be at very high energy, hence will be at
the level of quarks. Quarks far from the Fermi surface will be
in the perturbative regime of QCD, though near the fermi surface, degrees 
of freedom will be that of confined hadrons. 
Chiral symmetry breaking in this phase has non-trivial
behavior, as the sigma meson would be a bound state of a quark slightly
above the Fermi energy with the anti-quark being a hole slightly below the
Fermi energy, leading to non-zero net momentum for the bound state. The sigma
meson condensate having net momentum, this condensate will break rotational 
and translational invariance, \cite{McLerran:2020rnw, Kojo:2011cn}.
It has been argued that the quarkyonic  phase can lead to
very hard equation of state, with the sound speed exceeding the value
of $1/\sqrt{3}$ (in natural units with $c = 1$), at density which is 
3-4 times larger than the nuclear density \cite{McLerran:2020rnw}.

\subsection{Very high baryon density, low temperature regime, 
color superconductivity}

Discussions relating to this part of the phase diagram have intensified 
relatively recently. There have been insightful exchanges of ideas
between the usual condensed matter physics and this area, which is 
appropriately being called as {\it condensed matter physics of strongly 
interacting matter}. The phases in the part of the phase diagram are
typically termed as {\it color superconducting phases}. Note here
that for this part of the QCD phase diagram, we are not including the
so called {\it high baryon density QGP phase} which will be the part above
the solid yellow line with very large values of $\mu_B$. Though there
will be important differences in the physical properties of QGP with
high baryon density from QGP with very low baryon density 
(and high temperature), they are the same  thermodynamic phase,
with no phase boundary expected between these two regimes. That is why
we included high $\mu_B$ QGP phase in the preceding subsection.
Discussion below is primary taken from 
refs.\cite{Rajagopal:2000wf,Alford:2007xm,Rajagopal:2001ngu,Alford:2000sx}
which can be consulted for details.

\begin{figure}
\centering
\includegraphics[width=0.5\linewidth]{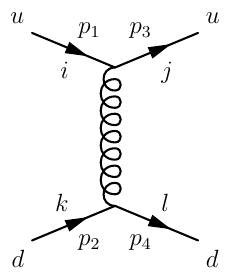}
\caption{Quark-quark scattering with one-gluon exchange}
\label{fig:qqscatt}
\end{figure}

High $\mu_B$ with low temperature changes the physics qualitatively
giving rise to new thermodynamic phases.  The basic physics of these 
phases lies in the realization that at very high values of
$\mu_B$, and relatively low temperatures, the physics will be governed
by low energy excitations near the Fermi level. Standard BCS 
Superconductivity has taught us that any attractive interaction between
fermions at the Fermi level destabilizes the Fermi level, forming
Cooper pairs of fermions. First we note that for the relevant
values of chemical potential, only relevant quark flavors are
$u,d,$ and $s$ quarks. (We do not discuss here heavier quarks as
there is no physical system known where one could reach quark chemical 
potential high enough so that heavier quarks can play any important 
role for these condensed phases of QCD.) At ultra high chemical potential,
of order 500 MeV for quarks, asymptotic freedom will make perturbative
calculations reliable, especially for $u$ and $d$ quarks.
Consider scattering of light quarks with one gluon exchange as shown
in Fig. \ref{fig:qqscatt}.

Here, $i,j,k,l = 1,2,3$ (or $r,g,b$) refer to colors of 
$u$ and $d$ quarks.
Now we note that the amplitude for this process is the same as that 
for the QED process $e^- \mu^- \rightarrow e^- \mu^-$ with the 
replacement of the electromagnetic coupling $e$ by the strong coupling 
$g_s$ (i.e. replacing $\alpha$ by $\alpha_s$), and inclusion of the
{\it color factor},
\begin{equation}
C_F(ik \rightarrow jl) = \frac{1}{4} \lambda^a_{ij} \lambda^a_{kl}
\end{equation}
quark-quark can combine in two ways, with $3 \times 3 = 3^* + 6$.
The color factors for the two representations are,
$C_F(3^*) = -2/3$  and  $C_F(6) = 1/3$.

As the EM potential between $e^-$ and $\mu^-$ is repulsive, the qq 
potential is attractive in the 3* channel, and repulsive in the 6 channel.
BCS pairing of quarks in the $3^*$ channel leads to the {\it
color superconducting phase} (as the Cooper pairs here are colored). 
Depending on the relative masses of the quarks forming the condensate, 
different phases are possible.

\vskip .1in
\textbf{Color-Flavor Locked (CFL) phase:}
\vskip .1in

With color antisymmetric ($3^*$ channel), spin antisymmetric
(for $^1S_0$ pairing), we conclude that the condensate should
have flavor antisymmetric.  For very high chemical potential
it may be reasonable to treat all three quarks ($u,d,s$) as
massless. Condensate for this most symmetric case, (with
$^1S_0$ spin pairing) will have the following structure
\begin{equation}
\langle q_i^\alpha q_j^\beta \rangle \sim \Delta_{CFL} (\delta^\alpha_i 
	\delta^\beta_j - \delta^\alpha_j \delta^\beta_i) = \Delta_{CFL}
	\epsilon^{\alpha\beta n} \epsilon_{ijn}
\end{equation}
where $\alpha\beta$ are flavor indices and $i j$ are color indices.
Note that the condensate is invariant under equal and opposite
rotations of color and (vector) flavor. Hence, it  leads to the
following spontaneous symmetry breaking
\begin{align}
SU(3)_{Color} &\times SU(3)_L \times SU(3)_R \times U(1)_B \nonumber \\
&\rightarrow SU(3)_{C+L+R} \times Z_2 \,.
\end{align}

Thus color symmetry of QCD $SU(3)_C$ is  spontaneously broken. Three
flavor Chiral symmetry is also spontaneously broken. Interestingly, QGP
phase restores the chiral symmetry (at least for $\mu_B = 0 $ case).
Here, at large $\mu_B$ also, we expect chiral symmetry restoration in
the {\it high density QGP phase} which occurs at higher temperatures.
But at low temperatures, high $\mu_B$ quark-gluon phase breaks the
chiral symmetry spontaneously. 
Spectrum of pseudoscalar mesons in this chiral symmetry
broken phase is very different from the conventional, low density case.
For example, the kaon is lighter than the pions, with mass possibly in
the range of 5 to 20 MeV at quark chemical potential of about 400 MeV
\cite{Rajagopal:2000wf}.

Spontaneous breaking of $SU(3)$ color symmetry implies that all gluons
become massive, (so there are no long range color forces even in this
deconfined phase), hence the name {\it Color superconductor}. As the
condensate here pairs quarks of different charges, the $U(1)$ symmetry of
electromagnetism is spontaneously broken. However, it turns out that a
linear combination  of EM $U(1)$, and one of the gluons (corresponding
to the generator $T^8$ in Eq.(2)), remains massless,
with its orthogonal combination becoming massive. (This is somewhat
similar to the electroweak symmetry breaking where a combination of weak
$SU(2)$ and the hypercharge $U(1)$ leaves the condensate invariant, and
remains a valid gauge symmetry in the broken phase, which becomes the
standard $U(1)$ symmetry of electromagnetism.) The surviving massless gauge
boson in the CFL case thus plays the role of {\it rotated} photon in the
broken phase. At the expected baryon densities, the gluons are strongly
coupled. With weak coupling of electromagnetism, this {\it rotated
photon} dominantly consists of the photon from electromagnetism, with
only a small admixture of the $T^8$ gluon, see \cite{Rajagopal:2000wf} 
for details.

\vskip .1in
\textbf{2SC Pairing:}
\vskip .1in

In the above, we discussed the symmetry in QCD with
three massless degenerate flavors ($u$, $d$ and $s$ quarks). Depending
on the realistic masses of individual quark flavors and the relative mass
difference across the flavors, various other exciting phases may appear
in the neutron star interior in the high baryon chemical potential
region. For two light $u$ and $d$ quarks with large quark chemical
potential, but not large compared to strange quark mass, the two flavor color
superconducting ($2SC$) phase is expected (See \cite{Rajagopal:2000wf}
for an extensive review of various QCD phases and the possibility of
their appearance in neutron star core.).  The resulting ($u,d$ quarks) 
Cooper-pair condensate
$<\epsilon_{ij3} \epsilon^{\alpha \beta} q_\alpha^i q_\beta^j>$ creates
a gap at the Fermi surfaces of quarks with two (out of three)
colors and breaks the color symmetry $SU(3)_C$ to $SU(2)_C$, giving mass
to five (out of eight) gluons. Here, $i,j = 1,2 $ (or r,g) refer to
colors of $\alpha,\beta = 1,2$ (or u, d) quarks. In the 2SC phase, no
global symmetries are broken, unlike in the CFL phase.

\vskip .1in
\textbf{Crystalline Color Superconducting  phase:}
\vskip .1in

The crystalline color superconducting phase is
similar to the so-called Larkin–Ovchinnikov–Fulde–Ferrell (LOFF) phase,
first explored by Larkin, Ovchinnikov \cite{Larkin:1964wok}, and Fulde,
Ferrell \cite{Fulde:1964zz}, independently, in the context of
electron superconductivity, where this phase occurs when the participating
electrons forming the Cooper pairs have nonzero net momentum. In the context
of QCD, the LOFF phase at very high baryon density may arise when the
quark chemical potential is not very large compared to the strange quark
mass. In this case, the relative mass difference between $u,d$ quark and
$s$ quark becomes significant. The Cooper pairing of different quarks with
different Fermi momenta leads to spatial modulation of the order
parameter. This results in the spontaneous breaking of translations and
rotation symmetries, leading to superconducting gaps that vary
periodically in a crystalline pattern. As an observable consequence,
there was a suggestion \cite{Alford:2000ze} that if the density of
quark matter core favors the formation of crystalline color
superconductivity, the standard pinning-unpinning mechanism of
superfluid vortices \cite{Anderson:1975zze} can be applicable here, and
the core can also contribute to the glitch phenomena (see the article
\cite{Alford:2000ze} and the review \cite{Rajagopal:2000wf} for
more details).

We are only giving brief summary of various phases here, just to give
an idea of the richness of the QCD phase diagram in different regimes.
There are excellent references available in literature, we have listed
some here in refs. \cite{Rajagopal:2000wf, Alford:2007xm,Rajagopal:2001ngu, 
Alford:2000sx}. We end this section with brief discussion of the important 
subject of chiral symmetry in QCD.

\subsection{Chiral Symmetry in QCD}

We here briefly recall the notion of chiral symmetry in QCD. For
details, literature can be consulted, e.g.  ref. \cite{Hosaka:2001ux}.
Note that we have been using the term {\it quark-hadron transition} in
the above discussion. This term can have two meanings. One is the
confinement-deconfinement (C-D) transition where a system of hadrons,
(with quarks/gluons being confined inside hadrons) makes a transition to
the deconfined phase of a plasma of quarks and gluons. The other meaning
can refer to the chiral transition. For 2 massless flavors, QCD
Lagrangian has exact $SU(2)_L \times SU(2)_R$ symmetry, called as
2-flavor chiral symmetry, which corresponds to independent
transformations of left and right components of $u$ and $d$ quarks.
Hadron spectrum does not show any such doubling of mass spectrum, but it
does show multiplet structure of $SU(2)_{isospin}$. This leads to the
conclusion that the chiral symmetry $SU(2)_L \times SU(2)_R$ is
spontaneously broken to the diagonal subgroup $SU(2)_{isospin}$, with
pions as the Goldstone bosons. For three massless flavors, all $SU(2)$
groups should be replaced by $SU(3)$ groups, leading to 3-flavor chiral
symmetry breaking. Of course, quarks are not massless, this leads to
explicit breaking of chiral symmetry, leading to small masses for the
Goldstone bosons, and contributing to mass differences within the
multiplets. The explicit symmetry breaking being relatively small,
especially for the 2-flavor case, the notion of chiral symmetry in QCD
has been immensely useful, especially in constructing effective field
theory models which capture physics at low energy scale. Chiral sigma
models, Nambu-Jona-Lasinio (NJL) models etc. are all bases on the notion
of chiral symmetry and have been the only tools for theoretical discussions
of QCD phase diagram at high baryon densities.

It is believed that chiral symmetry transition and confinement-deconfinement
(C-D) transition are the same. Lattice results support this idea
(though at times there are differing results also). Conceptually,
these two transitions are completely different. Indeed, different
order parameters characterize these two transitions, with expectation
value of the Polyakov loop characterizing the C-D phase, and the
${\bar \psi}\psi$ condensate charactering the chiral transition
($\psi$ being the quark field). We will have discussion of some of these
(especially the Polyakov loop condensate) later in Section 
\ref{sec:phase_transition}.
The conceptual difference between these two transitions becomes clear when we 
consider high $\mu_B$ phases, in particular the color superconducting 
phases. These phases appear in the regime where inter-quark separation
is so small that confining forces are irrelevant. This is like a plasma
of deconfined quarks and gluons, but not the usual QGP phase
as here thermal effects are insignificant compared to the quantum
statistics effects. Thus, we have chiral symmetry breaking here (as
discussed above for the CFL phase), even though we have a system of
deconfined quarks and gluons. Note that although we have massive gluons
here, there is no relevance  of color singlet objects  here. 

\section{Neutron stars}
\label{sec:neutron_stars}

This section  will discuss basic physics of neutron stars, including
the superfluid phase in its interior as well as the possibility of
exotic QCD phases in the inner core. We will also briefly discuss how
the recent gravitational wave detections have allowed to probe  NS
interior properties. 

{\bf Basics of a neutron star:} The existence of neutron stars as one of
the possible end states of massive stars was predicted by Walter Baade
and Fritz Zwicky in 1933 \cite{Baade:1934onh}, a long time before
the discovery of the pulsating neutron star ({\it little green man})
in late 1967 \cite{Hewish:1968bj}. Neutron stars are the remnants of the
supernova explosion of supergiant stars of mass in the range $10~\text{ M}
_{\odot}-20~\text{M}_{\odot}$.(see 
\cite{2004hpa..book.....L} for neutron star basics and 
\cite{2019gwa..book.....A} for the role of pulsars in gravitational-wave 
astronomy.) During its formation, gravity squeezes the
matter to achieve extremely high baryon density. Gravitational collapse is
counterbalanced by the neutron degeneracy pressure (along with repulsive
nuclear forces) leading to the formation of a stable neutron star.
The radius (R) of a neutron star lies in the range (10 - 14) km with mass
in the range $~\text{1.1 M} _{\odot}-2.1~\text{M}_{\odot}$.  Thus the
average mass density is of the order of nuclear matter's saturation
density $\rho_0 = 2.8 \times 10^{14}~\text{g cm}^{-3}$. The discovery
of the neutron star put the discussions of such compact astrophysical
objects on solid footing, with the realization that it provides the
opportunity for testing exciting physics at a high baryon density regime
that cannot be tested otherwise by terrestrial experiments.  However,
with the opportunity, new theoretical challenges arise, of understanding
the properties of dense interior materials, the inner core in particular,
and relating these properties to various testable observables.

There have been continuous efforts towards a theoretical understanding
of the internal structure of neutron stars keeping in view its
astrophysical implications and observational constraints.  These
include the mass-radius relations, the moments of inertia (MI) of
the star (including the fractional contribution of the solid crust
and superfluid/superconductor components), the extent of rigidity of
the outer crust, and the deformation from the sphericity etc. The above
quantities, in turn, determine the frequency of pulsar's free precession,
glitch size in the context of various pulsar glitch models (crustquake,
superfluid-vortex model etc.), or the feasibility of GW emission etc.

\begin{figure}[t]
\begin{center}
\includegraphics[width=0.9\linewidth]{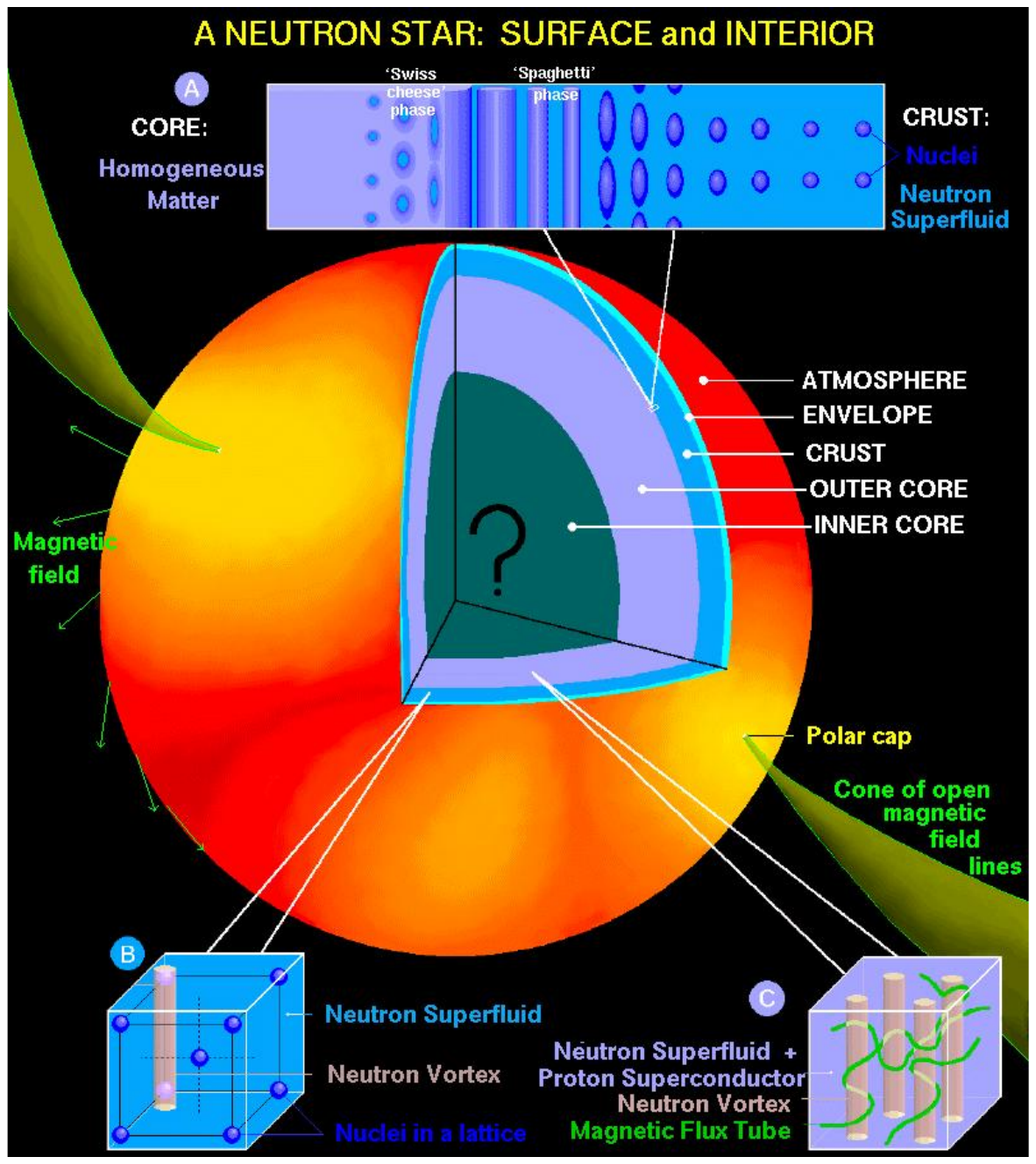}
\end{center} 
\caption{The expected internal structure of a neutron star. 
Credit : Dany P Page (https://phys.org/news/2015-09-neutron-star.html).}
\label{fig:structure}
\end{figure}

The standard approach of constructing a model of neutron star's internal
structure is to implement hydrostatic equilibrium of a gravitating fluid
system, resulting in the well known Tolman-Oppenheimer-Volkoff (TOV)
equations \cite{Tolman:1939jz,Oppenheimer:1939ne} (in the units with $c=1=G$),
\begin{eqnarray} \label{eq:tov1}
\frac{dP(r)}{dr} &=& - \frac{(P +\rho)(M + 4\pi r^3 P)}{r(r-2M)}; \\
\frac{dM(r)}{dr} &=& 4\pi \rho(r) r^2. \label{eq:tov2}
\end{eqnarray}
The solutions of the above set of equations provide the density profile
$\rho(r)$, including the mass and the radius of the neutron star,
provided the equation of state (EOS) $P= P(\rho)$ and the central
density $\rho(0)$ are supplied. The above procedure thus produces a
neutron star structure model for a fixed value of central density, which
finally determines various physical parameters, such as the mass,
radius, moment of inertia, etc., of the star. However, the main
challenge lies in providing the EOS, the most critical input for solving
the Eqs. (\ref{eq:tov1}) and (\ref{eq:tov2}). Although the EOS of
the outer crust and, to some extent, the inner crust are known
\cite{Chamel:2008ca}, the first principle calculations of the many-body
QCD interaction relevant to the inner core are unknown. Thus, one has to
resort to phenomenological models for extracting the EOS for the core
region. One can then use the experimental measured values of various
physical parameters to test the predictions based on constructed stellar
models. For most practical calculations, one uses a set of canonical
values $M = 1.4~\text{M} _{\odot}$, R = 10 km, and MI
$=10^{45}~\text{g~cm}^2$. For the above set of values, and from the
observations of various pulsar events, such as glitches (to be discussed
later), one broadly takes the neutron star's internal structure as shown
in Fig. \ref{fig:structure}. (See Refs.
\cite{Link:1999ca, Lattimer:2000nx, Ozel:2016oaf, Madsen:1999ci,
Jha:2007ej, Kunjipurayil:2022zah} for various models of neutron star
structure.) It it worth mentioning here the discovery of the most massive
neutron star \cite{NANOGrav:2019jur}, namely, the pulsar J0740+6620 of
mass, $ \text{M} \simeq 2.14~\text{M}_\odot$ using the Green Bank
Telescope and combining the data from the North American Nanohertz
Observatory for Gravitational Waves (NANOGrav) puts a strong
constraint on the neutron star interior EoS. In this context, see the
latest review \cite{Koehn:2024set} for an overview on constraining the
equation of state of neutron-rich dense matter.

Observation of glitches in the rotation of young pulsars indicates a solid crust 
containing $\ge 1.4 \%$ of total MI \cite{Link:1999ca}, the outer layer of which 
consists of crystalline solid iron nuclei and a sea of degenerate electrons with 
mass density $\rho \simeq 10^{6}~
\text{gcm}^{-3}$. Going deeper into the inner crust region, as the density
increases from $\rho \simeq 10^{11}~\text{g cm}^{-3}$ onward, protons
and electrons start combining to form neutrons for creating neutron-rich
nuclei. Eventually, it becomes energetically favorable for the neutrons
dripping out of the nuclei and forming a sea of free (unbounded)
neutrons. A few hundred meters thick inner crust region with the density
ranging from approximately  $10^{11}~\text{g cm}^{-3}$ to $10^{14}~
\text{g cm}^{-3}$ plays a crucial role in the theory of glitches. There
is a strong belief, supported by glitch observations, that a significant
fraction of free neutrons in this region (see Fig. \ref{fig:structure})
are in a $^1S_0$ neutron superfluid state.  Further deeper, below the
inner crust region, a more favorable $^3P_2$ neutron superfluid state
is believed to co-exist with a $^1S_0$ proton superconductor. 

The composition of the inner core is highly speculative. 
The states of matter at 
the high pressures in the deep interior can form various hadronic states, such 
as hyperons, condensed-pions etc. The central region of the neutron star can 
also accommodate different QCD phases, such as QGP, CFL, 2SC, etc. The core 
centre can become quite mysterious for  massive neutron stars. One needs 
well defined signatures that can be tested to confirm possibilities of any such
phases which remains an important theoretical challenge.

We will be discussing later various proposals and observational
techniques for probing physics of pulsar interior. But, it seems most
appropriate to close this section with a discussion of a truly exciting
recent development towards probing neutron star interior. This is the
direct detection of gravitational waves by LIGO/Virgo which, for GWs
originating from a BNS merger allows to put observational constraints
on deformation of neutron stars during last stages of coalescence,
thereby directly probing the state of matter inside the neutron star.

\vskip .1in
{\bf Probing NS interior with direct detection of gravitational waves 
from BNS merger}
\vskip .1in

Possibly the most important advancement in experimental General Relativity
in recent years has been the direct detection of gravitational waves
by LIGO and Virgo coming from distant events of binary black hole mergers 
and subsequently, binary neutron star mergers. Black hole merger
events allow the possibility of directly probing the dynamics of 
intense gravity regime of near horizon regions of black holes by analysis
of the GW waveform corresponding to the near coalescence regime of black 
holes. Similarly, for binary neutron star merger events, GW waveform contains
information about strong tidal deformations towards the end of spiral-in 
of the neutron stars, when separation between the two neutron stars
approaches neutron star size. In such regime, the coalescence is
accelerated by a quadrupolar deformation of NS by the tidal field of
the companion NS. 

Indeed, GW waveform observations have been used 
\cite{LIGOScientific:2017vwq} to  
put constraints on  the dimensionless tidal deformability $\Lambda
= (2/3) k_2 R^5/M^5$ (in gravitational units with G = c = 1)
where $k_2$ is called the second Love number. Value of $k_2$ and hence the
tidal deformability depends on the equation of state of NS matter, 
with value of $k_2$ typically in the range of  0.05 - 0.15.
($\Lambda$ is related to the tidal deformability $\lambda$ discussed 
later in Sec. \ref{sec:weber_detector}, by $\Lambda = \lambda/M^5$.) 
Tidal deformation of the neutron stars 
accelerates the coalescence.
(Interestingly, black holes are expected to have $k_2 = 0$, so this
effect will be absent for binary black hole merger events.) For GW170817
event, the probability density for the tidal deformability parameters of
the two neutron stars were inferred from the detected signal by assuming
the same equation of state for both the neutron stars with the
quadrupolar deformation of a Tolman-Oppenheimer-Volkoff solution
\cite{Hinderer:2009ca}. The constraints on the tidal deformabilities
disfavored equation of states which predict less compact stars.  GW
observations thus provide a completely independent probe of the equation
of state of NS matter, which is usually constrained by mass-radius
relationship of neutron stars. There have also been investigations of
the effects of presence of deconfined QCD phase in the cores of merging
neutron stars on the gravitational waveform \cite{Bauswein:2018bma,
Most:2018eaw}.

Direct observation of source identification
for BNS mergers by resulting electromagnetic radiation (which would be absent 
for black hole mergers) has started the new chapter of multi-messenger
astronomy in the exploration of cosmos. With a range of observations, from
gravitational waves, to electromagnetic radiation in a range of energies,
along with the possibility of neutrino bursts from such BNS merger events
will jointly give a powerful probe of the structure and property of neutron
stars.

\section{Pulsars}
\label{sec:pulsars}

This section will be devoted to pulsars recalling the extreme accuracy of 
pulsar timing observations. We will also briefly discuss here the first 
(indirect) detection of gravitational waves (GWs) by pulsar observations, as 
well as ongoing attempts of pulsar timing arrays for detection of ultra low 
frequency GWs.

\vskip .1in
{\bf Pulsars Timing:} 
\vskip .1in

The atomic clock-like stability of pulsar rotation
period allows one, through monitoring of the pulsar rotation, to study a rich 
variety of phenomenon affecting the propagation of pulses while reaching 
earth. Most applications of pulsars involve a powerful technique known as the 
pulsar timing. Measurement of a sequence of time of arrival (ToA) of pulses 
over intervals ranging from hours to decades is the basis of pulsar timing
\cite{Manchester:2017ykr, 2004hpa..book.....L, 
2019gwa..book.....A}. These ToAs are first transferred, normally
to the solar-system barycentre, to remove the effects of rotation and
orbital motion of the Earth. The amount of useful information that can
be extracted critically depends on the precision at which the pulse
arrival times are measured.  In order to understand the pulsar timing,
we will take the example of isolated pulsars and describe the rotation in
the pulsar's comoving frame. 

For an isolated pulsar, one can express
the spin frequency $\nu$ (or time period $P$) in a Taylor series about
some reference epoch $t_0$ \cite{Manchester:2017ykr}
\begin{equation} \label{eq:toa}
\nu (t) = \nu_0 + \dot{\nu} (t-t_0) + \ddot {\nu} \frac{(t - t_0)^2}{2} + ...
\end{equation}
where $\nu_0 = \nu(t_0)$ is the pulsar's spin frequency at $t_0$ and
$\dot{\nu}, \ddot{\nu}, ...$  are the higher order time derivatives
of $\nu$ to be evaluated at $t_0$. These parameters are associated with
some physical process, knowledge of which provides valuable information
about the underlying process. For a normal (i.e.,
rotation-powered) pulsar, the period $P$ ($\sim \text{0.3 s - 3 s}$)
and its first derivative $\dot{P}$ ($\sim 10^{-15}~\text{s~s}^{-1}$) are
observed with high accuracy through timing measurements. These parameters
capture the spin-down history of the isolated pulsar. On the other hand,
the millisecond pulsars (MSP) have the most exotic applications, including
the detection of low-frequency gravitational waves (see below) because
of their extreme stability in their periodicity ($P \sim 3$ ms with
$\dot P \sim 10^{-20}~\text{s~s}^{-1}$) compared to the normal pulsar.

\vskip .1in
{\bf Indirect detection of GWs:}
\vskip .1in

\begin{figure}[t]
\begin{center}
\includegraphics[width=0.9\linewidth]{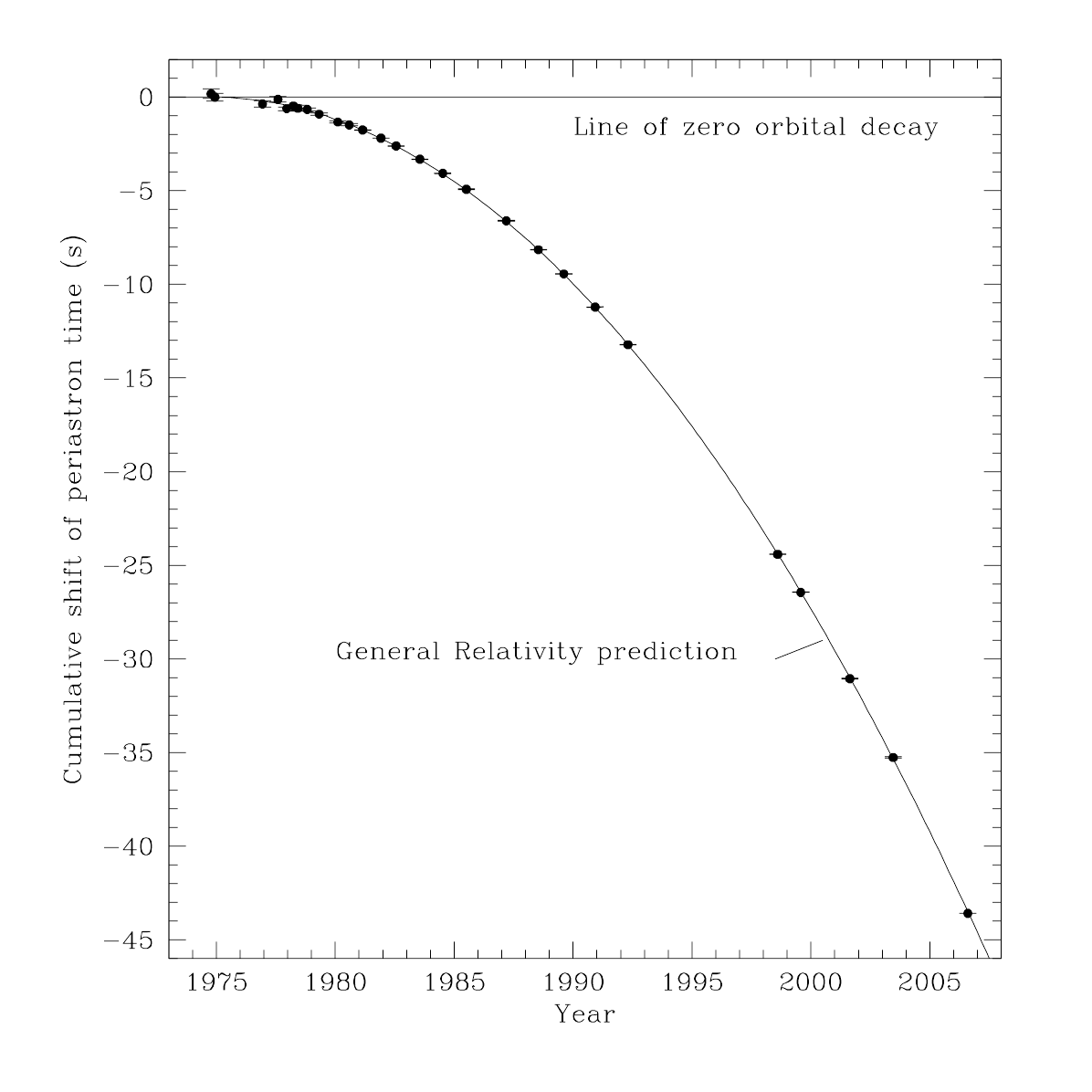}
\end{center} 
\caption{The shift in the periastron passage of the binary 
pulsar PSR B1913+16 with time, caused by gravitational radiation.
Figure taken from Ref. \cite{Weisberg:2010zz}}
\label{fig:orbital-decay}
\end{figure}

The celebrated discovery of the pulsar PSR B1913+16 in binary star
system by Russell Hulse and Joseph Taylor in 1974 \cite{Hulse:1974eb}
using the data from the Arecibo radio telescope opened up the
possibilities for the study of relativistic gravity in moderately
strong field regime.  The above discovery provided the first
indirect quantitative confirmation test in favor of the existence of
gravitational waves within the framework of Einstein's theory of gravity
\cite{Weisberg:1981mt,Taylor:1982zz,Taylor:1989sw}. The Hulse–Taylor
pulsar (PSR B1913+16) is a binary star system composed of a pulsar
of mass $\simeq 1.44~\text{M}_\odot$ and the silent companion neutron
star of mass $\simeq 1.39~\text{M}_\odot$ \cite{Weisberg:2010zz}, moving
around in elliptical orbits about their center of mass. As per Einstein's
theory of gravity, the orbital period of this binary system is expected
to decay with time. The heartening agreement between the observed data
with the theoretical prediction (see Fig. \ref{fig:orbital-decay}) not only
provided conclusive evidence for the existence of gravitational waves;
it laid the foundation for the belief of the possibility of {\it direct}
detection of GWs.  The remarkable first-ever direct detection of GWs in
2015 by LIGO \cite{LIGOScientific:2016aoc}, arising from a binary black
hole merger, fulfills that belief and opens a new era in gravitational
wave astronomy. Since then, there have been quite a few significant
detections of GWs. The peak strain amplitude ($h_0$) for all these
detections has been in the range $10^{-21}~-~10^{-22}$. 

There were several theoretical studies \cite{Bildsten:1998ey,
Jones:2001ui, Bagchi:2015tna, Layek:2019ede}, where the authors
suggested that isolated pulsars can be a potential source of GWs. A
few other neutron star activities, such as the neutron star flaring,
the formation of hyper-massive NS following binary coalescence, etc.,
are capable of exciting quasi-normal modes of a pulsar, resulting
in the GWs emission. The primary purpose of such studies is to probe
the internal structure of the pulsars. Hopefully, the more sensitive
ground-based advanced detectors, namely, aLIGO, VIRGO, Einstein Telescope
(ET) etc., will be able to measure the gravitational waves produced by
isolated pulsars shortly and answer a few questions on pulsar physics. In
fact, with the above purpose, even before the direct detection of GWs,
there were a few attempts for GWs searches from isolated pulsars. The
GWs associated with the timing glitch in the Vela pulsar in August
2006 \cite{LIGOScientific:2010epv} was one among those searches worth
mentioning. The observed timing noise, i.e., glitch, can be one such
activity that can excite quasi-normal modes in the pulsar and cause
GWs. Although the searches during the August 2006 Vela pulsar timing
glitch produced no detectable GWs \cite{LIGOScientific:2016aoc}, with
the improving sensitivity of advanced detectors, continuous attempts in
this direction may produce more conclusive results shortly.

\vskip .1in
{\bf Pulsar Timing Array (PTA):}
\vskip .1in

Pulsar timing arrays \cite{2019gwa..book.....A} are the GW 
detector setups provided by nature itself, in the form of a population of 
highly stable millisecond pulsars, with timing accuracies of $\sim 10$ ns 
over several years, with arm lengths of galactic scale. They are sensitive to 
much lower frequencies than ground-based instruments. With precise pulsar 
timing, an array of pulsars could detect extremely low-frequency sources with
typical frequencies less than $10^{-6}$ Hz. The origins of GWs with such
low frequency can be supermassive black hole mergers with masses in the
range $(10^9 - 10^{10})~\text{M}_\odot$, the stochastic background of
GWs arising from cosmological phase transition during the early stages
of the Universe, or otherwise, more exotic objects such as cosmic
strings. The basic principle is that the set of stable MSPs serves as
an array of clocks whose {\it time }, as observed on earth, would be
modulated by gravitational waves passing through the space between MSP
and Earth.  With observations of many pulsars, phenomena which affect
all pulsar periods in a correlated way can be separated from phenomena
which affect different pulsars differently. For example, a stochastic
gravitational wave background can be separated from errors in the time
standard because of their different dependence on pulsar sky position. For
example, clock errors will lead to all pulsars having the same Time of
Arrival (TOA) variations (monopole signature), solar-system ephemeris
errors will lead to a dipole signature. In contrast, gravitational waves
will lead to a quadrupole signature. These effects can be separated
if one has sufficient number of widely distributed pulsars.  With this
purpose, there is a worldwide effort to search for and observe the set
of stable millisecond pulsars for detecting gravitational waves through
pulse modulation. There are four current efforts in this direction,
operating under the joint umbrella of the International Pulsar Timing
Array (IPTA \cite{Manchester:2013ndt}); North American Nanohertz
Observatory for Gravitational Waves (NANOGrav) in the USA, the Parkes
Pulsar Timing Array (PPTA) in Australia, the European Pulsar Timing Array
(EPTA), and the Indian Pulsar Timing Array Project (InPTA). IPTA aims to
construct the most sensitive low-frequency gravitational wave detector,
which can be achieved through sharing resources among the stakeholders
and creating combined pulsar timing data sets.  The current sensitivity
of the experiments is exciting from the perspective of the potential
detection of GWs through PTA.

\section{Observational aspects of neutron star interiors}
\label{sec:observ}

Here we will discuss various proposals from the literature for
possible observational signatures of various phases in NS interiors.
Among these, glitches take a prominent role as well established signals for
the existence of the superfluid phase in NS interior. We will discuss here 
difficulty of this explanation in accounting for relatively recent 
observations of anti-glitches. We will also discuss various proposals for 
detection of the exotic color superconducting phases of QCD in NS interior. 
We again emphasize here that, as we mentioned above, neutron star cores 
typically have very low temperatures, of order few keVs (except for brief 
hot period during its formation). Thus, high baryon density matter at higher 
temperatures, e.g. temperature driven phase transitions in this regime cannot 
be probed there. Interestingly, this is precisely the regime which is mostly
out of bound in low energy relativistic heavy-ion collisions. Thus, these
systems together (also possibly including newly born neutron stars which
may allow even high temperature regime to be probed) allow the possibility 
of probing a large, interesting, part of the QCD phase diagram.

\vskip .1in
{\bf Pulsar glitches:} 
\vskip .1in

Pulsars are magnetized rotating neutron stars that emit periodic
short pulses of electromagnetic radiation with periods between 1.4 ms -
0.3 s. The non-alignment of the rotation axis with the magnetic axis
causes the light-house-like appearance of the pulses to the observer at
Earth. Amid the pulsars' extraordinarily stable rotational frequency,
many pulsars show sudden spin-up events (glitches) followed by a period
of slow  recovery.

\begin{figure}[t]
\begin{center}
\includegraphics[width=0.7\linewidth]{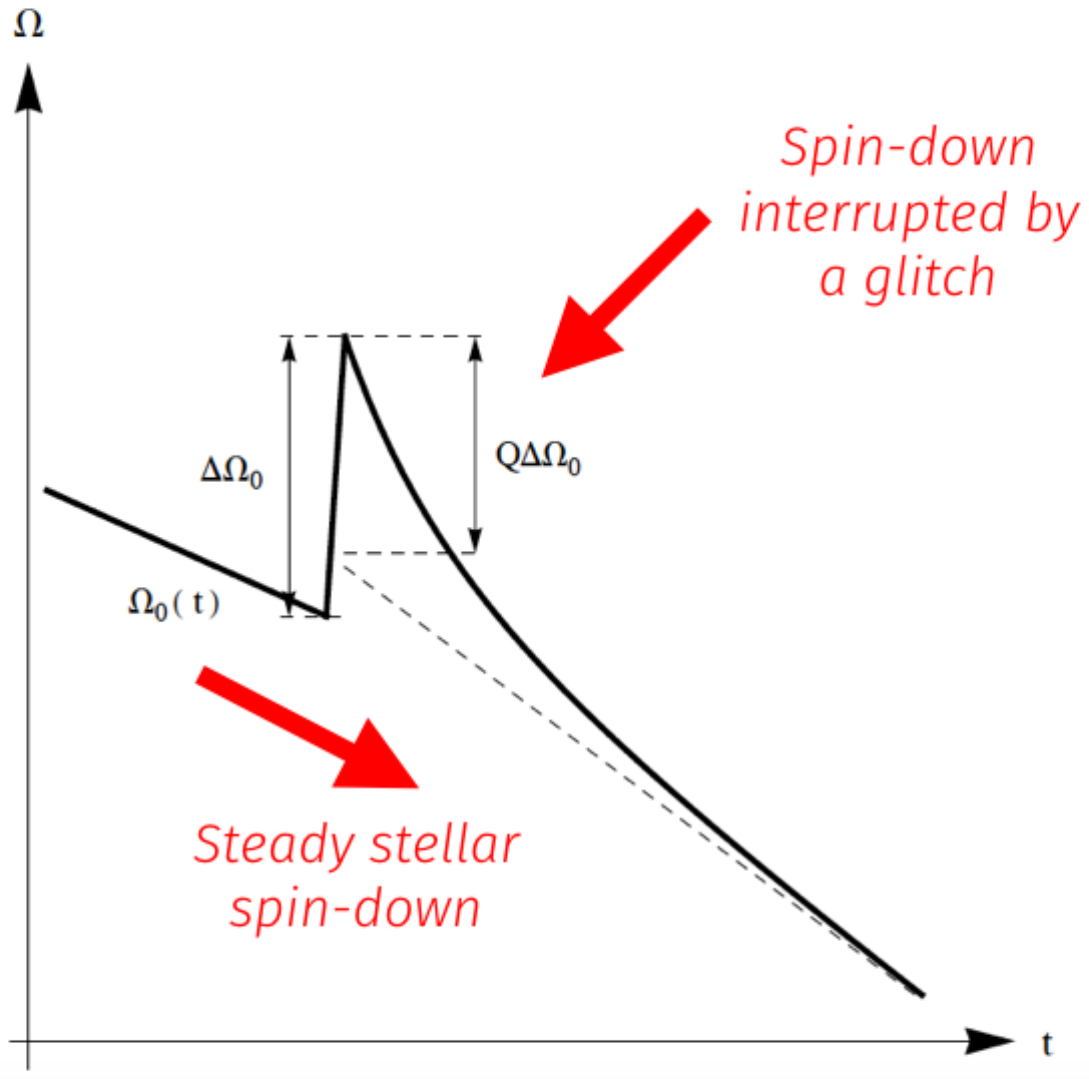}
\end{center} 
\caption{A schematic representation of a typical glitch 
pattern of a pulsar. Here Q is the recovery fraction, which measures the part 
of $\Delta \Omega$ which decays. The post-glitch recovery time scale 
typically ranges from a few days to weeks. (Taken from: 
http://hdl.handle.net/11343/36537.)}
\label{glitch}
\end{figure}
Since the discovery of the first glitch in the Vela pulsar
\cite{1969Natur.222..228R}, many glitches have been observed and reported
to date \cite{Espinoza:2011pq}.  A typical glitch pattern is shown in
Fig. \ref{glitch}. The fractional change of rotational frequency, i.e.,
the glitch size $(\Delta\Omega/\Omega)$ lies in the range $10^{-11}
- 10^{-5}$ with an average inter-glitch time of a few months to a few
years. The pulsar recovers from weeks to months to a period close to
the pre-glitch value. The oldest theoretical model for pulsar glitches
\cite{1969Natur.223..597R}, namely, the crustquake model, assumes
the existence of a deformed solid crust of a pulsar.  The oblateness
can be characterized by the parameter 
$\epsilon = (I_{zz} - I_{xx})/I_{0}$, where $I_{zz}$, $I_{xx}$ and 
$I_0$ are the moments of inertia about the z-axis (rotation axis), x-axis, and 
the spherical star, respectively \cite{1971AnPhy..66..816B}. The sudden 
decrease of
oblateness $\Delta \epsilon$ decreases the MI, resulting in a spin-up
event (following angular momentum conservation), i.e., 
$(\Delta\Omega/\Omega) = - (\Delta I/I_0) = \Delta \epsilon$.  
It was immediately realized \cite{1971AnPhy..66..816B} that inter-glitch time
being proportional to $\Delta \epsilon$, successive large-size glitches
need longer waiting periods.  Thus, crustquake might be responsible
for producing smaller Crab-like glitches $\Delta\Omega/\Omega \simeq
10^{-8}$; the model is incompatible with Vela-like large-size glitches
$\Delta\Omega/\Omega \simeq 10^{-6}$.

Anderson \& Itoh proposed the most popular superfluid-vortex
model in 1975 \cite{Anderson:1975zze}.  The model assumes the
existence of neutron superfluidity in the inner crust, with the
core being in a superfluid/superconducting state. Interestingly,
superfluidity in a neutron star was hypothesized \cite{1959NucPh..13..655M}
a long time before the above-proposed glitch model. The basic idea
of the vortex model is that the vortices (spinning neutron star
causes an array of quantized vortices for stability) act as an
angular momentum reservoir while being pinned to the nuclear sites
\cite{Haskell:2015jra, Layek:2019ede, Layek:2020ocz, Layek:2022kja}.  As the
neutron star slows down due to radiation loss, differential angular
velocity $\delta \Omega$ between the inner crust superfluid and the
rest (crust+core) is built until it reaches a critical value $(\delta
\Omega)_c$.  Once the Magnus force overcomes the pinning force, just
above $(\delta \Omega)_c$,  many vortices ($\sim 10^{18}$) are released
from the pinning site and share the excess angular momentum to the
rigid crust-core co-rotating system, causing the pulsar to spin-up. The
vortex model not only can account for large-size Vela-like glitches
($\sim 10^{-6}$), the long-relaxation time scale (weeks to months)
associated with the post-glitch recovery phase also arises naturally
in this picture, thus providing indirect evidence of superfluidity in
the interior \cite{1969Natur.224..673B}.  In this context we mention
that an interesting spin-down event was observed a decade ago in
the magnetar 1E 2259+586. As per the report \cite{Archibald:2013kla},
X-ray timing observation \cite{Campana:2011ht} of the magnetar clearly
shows an anti-glitch. Evidence for anti-glitches in the X-ray pulsar
NGC 300 ULX-1 is also reported recently \cite{Ray:2018vuy}. Such events
challenge the standard glitch theory and suggest the need for rethinking
this issue \cite{Bagchi:2015tna}.

\vskip .1in
{\bf Signals of mesonic condensate phases:}
\vskip .1in

It has been suggested that at high baryon densities in the core of neutron
stars, apart from the neutrons and protons, various mesonic condensates
may form. Following the suggestion of pion
condensation in neutron stars by Migdal \cite{Migdal:1971cu}, and Sawyer
\cite{Sawyer:1972cq}, there were various studies on understanding the
possibility of pion condensation and its consequences. The main interest
along this line was to theoretically explore the possible appearance
of p-wave pion condensation \cite{Sawyer:1972cq} in nuclear matter,
i.e., the pionic excitations with zero energy and finite momentum. It
has been argued that pion condensation, whether p-wave or s-wave, may
lead to significant modification of the equation of state
\cite{Sawyer:1972cq}. It has been argued that 
the possibilities of p-wave pion condensation at high densities
is improbable in view of the universal repulsion assumption
(see \cite{Ohnishi:2008ng} and the references therein). However,
the authors of Ref. \cite{Akmal:1998cf} suggested the possibility of
$\pi^0$ condensation in symmetric nuclear matter at high densities
$n_B > 0.2~\text{fm}^{-3}$. Thus, one can not completely rule out the
possibility of pion condensation in dense matter. 

Kaon condensates
have also been considered, and their effects on softening the equation
of state, hence constraining the maximum mass of neutron stars, has
been discussed \cite{Li:1997zb, Kaplan:1987sc}. 
Thus, observations of highest mass
neutron star forces re-evaluation of equation of state of the neutron
star matter. Along with such condensates, dominant presence of hyperons
in the NS interior has been discussed, and constraints on maximum mass
of NS have been derived \cite{Glendenning:1984jr}. (Effect of hyperons on
the possibility of pion condensates has been discussed in
\cite{Ohnishi:2008ng} and it has been argued that with hyperons, it is
unlikely that s-wave pion condensate could form in the NS interior.) 
These condensates may also affect the cooling rate of the neutron star.  
As we discussed in Sect.III, analysis of gravitational wave form resulting
from binary neutron star mergers allows direct probe of the equation of
state of the matter inside colliding neutron stars. Thus, one hopes to
be able to put strong constraints on the possibility of such condensates
from GW observations.

\vskip .1in
{\bf Signals of quark matter core:}
\vskip .1in

At high baryon densities, the core of the neutron star
is likely to convert to quark matter \cite{Baym:2017whm,Weber:2010zza}.
This is expected from very general considerations as with high enough
density, baryons will percolate so that the quarks inside baryons can no
longer be associated with a given baryon. (Though, recall the discussion
of the quarkyonic phase in this context.) High baryon density transition
form hadronic phase to the deconfined quark matter phase is expected to
be a first order transition. With that, neutron star matter with a core of 
quark matter is described by a hybrid equation of state, with hadronic phase
at lower density in the outer region separated from the quark matter
core by the intermediate region in the mixed phase. In the absence of
reliable lattice calculations for high baryon density, the description
of the interacting quarks in this high density regime is done using
phenomenological models of QCD. The Nambu-Jona-Lasinio (NJL) model is
frequently used for this with mean field approximation for constructing
the equation of state for dense quark matter. Vector type of interactions  
yield stiff equation of state as required by the existence of neutron stars
with masses larger than $2~\text {M}_\odot$  \cite{Yamamoto:2021htv}.

With the outer regions being in the hadronic phase, there will be a phase
boundary separating the quark core from the hadronic region. With quark
matter equation of state for the core, the mass and size relationship
of the neutron star is affected \cite{Baym:2017whm} which can be
probed by observations, e.g. using thermonuclear X-ray bursts from
matter accretion on NS surface in binary systems \cite{Ozel:2016oaf}.
The compactness of the neutron star ($M/R$) is directly probed by the
tidal deformability from gravitational wave detection from binary neutron
star mergers. Detailed analysis of the wave form of gravitational waves
from BNS merger have been used to constrain equation of states of the
matter in the core, as we discussed above. With a hybrid equation of
state, in NS mergers, the core may lead to shock waves reflecting from the
quark-hadron interface inside hypermassive neutron stars.  In addition,
the formation of high density region and its transition to quark matter
during the merger may give rise to dissipation during the merger leading
to an enhanced damping of the ringdown \cite{Annala:2019puf}.  Damping of
global oscillation modes, like the r-mode, is an important observational
tool to probe the existence of quark matter core using gravitational
wave signals, or indirect electromagnetic probes. Required damping is
difficult to account for in the minimal neutron star models, and quark
matter core can provide for the enhanced damping \cite{Alford:2019oge}.

Possibility of strange quark matter in the high density core of  neutron
star has also been discussed where three-flavor quark matter is assumed
more stable than nuclear matter at low density \cite{Witten:1984rs,
Farhi:1984qu, Alcock:1986hz, Horvath:1992wq, Weber:1994yx}.  
The composition of the core also affects the cooling of neutron star by
neutrino emission, though this is more effective as a signal for very
high baryon density phases, namely the color superconducting phases.

\vskip .1in
{\bf Signals of color superconducting phases:}
\vskip .1in

The quark matter core itself can have very rich phase structure at
very high baryon densities, as we discussed earlier in the 
Section \ref{sec:qcdphase}.
In the QCD phase diagram, we have seen that theory allows the
appearance of various exotic phases in the low-temperature ($T$) and very
high baryon chemical potential ($\mu_B$) regime. The neutron star's core
can provide such conditions to accommodate those phases. One such phase,
the color superconducting phase, may arise due to spontaneously broken
$SU(3)_C$ color and chiral symmetry for three light quark flavors (u,
d \& s). For two light quarks, color group $SU(3)_C$ is broken down
to color $SU(2)_C$ causing the so-called $2SC$ phase. Observational
signatures of these phases have been discussed in the literature
\cite{Alford:2007xm, Rajagopal:2001ngu}. For example, central core with
CFL phase leads to suppressed cooling by neutrino emission, and also
has smaller specific heat. Thus the total heat capacity and neutrino
emission of NS with CFL core will be dominated by outer layer which are
in the standard nucleonic phase.  For somewhat lower baryon densities,
crystalline color superconductivity may arise in the NS core. It has
been suggested that the rigidity of such a phase (possibly with suitably
misaligned magnetic field form the rotation axis) will lead to non-zero
quadrupole moment \cite{Alford:2007xm}. This will be a significant source
of gravitational waves, which can be  observationally constrained with
the present generation of gravitational wave detectors.

With the realization that many of these possible signatures of the exotic
QCD phases are subject to model uncertainties, an attempt was made by
some of us in \cite{Bagchi:2015tna} focusing on the fact that these 
symmetry breaking phase transitions will lead to density fluctuations 
in the core, especially through the formation of topological defects, 
leading to a transient change
of MI tensors components.  It was shown in Ref. \cite{Bagchi:2015tna}
that the change of diagonal components of the MI tensor may lead to the
change of the spin frequency of the pulsars and may be responsible for
glitches and/or anti-glitches. As the fluctuations are random, there is a
possibility of the generation of quadrupole moment leading to the emission
of GWs. Development of the non-zero off-diagonal components of the MI tensor 
may also lead to modulations of pulse profile \cite{Bagchi:2021etv}. (We 
mention that there was also a view in Ref. \cite{Madsen:1999ci}
that the color superconducting phase associated with the three light
flavors might not exist. The authors also claimed that two-flavor color
superconductivity (2SC) might be marginally inconsistent with pulsar data.)

\subsection{Possibility of ultra-high baryon densities in matter collapsing
to a black hole}

Neutron star core is generally believed to be the place where the
highest baryon density can be achieved in nature. Thus the possibility
of quark matter and other exotic QCD phases are usually discussed in
that context (here we include quark star, strange star etc. in the same
category). These are supposed to be equilibrium configurations achieved
by using suitable equation of state for the interior. What happens when
such compact objects accrete matter? Beyond a limit, the equilibrium is
broken and the collapse to black hole occurs. Typically, the collapse is
very rapid, occurring in milliseconds for a typical stellar mass object.
During this collapse, density keeps increasing, while an event horizon
starts forming near the centre, which grows outwards. It is reasonable to
expect that during this dynamical evolution, density will be significantly
larger than the initial density of the equilibrium configuration.
Even before the event horizon forms at the centre, the central density
will keep growing, and after the formation of horizon, the region outside
it will also have large density before it gets engulfed by the growing
horizon. 

Largest density will be expected for collapse forming smallest mass
black hole. Upper mass limit of a neutron star is believed  to be less
than 2.5 solar mass (though there are theoretical uncertainties in
estimates of this limiting mass). The Schwarzschild radius $r_s =
2GM/c^2$ of sun is about 3 km. Thus, a black hole of mass $2.5~\text
{M}_\odot$ will have Schwarzschild radius of about  7.5 km. For
simplest, uniform density collapse forming such a black hole (by
possibly accretion on critical mass neutron star) it may then lead to
baryon densities of about $ (2.5/1.4) (10/7.5)^3 \simeq 4$ times larger
than the average mass density of a canonical neutron star of mass
$1.4~\text{M}_{\odot}$ and radius 10 km. With proper density profile of
the collapsing star, with event horizon forming near the center, and
growing outward, much larger densities will be expected due to very
small size of initial (growing) black hole.

Thus, such collapsing {\it proto black holes} may provide
possibly the highest possible densities available in nature, larger than
any compact equilibrium stars.  This should be most optimistic place for
looking for extreme exotic QCD phases, such as CFL phase which require
very high baryon densities. Even though such high densities will last
for a very short times, less than a millisecond, it is a very large
time for the QCD processes which occur at time scale of fm/c. Thus
observational signals from phase transitions to different high density
QCD phases may be detectable when such collapse occurs. Possibility of
transition to quark matter core in core-collapse supernova simulations
was discussed in \cite{Kuroda:2021eiv}. What we are proposing is that
even more exotic QCD phases requiring extreme baryon densities may show
up transiently in matter collapsing to black hole, for suitable masses
of collapsing object. Numerical simulations of neutron stars accreting
matter and collapsing to black hole show that central density can
significantly increase even before event horizon forms at the centre
\cite{Stark:1985da,Giacomazzo:2012bw,Giacomazzo:2011cv}. The signatures
from such collapse should be very unique, as there will be a succession
of phase transitions to various high baryon density QCD phases, all
occurring within a time span of order milliseconds. Apart from signaling
exotic QCD phases, it can provide unique signature of such a gravitational
collapse to black holes.

\section{Detecting phase transition occurring inside a pulsar}
\label{sec:phase_transition}

Here we will discuss observational implications of  phase transitions
occurring inside a pulsar, in particular, on the nature of its pulses.
The consequences of phase transition (for example from nuclear matter to
QGP) occurring in the core of a neutron star on its rotational dynamics
has been discussed in \cite{Glendenning:1997fy,Heiselberg:1998vh}.
Basic physics of the model is that as pulsar rotation slows down
due to radiation braking, the density of the core steadily increases
(with reduction of centrifugal force).  If the density was initially
below the critical density of the pulsar core (corresponding to a phase
transition), then at some stage central density crosses the critical
value leading to phase transition. It is assumed that as the high density
core grows in size (slowly, over the time scale of millions of years),
it continuously converts to the high density QGP phase (even when the
transition is of first order).  The dependence of pressure on density
will determine the manner in which the phase transition will affect the
moment of inertia of the neutron star, hence the angular velocity of
the star. It was emphasized in these works that the quantity of special
interest is the {\it braking index} defined as 
$n(\Omega) \equiv \Ddot{\Omega}\Omega/\dot{\Omega^2}$. It is argued that
while usually $n(\Omega)$ will be equal to the intrinsic index n of the
energy-loss mechanism (with the energy loss 
$dE/dt = \frac{d}{dt}(I \Omega^2/2) = -C \Omega^{{\rm n}+1}$), 
during phase transitions, it can differ markedly from this value, possibly 
even by orders of magnitude. Thus, even if the changes in moment of inertia, 
and hence the spin rate changes are not directly observable (due to very 
large time scale), braking index may provide a more promising signal for
phase transitions occurring inside pulsars.

The discussions in refs. \cite{Glendenning:1997fy,Heiselberg:1998vh}
primarily focused on the change in the equation of state during the
phase transition. Main consequence of the phase transition was thus
related to change in the diagonal components of the moment of inertia
tensor  affecting the spin rate of the neutron star.  However, all
phase transitions necessarily produce density fluctuations. 
Typical length scale of the fluctuations is governed
by the correlation length of the fluctuations in the magnitude of the
relevant order parameter field. For certain cases, such as for
topological defects even the correlation length of the orientational
fluctuations of the order parameter determines the length scale of the
density fluctuations. It is important to realize that even for very slow
phase transition, density fluctuations will result at the transition
point as long as relevant correlation length remains smaller than the
system. As the correlation domains grow in size, the system becomes more
homogeneous. It is important to note that for continuous transitions,
the divergence of correlation length at the critical point leads to a
scale invariant distribution of density fluctuations (up to the system
size).

\vskip .1in
A rich spectrum of physics is encoded in the distribution of density
fluctuations relating to the nature of phase transition (first order,
second order), and in particular the symmetry breaking pattern (if any)
associated with the phase transition.  Density fluctuations perturb entire
moment of inertia tensor, including its off-diagonal components. This
was discussed  by some of us in \cite{Bagchi:2015tna,Srivastava:2017itj}
for the situation when phase transition occurs rapidly in a large core
of neutron star.  It was pointed out that as phase transitions induced
density fluctuations modify the entire moment of inertia
tensor of the pulsar, the resulting off-diagonal components will lead
to wobbling of star (in addition to any previously present) which will
induces modulations of the pulse profile. Thus, it was argued that the
detailed modification of the pulses carries the information of statistical
nature  of density fluctuations, and hence the precise nature of phase
transition occurring inside the NS interior. An important aspect of this
model is that it predicts that off-diagonal components of the MI tensor
components necessarily become non-zero along with its diagonal components,
with all perturbations being of similar order, due to statistical nature
of the phase transition induced density fluctuations. Thus, in this model,
spin rate changes will be necessarily associated with the modulations
of pulse profile. This can be used to test this model for any observed
pulse modifications, as if spin rate  changes occur due to de-pinning
of vortices, then dominant changes only occur in the diagonal components
of the MI tensor (as all vortices are aligned along the rotation axis).

\subsection{Effects of density fluctuations} 
\label{sec:1st_order}

The earlier discussions in
refs. \cite{Glendenning:1997fy,Heiselberg:1998vh} related to a scenario
of slow transition as applicable for slowly evolving star (e.g.
by accretion), with a transition which is either a weak ﬁrst order,
or a second order, (or a crossover). For a strong
ﬁrst-order transition a different possibility arises as discussed in
\cite{Bagchi:2015tna,Srivastava:2017itj}. Strong supercooling can lead
to a highly suppressed nucleation rate, so that no bubble nucleation
occurs for a very long time. The transition can then occur suddenly,
possibly due to some inhomogeneity, after the supercritical core becomes
macroscopic in size. The phase transition thus occurs rapidly over a
macroscopically large core. This scenario would be quite similar to the
one discussed by Witten \cite{Witten:1984rs}  where a very low nucleation
rate could lead to macroscopic length scales, of order of meters, for
bubble nucleations for  the quark-hadron transition where the typical
length scale would be of order fm. (The discussion in \cite{Witten:1984rs}
assumed a first order transition. With lattice results, now one knows
that for very low baryonic chemical potential the quark-hadron transition
is a cross-over).  Rapid phase transition occurring over a large core
can occur in other situations also, for example during early hot
stages of the neutron star undergoing rapid cooling. The discussion
in ref. \cite{Bagchi:2015tna,Srivastava:2017itj} related to a general
such situation and argued that, along with the expected change in the
moment of inertia, and hence the spin rate, (which could be directly
observable), density fluctuations will be produced in the entire large
core region undergoing this rapid phase transition. This will then affect
off-diagonal MI components and hence induce wobbling of the neutron star.

The situation considered in ref. \cite{Bagchi:2015tna, Srivastava:2017itj} related 
to the case when phase transition induced spin rate remains small, (say, within 
the range set by observations of glitches), and determined the effects of density
fluctuations on the MI tensor. In a simplified, two density picture of the phase
transition occurring in the NS, it is assumed that the phase transition converts 
the core of radius $R_0$ to the new phase with density $\rho_2$, while the rest 
of NS remains in the old phase with density $\rho_1$. The resulting fractional 
change in the moment of inertia \cite{Heiselberg:1998vh} is
 
\begin{equation}
\frac{\Delta{I}}{I} \approx \frac{5}{3}
\left(\frac{\rho_2}{\rho_1}-1\right)
\frac{R_0^3}{R^3} 
\label{eq:MI_change}
\end{equation}

Where $R$ is the radius of the star (taken to be spherical) in absence of
the dense core.  If we consider the possibility that  phase transitions
could have occurred in the pulsars which have been regularly monitored,
then observations of glitches can be used to constrain the size of the
core undergoing phase transition. One would then constrain the largest
fractional change of moment of inertia to be  less than $10^{-5}$,
relating to strongest glitches observed so far. Various  phase transition
cases can then be considered.  For example, a sample value of change in
density due to phase transition can be taken to be
about $30 \%$ (which could be appropriate for QCD transition where density 
change can be of order one). This will constrain the value of $R_0$ to be less
than about $300$ m (taking $R = 10$ km). Another important case is that
of superfluid transition where the density change can be taken to be
of order of the superfluid condensation energy density ($\approx 0.1$
MeV/fm$^3$) \cite{Richardson:1972xn,Yang:1971oux}. In this case, glitch
constraint will implying that $R_0$ can be as large as $5$ km.

With the size of the core undergoing phase transition being constrained,
one can then discuss effects of density fluctuations occurring in this core
\cite{Bagchi:2015tna,Srivastava:2017itj}. First it is useful to get
generic estimates which depend simply on the nature of phase transition.
First, we focus on density fluctuations  due to the nucleation of
bubbles. As discussed above, for a strong first-order case, a core of
size a few hundred meters (or larger) can undergo rapid phase transition.
However,  in general, the core region will be expected to have minute
nonuniformities, even of purely statistical origin, so that one can
consider a situation where many bubbles may nucleate in different parts
of the supercritical core.  The bubbles  nucleated with a critical size
of the order of tens of fm, will expand and coalesce.  At the time of
coalescence, the supercritical core region will consist of close packing
of bubbles of the new phase, embedded in the old phase.  We carried
out simulation of random spherical bubble nucleation of radius $r_0$
(at the coalescence stage) filling up a spherical core of radius $R_0 =
300$ m. The density change in bubble nucleation is taken to $\sim 160$
MeV/fm$^3$, as appropriate for the QCD transition.  We find fractional
change in moment of inertia $\Delta{I}/I \approx 4\times 10^{-8}$ for
$r_0 = 20$ meters and remains of same order when $r_0$ varies from
$5$ meters to $20$ meters.  Due to random nucleation of the bubble,
the off-diagonal component of components of the moment of inertia, as
well as the quadrupole moment become nonzero. The ratio of both to the
initial moment of inertia are found to be of same order, $Q/I_0
\simeq I_{xy} / I_0 \simeq 10^{-11} - 10^{-10}$.

\subsection{Density Fluctuation From Topological Defects}

The formation of topological defects routinely occurs in
symmetry-breaking transitions and has been extensively discussed in
the literature, from the early Universe to condensed matter systems.
Depending on the relevant energy scales these defects can be a source
of large density fluctuations.  The underlying dominant mechanism
for their formation in a phase transition is the so-called Kibble
mechanism \cite{Kibble:1976sj,Kibble:1980mv} which predicts a defect
density proportional to the number density of correlation domains, with
proportionality constant determined using universal arguments depending
only on the specific symmetry breaking pattern and dimensionality of space
under consideration. Thus, the random network of defects arising in any
phase transition and resulting defect distribution can be determined
entirely using the symmetry-breaking pattern. For example, a random
network of vortices arises from superfluid transition. A network of
domain walls and global strings arises from the spontaneous breaking of
$Z(3)$ center symmetry for confinement–deconfinement QCD transition
\cite{Gupta:2010pp}. Some QCD transitions (e.g. the color flavor locked
(CFL) phase, expected to arise at very large values of baryonic chemical
potential) may give rise to only global strings, with $SU(3)_c\times
SU(3)_L \times SU(3)_R\times U(1)_B$ symmetry broken down to the diagonal
subgroup $SU(3)_{c + L + R}\times Z_2$ \cite{Alford:2007xm}. 

The evolution
of such a defect network shows universal characteristics. Starting with
initial defect densities (basically determined using correlation length
and topological probability calculations), the later evolution of string
defects and domain wall defects shows scaling behavior. This has important
implications for  predictions of changes in the moment of inertia (hence
glitches/anti-glitches), quadrupole moment, and subsequent relaxation
to the original state of rotation in a reasonably model-independent
way.  An important aspect of topological defect sourced density
fluctuations relates to the manner in which density fluctuations evolve
in time. Eventually density fluctuations decay away, leaving uniform new
phase (in the core region undergoing phase transition). But the manner of
this decay, and the time duration, depend crucially on the specific nature
of the defect, and hence on the symmetry breaking pattern.  For example,
while bubble-generated density fluctuations decay away quickly in the
time scale of coalescence of bubbles, the domain wall network and the
string network coarsen on much larger time scales (with specific scaling
exponents) which are characteristic of specific type of defect. Thus the
precise measurement of the pulsar spinning rate and modification of pulse
profile, and its time evolution, can also provide important information
about the specific transition occurring inside the neutron star.

\subsubsection{Lattice simulation of string network}

Due to topological nature of these defects, generic features of the defect
network can be determined using simple lattice picture with lattice size
representing the correlation length. An estimate  of the change in MI
due to string and wall formation can be made by producing a network of
defects inside the core of the pulsar by modelling the correlation domain
formation in a cubic lattice, with lattice spacing $\xi$ representing
the correlation length \cite{Vachaspati:1984dz}. To model U(1) global
string formation each lattice site is associated with an angle $\theta$
(randomly varying between $0$ and $2\pi$), or two discrete values $0,
1$ while modelling $Z_2$ domain wall formation. ($Z_2$ domain wall is
considered instead of $Z_3$ walls of QCD just for simplicity). For string
case,  geodesic rule \cite{Kibble:1976sj,Kibble:1980mv,Vachaspati:1984dz}
is used to determine  the winding of $\theta$ on each face of the cube.

Starting with correlation length of order of fm to simulate the
network of order few hundreds meter is numerically not possible. So
the estimates were made in \cite{Bagchi:2015tna,Srivastava:2017itj}
by considering spherical star of size $R$ and a spherical core of
radius $R_c = (0.3/10)R$. Then, considering $\xi = 10$ fm,
$R_c$ is increased from $5 \xi$ to $400 \xi$, and it was found that
$\delta{I}/I_i\simeq$ appears to stabilize at $10^{-13}-10^{-14}$.
Using this numerical results, it was suggested that the same fractional
change in the MI may also be possible for realistic value of  $R =
10$ km, especially when one  accounts for statistical fluctuations in
the core. For the case of domain wall formation, one finds fractional
change in  the off-diagonal components of MI  (as well as quadrupole
moments) to be larger by a factor of $40$.  For the case of superfluid
transition, a rapid superfluid transition can take place due to
transient heating, and subsequent cooling  of star. This may occur
either due to another transition releasing latent heat, or due to
accretion, etc.  Taking vortex energy per unit length to be $100$
MeV/fm and correlation length for vortex formation of order $10$ fm
(\cite{Richardson:1972xn,Yang:1971oux}) it is found that the superfluid
vortices induced transient fractional change in MI is of order $10^{-10}$
(compared to net fractional change in MI of order $10^{-5}$ as discussed
in Section \ref{sec:1st_order}). Also it was found that the ratios of
the quadrupole moment and off-diagonal components of MI to the net MI
of the pulsar are  of order $10^{-10}$.

\subsubsection{Field theory simulation of strings and domain 
walls}

The technique proposed in \cite{Bagchi:2015tna,Srivastava:2017itj} has
potential of using detailed observations of pulse modification to learn
about precise nature of phase transition occurring inside the pulsar. For
this, one would need to know specific nature of density fluctuations for
different cases and their precise time evolution. As the relevant cases
here refer to quantum field theory phase transitions, one has to resort
to field theory simulations to get such details. Unfortunately, with
typical length scales of such QFT phase transitions being microscopic
(e.g. of order fm for QCD transition), one can only hope to do these
simulations in very small spatial regions. (This is different from the
case of bubble nucleation where generic arguments of strong supercooling
were invoked to determine density fluctuations over macroscopic regions
even for QCD transition. even for lattice modelling of defect networks
as discussed above, one could consider relatively large lattice sizes.)

This was achieved in \cite{Bagchi:2015tna,Srivastava:2017itj} by
studying string and wall formation in a field theory simulation
for confinement-deconfinement (C-D) QCD transition using effective
field theory Polyakov loop model. The expectation value of the
Polyakov loop, $l(x)$, is the order parameter for the C-D transition
\cite{PhysRevD.24.450,Svetitsky:1985ye}. $l(x)$ vanishes in the
confined phase and is non-zero in the deconfined phase where  $Z(3)$
center symmetry (for the $SU(3)$ color group) is spontaneously broken
as $l(x)$ transforms non-trivially under $Z(3)$. This gives rise to
three different vacua for different value of $l(x)$ in QGP phase,
leading to topological domain wall defects (which interpolate between
different $Z(3)$ vacua) \cite{Bhattacharya:1992qb}, and also string
defects (QGP strings) forming at the junction of these $Z(3)$ walls
\cite{Layek:2005fn,Gupta:2010pp,Gupta:2011ag}. (Note, we have used the
notations $Z(3)$ and $Z_3$ interchangeably.)  

Defect formation is studied
using field theory simulation of the evolution of $l(x)$ from an initial
value of zero (appropriate for the confining phase) as the system is
assumed to undergo a rapid transition (quench) to the deconfined phase
(as in \cite{Mohapatra:2012ck}). Use of quench is not an important point
here as the formation of defects only requires formation of uncorrelated
domains, and the size of the domains in this model has to be treated as a
parameter as it is not possible to cover length scales of km (for star)
to fm (QCD scale). Hence these simulations are necessarily restricted
to system sizes of tens of fm only. The physical size of the lattice
is taken as ($7.5$ fm)$^3$ and ($15$ fm)$^3$.  Another possible phase
transition is to the so called color flavor locked (CFL) phase inside
the core of a pulsar where the QCD symmetry for three flavors (for very
large baryon chemical potential so that mass differences between these
quarks become unimportant), $SU(3)_c\times SU(3)_L\times SU(3)_R\times
U(1)_B$ is broken down to the diagonal subgroup $SU(3)_{c+L+R}\times
Z_2$ \cite{Alford:2007xm,Alford:1997zt}. This transition will give
rise to global strings. To roughly estimate resulting change in MI, a
simplified case is considered by removing the cubic term from effective
potential. This modification in the potential gives rise to string defects
only without any domain walls, as appropriate for the transition from,
say, QGP phase to the CFL phase, while ensuring that one has the correct
energy scale for these string defects.

\begin{figure}
\centering
\begin{tabular}{c}
\includegraphics[width=\linewidth]{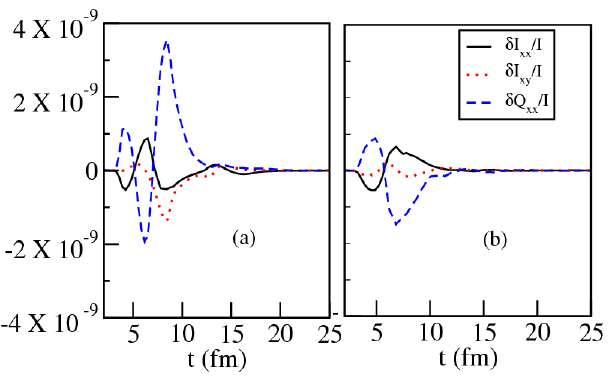}\\%
\includegraphics[width=\linewidth]{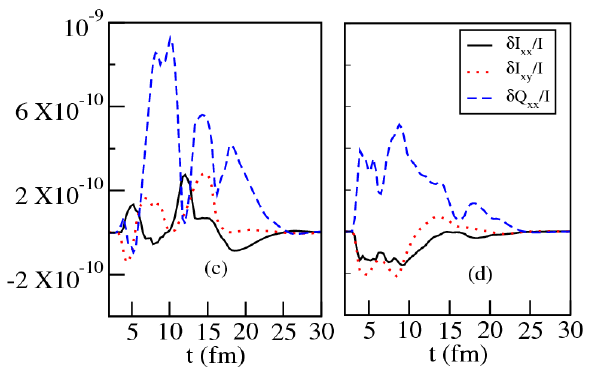} 
\end{tabular}
\caption{Time evolution of fractional changes in the
moment of inertia as well as the quadrupole moment during the process
of phase transition. (a,b), and (c,d) correspond to lattice sizes ($7.5$
fm)$^3$ and  ($15$ fm)$^3$ respectively. Plots in (a) and (c) correspond
to the confinement-deconfinement phase transition which results in the
production of Z(3) walls and associated strings, while plots in (b) and
(d) correspond to the transition where only string defects form which will
be the case for the CFL phase. (Fig. taken from \cite{Bagchi:2015tna}.)}
\label{fig:fluc_glitch}
\end{figure}

Fig. \ref{fig:fluc_glitch} shows how the induced fractional off-diagonal
MI components and ratio of quadrupole moment to MI change in time. Even
though the exact values of these quantities remain to be determined for
the truly macroscopic core sizes (as for the lattice simulation of defect
network discussed above), it is clear that in principle, exact temporal
profile of these quantities can be determined. Thus resulting pulse
modifications can be predicted in complete detail, in principle. One
important, completely robust feature of these results is that density
fluctuation induced contributions to even the diagonal components of the
MI tensor can be positive as well as negative (with equal probability).
These changes in diagonal components can then account for glitches as
well as anti-glitches in the same unified framework. For
this it is important to note that density fluctuations arise during
phase transition to a new phase which by itself leads to change in the
spin rate.  Thus, when density fluctuations die away, the original spin
rate is only partially restored (depending on the density difference
between the two phases). 

\section{Pulse Modification for Random Density Fluctuations}
\label{sec:pulse_modification}

As we discussed above, in principle, a detailed analysis of perturbed
pulses should be able to reveal a wealth of information about underlying
density fluctuations occurring inside a pulsar. The order of phase
transition, the symmetry breaking pattern, the time scale of phase
transition, etc. will leave distinctive characteristic perturbations on
the pulses. For this, one should consider specific source of density
fluctuation, with well defined distribution of density fluctuations,
and calculate in detail the resulting pulse modification. However,
certain aspects of generic density fluctuations, in particular those
resulting from any phase transition, will have qualitative implications
for pulse modifications. Such density fluctuations will be statistical in
nature, and will lead to random contributions to the components of the MI
tensor. A general study of such fluctuations was carried out by some of
us in \cite{Bagchi:2021etv} where the density fluctuations were modeled
in terms of Gaussian distributed random components of MI tensor added to
the unperturbed diagonal MI tensor of the neutron star. With the form of
perturbed MI tensor prescribed, the resulting effects on pulse timings as
well as specific nature of pulse  modulations was then calculated. 

An important finding in \cite{Bagchi:2021etv} was that even for very
tiny density fluctuations, with resulting changes in pulsar timings
being extremely small, pulse profile modification were found to become
relatively large due to modulations resulting from wobbling of NS. Main
reason for this was that while pulse timing changes remain proportional
to typical density fluctuation magnitude $\epsilon$, the pulse profile
modification  were found to be proportional to $\epsilon/\eta^2$ where
$\eta$ is the NS deformation parameter. ($\eta$ is typically very small
$\sim 10^{-8} - 10^{-4}$.  This observation will play an important role in
later discussion when we discuss possible detection of external GWs using
NS deformations.) We first discuss this case of random MI components from
\cite{Bagchi:2021etv}.  Subsequently, we will also briefly mention how
this technique can be applied to cases where where modified MI tensor is
precisely known, for example, the case of specific density fluctuations
occurring in neutron stars during collision with an asteroid.

Basic physics of calculations in \cite{Bagchi:2021etv} is to consider
a reasonably symmetric initial configuration of neutron star with
homogeneous density, and then incorporate effects of random density
fluctuations on its rotational dynamics 
by calculating the perturbed moment of inertia tensor. The calculation
of moment of inertia of neutron stars has been extensively discussed in
the literature \cite{Lattimer:2000nx, Haensel:2007yy,
Heiselberg:1998vh}. The effects of General Relativity appear in terms of
the compactness parameter $x_{GR} = r_S/R$ where $R$ is the radius of
the neutron star and $r_S = 2GM/c^2$ is its Schwarzschild radius, $M$ being
the mass of the neutron star. 
For all observed pulsars, slow rotation approximation holds where
centrifugal forces are small as compared to the gravitational force
\cite{Haensel:2007yy}. Detailed expression of the moment of inertia
depends on the equation of state. To illustrate General Relativity
corrections, we give here the expression for the moment of inertia of
the crust of the neutron star in terms of $x_{GR}$ and the neutron star
moment of inertia $I$ (in thin crust approximation, with equation of
state where pressure is negligible compared to the mass energy density,
see \cite{Haensel:2007yy} for details):
\begin{equation}
I_{crust} \simeq {2 \over 3} M_{crust} R^2 (1 - x_{GR} {I \over MR^2})
\end{equation}
The $x_{GR}$ term gives the lowest order General Relativity correction
to the Newtonian approximation for the crustal moment of inertia. If we
take the typical value of $x_{GR} \simeq 0.3$ for a neutron star, then
various calculations give estimates of $I/(MR^2)$ of about 0.2-0.3,
leading to General Relativity corrections being of order 10 \%
\cite{Haensel:2007yy, Heiselberg:1998vh}.
   
The calculations in \cite{Bagchi:2021etv} were aimed at showing 
qualitative aspects of the effects of density fluctuations, with order of
magnitude estimates. For this purpose, Newtonian approximation for the moment 
of inertia was sufficient, and calculations in this (and following) sections
will be given within this approximation.

For the initial configuration of the unperturbed pulsar, an oblate spheroid 
shape was taken, with the pulsar rotating about the $z$-axis with angular 
frequency $\omega$ and angular momentum $L_z = L$ ($L_x = L_y = 0$).
The principal moment of inertia components were taken as $I^0_{11}
\equiv I^0_1 = I^0_{22} \equiv I^0_2$ and $I^0_{33} \equiv I^0_3 = I_0$
with $I_0 > I^0_1, I^0_2$. The oblateness is parameterized through $\eta =
(I_0 - I^0_1)/I_0$, the value of which depends on the neutron star's mass,
the rigidity of the crust and magnetic field etc. Theoretical studies
by \cite{Horowitz:2009ya,Baiko:2018jax} put the upper bound of $\eta$ as
$\simeq 10^{-6}$, whereas work of \cite{Makishima:2014dua} puts the upper
bound limit as $10^{-4}$.  Observational studies put the upper bound close
to $10^{-5} - 10^{-4}$ \cite{LIGOScientific:2020gml} with some pulsars
having $\eta \sim 10^{-2} - 10^{-3}$ \cite{LIGOScientific:2013rhu}.

\begin{figure}[!htp]
\begin{center}
\includegraphics[width=0.45\textwidth]{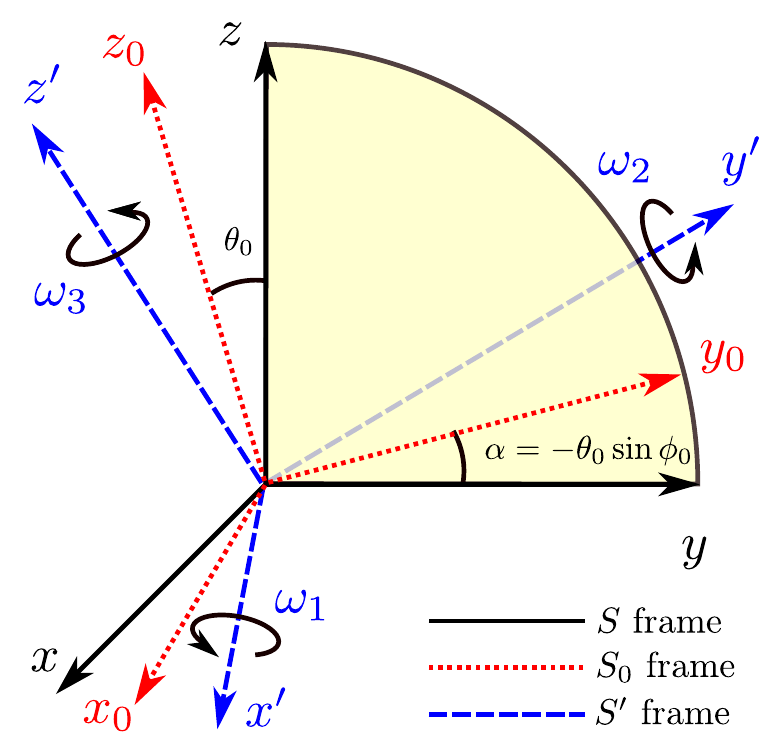}
\caption{
An oblate shape unperturbed pulsar is initially (i.e. before any phase
transition) rotating about the z-axis relative to a space-fixed frame S
(solid black lines). The red dotted lines show the principal axes ($x_0$,
$y_0$, $z_0$) of $S_0$ frame immediately after the phase transition (at
$t = 0$). The body-fixed $S^\prime-$frame at any arbitrary time $t>0$
is shown with blue dashed lines (taken from \cite{Bagchi:2021etv}).}
\label{fig:axisrot}
\end{center}
\end{figure}

Immediately after a phase transition (at $t=0$), density
fluctuation will alter the MI tensor of the pulsar. The $S_0$
frame in Fig. \ref{fig:axisrot} shows the set of principal axes
immediately after the phase transition. The subsequent dynamics of
the perturbed pulsar are governed by the set of Euler equations,
\cite{Kleppner:2014int,Goldstein:2002cla}
\begin{align}
I_1 \dot \omega_1 - (I_2 - I_3) \omega_2 \omega_3 & =  0 
\label{eq:euler1} , \\
I_2 \dot \omega_2 - (I_3 - I_1) \omega_1 \omega_3 & =  0 
\label{eq:euler2} , \\
I_3 \dot \omega_3 - (I_1 - I_2) \omega_1 \omega_2 & =  0 
\label{eq:euler3},
\end{align}
where $I_i~(i =1,2,3)$ denote the principal MI tensor components
relative to the body fixed frame ($S^\prime$ in Fig. \ref{fig:axisrot})
and $\omega_1(t), \omega_2(t)$, and $\omega_3(t)$ are the angular
frequencies of the star with respect to space fixed frame (which
momentarily coincides with $S^\prime$). As $\omega_1$, and $\omega_2$
are expected to be very small compared to $\omega_3 \simeq \omega$,
one can write the equation of motion for $\omega_1$ as
\begin{align}
\ddot{\omega_1} &+ \Omega^2~ \omega_1 = 0 \,,
\label{eq:diff} \\
\textrm{where,}\qquad
\Omega &= \omega_3 \left[\frac{(I_3 - I_1)(I_3 - I_2)}
{(I_1 I_2)}\right]^{1/2} \,.
\label{eq:precfreq}
\end{align}
Here, $\Omega$ is the precession frequency due to the perturbation. We 
consider the situation when perturbations are tiny in comparison to the 
oblateness parameter ($\epsilon \ll \eta$), hence the condition that 
$I_3 > I_1, I_2$ is still valid (as we will see below).  The solution of the 
above equation is then given by
\begin{equation}
\omega_1(t) = A~\cos (\Omega~t) + B~\sin (\Omega~t).
\label{eq:omega1}
\end{equation}
$A$ and $B$ are two arbitrary constants determined from the initial conditions. Similarly using 
Eq. (\ref{eq:omega1})) and Eq. (\ref{eq:euler2}), one obtains the time evolution of $\omega_2$
\begin{equation}
\omega_2 (t) = k [ A \sin (\Omega~t) - B \cos (\Omega~t)],
\label{eq:omega2}
\end{equation}
where the factor $k$ is given by
\begin{equation}
k = \left[\frac{I_1(I_3 - I_1)}{(I_2(I_3 - I_2))}\right]^{1/2}.
\label{eq:kfactor}
\end{equation}
The time $t=0$ is assumed to be the onset of the pulsar's precession 
immediately after the completion of phase transition. Denoting the respective 
angular velocities at $t=0$ by $\omega^0_1$ and $\omega^0_2$ 
respectively, Eqs. (\ref{eq:omega1}) and (\ref{eq:omega2}) can be 
rewritten as 
\begin{align}
\omega_1 (t) &=  \omega_1^0 \cos (\Omega~t) - \frac{\omega_2^0}{k}  
\sin (\Omega~t) \label{eq:omega10} \\
\omega_2 (t) &= k \omega_1^0 \sin (\Omega~t) + \omega_2^0  
\cos (\Omega~t). \label{eq:omega20}
\end{align}
Here, the arbitrary constants $\omega^0_1$ and $\omega^0_2$ can be 
fixed by initial conditions.

\subsection{Initial conditions and analytical estimates of various parameters}

Immediately after the phase transition, the orientation of the principal
axes ($S_0$ frame) is shown in Fig. \ref{fig:axisrot}. The standard
polar angle and the azimuthal angle of $z_0$ axis with respect to the
$S$ frame are denoted by $\theta_0$ and $\phi_0$. These angles are
obtained by diagonalizing the perturbed moment of inertia tensor and
finding the eigenvalues and the corresponding set of eigenvectors (see
\cite{Bagchi:2021etv} for details). In the absence of external torque,
the angular momentum of the pulsar must be conserved. Thus, after the
phase transition, the components of the angular momentum along $x_0,
y_0$, and $z_0$ are given by

\begin{align}
L_{x_0} (t=0) &= I_1 \omega_1^0 = - L \theta_0 \cos \phi_0 
\label{eq:Lx0} \\
L_{y_0} (t=0) &= I_2 \omega_2^0 = - L \theta_0 \sin \phi_0 
\label{eq:Ly0} \\
L_{z_0} (t=0) &= I_3 \omega_3^0 = L.
\label{eq:Lz0}
\end{align}

Because $\omega_3$ is approximately constant and the angle $\theta_0$
is small for tiny density fluctuations, one can use the approximation
$L/I_1 \simeq L/I_3 \simeq \omega$.  The above set of equations can be
used now to express Eqs. (\ref{eq:omega1}) and (\ref{eq:omega2}) in terms
of $\theta_0$ and $\phi_0$. Since the phase transition fluctuations are
small, the perturbed MI tensor can be written as
\begin{align}
I_{1,2} &= I_0 (1-\eta + \epsilon_{1,2})\,, \\
\textrm{and} \qquad I_3 &= I_0 (1+\epsilon_3)\,,
\end{align}
with $\mathcal{O}(\epsilon_i) \simeq \mathcal{O}(\epsilon)$ for 
$i = 1,2,3$. Therefore the precession frequency (Eq. (\ref{eq:precfreq})) 
can be written as
\begin{equation} 
\Omega \simeq \left(\frac{\eta+\epsilon}{1-\eta+\epsilon}\right) \omega 
\simeq \eta ~\omega.
\label{eq:omegaapprox} 
\end{equation} 
Here it is assumed that $\epsilon,~\eta \ll 1$ and $\epsilon \ll \eta$. 
The initial angle $\theta_0$ is obtained by diagonalizing the 
perturbed MI tensor $I_{ij}$. For an order of magnitude estimate, the 
perturbation $\epsilon_{ij} \equiv \delta I_{ij}/I_0$ can be taken as 
$\epsilon_{ij} = \epsilon/I_0$. This results in 

\begin{align} 
\cos \theta_0 &= \left(1+ 2\left(\frac{\epsilon I_0}{I_3 - 
I_1-\epsilon I_0}\right)^2 \right)^{-1/2} .
\end{align}

Assuming $\epsilon \ll \eta$, one gets $\theta_0 \simeq \sqrt{2} 
\left(\epsilon/\eta\right)$. On similar argument, one obtains 
the factor $k = 1 + \epsilon/2\eta$. Thus, the
Eq.(\ref{eq:omega10}) and Eq.(\ref{eq:omega20})can be expressed as
\begin{align}
\omega_1 (t) &= - \omega \theta_0 \left[\cos (\Omega t+\phi_0) + 
\frac{\epsilon}{2\eta} \sin\phi_0 \sin (\Omega t)\right] 
\label{eq:omega1approx} \\
\omega_2 (t) &= - \omega \theta_0 \left[\sin (\Omega t+\phi_0) + 
\frac{\epsilon}{2\eta} \cos\phi_0 \sin (\Omega t)\right] 
\label{eq:omega2approx}.
\end{align}
The corresponding rotational angles $\theta_1(t)$ and $\theta_2(t)$ 
can also be re-expressed as 
\begin{align}
\theta_1 (t) &= - \frac{\omega \theta_0}{\Omega} 
\left[\sin (\Omega t+\phi_0) - 
\frac{\epsilon}{2\eta} \sin\phi_0 \cos (\Omega t)\right] 
\label{eq:theta1approx} \\
\theta_2 (t) &=  \frac{\omega \theta_0}{\Omega} 
\left[\cos (\Omega t+\phi_0) + 
\frac{\epsilon}{2\eta} \cos\phi_0 \cos (\Omega t)\right] 
\label{eq:theta2approx}.
\end{align}
Putting the values of $\Omega$ and $\theta_0$, the
amplitude $\omega_m$ (From Eqs. (\ref{eq:omega1approx}) and 
(\ref{eq:omega2approx})), and $\theta_m$ (From Eqs.
(\ref{eq:theta1approx}) and (\ref{eq:theta2approx})) are determined 
as 
\begin{align}
\omega_m &= \omega \theta_0 \simeq \sqrt{2} 
\left(\frac{\epsilon}{\eta}\right) \omega \,,
\\
\theta_m &= \left(\frac{\omega}{\Omega}\right) \theta_0 
\simeq \sqrt{2} \left(\frac{\epsilon}{\eta^2}\right) \,.
\end{align}
Thus for $\eta = 10^{-3}$, the oscillation amplitude $\theta_m$
is  of order $10^{6}~\epsilon \simeq 1^\circ$ for 
$\epsilon = 10^{-8}$. It was shown in  ref. \cite{Bagchi:2021etv} 
that the above analytical estimates approximately match the results obtained 
from the simulation.

\subsection{Numerical Results} 

\begin{figure}[!htp]
\begin{center}
\includegraphics[width=0.95\linewidth]{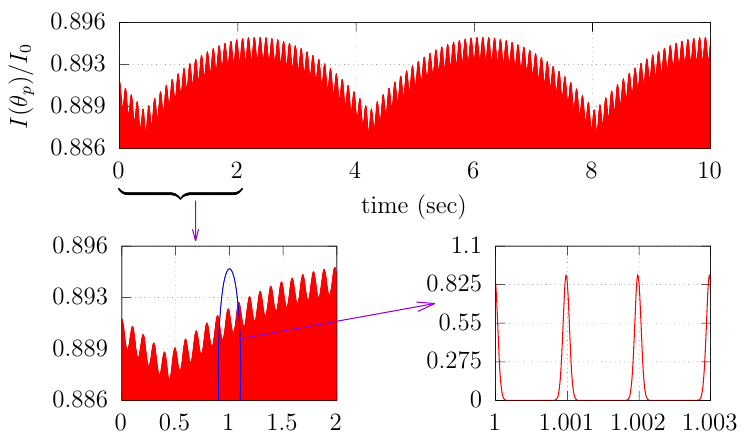}
\caption{The figure shows the time evolution of the pulse profile 
$I(\theta_p)/I_0$ of a millisecond pulsar for $(\eta,\epsilon) = 
(10^{-2}, 10^{-5})$. The top plot shows the two different modulations 
at different time scales. Bottom-left shows the same plot with better 
resolution. Bottom-right shows a few pulses for a typical millisecond 
pulsar. The figures are taken from \cite{Bagchi:2021etv}.}
\label{fig:case1a}
\end{center}
\end{figure}

\begin{figure}[!htp]
\begin{center}
\includegraphics[width=0.95\linewidth]{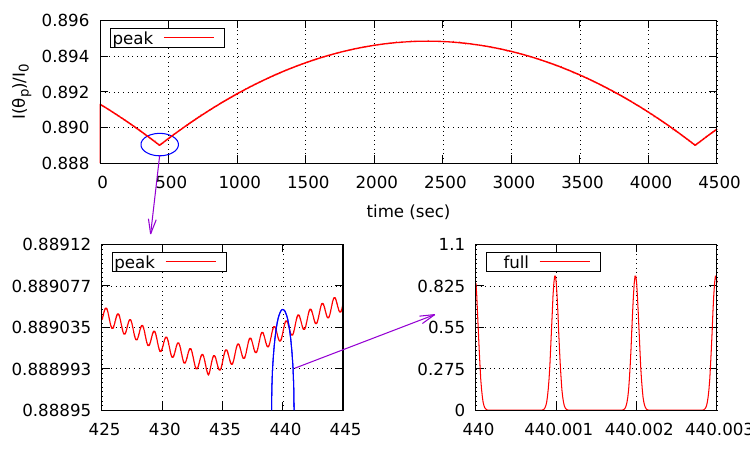} \caption{(Taken from
\cite{Bagchi:2021etv}) Same plot as Fig. \ref{fig:case1a}, however for
$(\eta,\epsilon) = (10^{-3}, 10^{-8})$.  An apparent `kink' at the top
figure appears just below 500 s smoothed out with improved resolution
(bottom-left).  For clarity, here only the top part of the pulses is
shown compared to Fig. \ref{fig:case1a}.} 
\label{fig:case2}
\end{center}
\end{figure}

Here we briefly review the simulation results of Ref.
\cite{Bagchi:2021etv}.  There, two sets of values $ (\eta, \epsilon) =
(10^{-2}, 10^{-5})$ and $(10^{-3}, 10^{-8})$ were chosen to observe the
impact of perturbation on pulse modulations (See Table 1 in 
\cite{Bagchi:2021etv} for the values of other parameters). Note that
Eqs. (\ref{eq:omega1approx}) and (\ref{eq:omega2approx}) can be
approximately written as $\omega_{1,2} \sim \omega_m \cos (\Omega t)$.
Since $\Omega \sim \eta \omega$ (Eq. (\ref{eq:omegaapprox})), the time
period $T_\Omega = T_\omega/\eta$. Thus $T_\Omega = 0.1$ for $\eta =
10^{-2}$ and $1$ sec for $\eta = 10^{-3}$. This matches the numerical
simulation results in Ref. \cite{Bagchi:2021etv}. The results are shown
in Figs. \ref{fig:case1a}, and \ref{fig:case2}.

Other than the above (first) modulation, another (second) modulation is
also expected since $\omega_1$ and $\omega_2$ also oscillates about the
$x$ and $y$ axis, respectively.  From the frequency oscillation amplitude
$\omega_m \sim (\epsilon/\eta)\omega = (2\pi \epsilon/\eta)1000$ /sec.,
the second modulation time scale can be approximately determined as $T_m
\simeq 10^{-3}~(\eta/\epsilon)$ sec. Thus, the time scale $T_m$ varies
from a few seconds [for $ (\eta, \epsilon) = (10^{-2}, 10^{-5})$] to a
few hundred seconds [for $(\eta, \epsilon) = (10^{-3}, 10^{-8})$]. The
figures \ref{fig:case1a}, and \ref{fig:case2} indeed show that there
is a second modulation, though the numerical results show time scale
is larger compared to the analytical estimate $T_m$. (For clarity, in
Fig. \ref{fig:case2}, only the top part of the pulses is shown compared
to Fig. \ref{fig:case1a}.)  Of course, considering the complexity of
the rigid body dynamics, and the order of magnitude estimates used,
one can expect uncertainty in the analytical estimates of the concerned
quantities.

An important feature of these results can be termed as {\it the
memory effect}. This relates to the fact that, even after the density
fluctuations fade away leading to vanishing off-diagonal components,
thereby restoring the original rotation axes, and hence the original pulse
profile, there will in general be a net shift of the angular position
of the pulse. (This is apart from the effect of any possible net change
in the spin rate. So, if net spin rate change remains unobservable,
this net shift of the angular position may still be observable as it
originates from the transient pulse modification from intermediate stage
of wobbling of star.) Therefore, a net, residual shift of the angular
position of the pulse could signal a missed phase transition.

\subsection{Pulse modification due to asteroid impact on neutron star}

Impact of asteroids, comets etc. on astrophysical bodies frequently occur
and with intense gravity of neutron stars, such impacts have dramatic
effects. For example, it has been proposed that certain specific gamma
ray burst event may have origin in the impact of a solid body (comet
or asteroid) with mass of order 10$^{18}$ g colliding with NS surface
\cite{1981ApJ...248..771C}. The tidal distortion of the body during last 
stages of impact, with intense magnetic field of neutron star leads to strong
compression between magnetic longitudes. The interaction of this with
NS surface material, with exploding material falling back at magnetic
conjugate points was studied in detail in \cite{1981ApJ...248..771C} and it 
was proposed that it could explain specific gamma ray burst events.  In view
of discussions in previous sections, it becomes natural to expect that
such a collision occurring on the surface of a pulsar should lead to
perturbation in the MI tensor of the pulsar, and hence should leave
imprints on the pulses. This interesting possibility is explored in
\cite{asteroid}.  Special feature of this case is that, in this case, one
can determine exact modification of the MI tensor of the pulsar. This is
done by following detailed impact dynamics of the body on the NS surface
as calculated in \cite{1981ApJ...248..771C}.  With the perturbed MI tensor 
known, it is then straightforward to apply the technique discussed above, and
determine the detailed nature of perturbed pulses.  One important unknown
in this case is the NS deformation parameter $\eta$ which determines
the pulse modulations (as discussed above). As resulting change in MI
tensor will have very small magnitude in this case (with $\epsilon$
of order 10$^{-15}$ for a 10$^{18}$ g body impacting on a solar mass
NS), observation of pulse modulations will only be possible for NS
with very small values of $\eta$. Still, this suggests an interesting
technique to probe impact of bodies on pulsar surfaces by observing
pulse modifications.  In particular, for any proposed explanations on
gamma ray bursts etc., one will expect an accompanying observation of
pulse modification (depending on value of $\eta$).

\section{Gravitational waves due to phase transition induced 
density fluctuations}
\label{sec:grav_waves}

Here  we discuss how phase transitions occurring inside neutron stars
may provide a new {\it high frequency} source of gravitational waves
through density fluctuation induced rapidly changing quadrupole moment
of the star.

Neutron stars are considered to be one of the most sought after
sources of gravitational waves. As the emission mechanism from
such a source is governed by its internal structure, searches
for gravitational waves from such a source can provide valuable
information. For example, through a series of works, Thorne et al.
\cite{1967ApJ...149..591T,1969ApJ...155..163P,1969ApJ...158....1T,Thorne:1969rba}
developed a non-radial pulsation theory for a general relativistic
static neutron star model. The author suggested that various mechanisms
can excite quasinormal modes of the parent neutron star and may result
in gravitational wave emission. The mechanisms capable of exciting
quasinormal modes include flaring activity, the formation of hypermassive
neutron stars following the coalescence of binary neutron stars etc.
The mechanism associated with the timing glitch can excite quasinormal
modes in the parent pulsar also. With this motivation, the searches for
the emission of GWs from isolated pulsars began even before the first
ever direct detection of GWs from the binary black hole merger. In this
context, an attempt for GW search associated with the timing glitch in
the Vela pulsar in August 2006 \cite{LIGOScientific:2010epv} is worth
mentioning. Although the above search produced no detectable GWs, with
improving sensitivity of the detectors, continuous attempts in this
direction are believed to yield fruitful results soon.

In the literature, there are a few other theoretical studies where the authors
\cite{Zimmermann:1979ip, Bildsten:1998ey, Jones:2001ui, Horowitz:2009ya, 
Layek:2019ede} have explored the feasibility of emission of gravitational 
waves from isolated pulsars (see the review \cite{Lasky:2015uia} for more
details). These studies considered various reasons for GW emissions. These
include the existence of deformity of neutron stars in the form of
crustal mountains on the surface \cite{Bildsten:1998ey}, or permanent
triaxiality \cite{Jones:2001ui} of the star. This mechanism, as mentioned
above, produces monochromatic GWs, the frequency of which is determined
by the spin frequency of the star. There was also a suggestion that
crustquake (see \ref{sec:observ}) may have a role in generating GWs
\cite{Keer:2014uva,Layek:2019ede}. In reference \cite{Keer:2014uva},
the authors considered the crustquake as a trigger which can excite
various oscillatory modes of the pulsar \cite{Keer:2014uva}, causing
emission of GWs. The role of crustquake as a possible source of GWs was
also suggested in \cite{Layek:2019ede}, where the authors noticed that
the sudden change of the oblateness corresponded to a change of the
quadrupole moment within a very small time scale and crustquake could
indeed cause the bursts of GWs \cite{Layek:2019ede}.  In the context
of bursts of GWs from an isolated pulsar, we will discuss the case
discussed in \cite{Bagchi:2015tna} where density fluctuations induced
during phase transitions were discussed. These density fluctuations were
found to perturb the entire MI tensor of the neutron star. Along with
that, it also necessarily produced non-zero quadrupole moment. With a
very short time scale associated with the phase transition, e.g. for
a QCD phase transition, in \cite{Bagchi:2015tna}, it was suggested
the rapidly evolving quadrupole moment can provide a new source of
gravitational waves.

In \cite{Bagchi:2015tna}, the authors have discussed how density
perturbations due to bubble nucleation or formation of topological defects
during phase transitions change the moment of inertia (MI) tensor of the
star, in general with the addition of off diagonal components in the MI
tensor. In other words, the rotation axis of the star no longer remains
aligned with one of its principal axes. As the shape of the star diverges
from sphericity during phase transition, it develops a quadrupole moment
$Q$ also. The quadrupole moment tensor $Q_{ij}$ ($i,j=1,2,3$) varies
rapidly during phase transition, thereby giving rise to quadrupolar
gravitational radiation. In the weak field limit, the power emitted by
the star through gravitational radiation is given by \cite{Riles:2012yw}
\begin{align}
\frac{dE}{dt} & \,= - \frac{32G}{5c^5} (\Delta Q)^2 \omega^6 \nonumber \\
& \,\approx - (10^{33} J/s) \left( \frac{\Delta Q/I_0}{10^{-6}} \right)^2 \left( \frac{10^{-3}s}{\Delta t} \right)^6 .
\end{align}
Here, we use $\Delta Q$ to represent generic value of $\Delta Q_{ij}$
which is the change in the $ij^{\textrm{th}}$ component of the quadrupole
moment tensor during a time interval $\Delta t$ and $I_0$ is the initial
moment of inertia of the spherical NS.  (The above expressions were
derived in \cite{Riles:2012yw} for a periodically varying quadrupole
moment with angular frequency $\omega$. As our aim here is only to provide
the possibility of phase transition induced density fluctuations being a
new source of GWs, we use these expressions with the typical time scale
of the phase transition as playing the role of the period associated with
$\omega$. The important factors to be noted above are $(\Delta Q)^2$ and
$\omega^6$ leading to $(\Delta t)^{-6}$ dependence for the GW power with
$\Delta t$ being the time scale of the phase transition.)  According to
\cite{Bagchi:2015tna}, $\Delta Q_{ij}/I_0$ lies within the range
$10^{-11} - 10^{-10}$ for phase transition through bubble nucleation and
within the range $10^{-14} - 10^{-10}$ when inhomogeneities in density of
the star arise due to formation of topological defects (see for example
Table 1 of \cite{Bagchi:2015tna}). Though this value is much lower than
the typical value of $10^{-6}$ used for deformed neutron stars, the fact
that the change in $Q$ occurs over a time scale of microseconds (as a
conservative estimate) makes the power radiated by gravitational waves
in this situation significant. An estimate of the strain amplitude $h$
arising from a pulsar at a distance $r$ from the field point is given by
\begin{align}
h & \,= \frac{4\pi^2 G \Delta Q f^2}{c^4 r}  \nonumber \\
& \, \approx 10^{-24} \left( \frac{\Delta Q/I_0}{10^{-6}} \right)^2 \left( \frac{10^{-3}s}{\Delta t} \right)^6 \frac{1kpc}{r} .
\end{align}
Taking $\Delta Q/I_0 \approx 10^{-10}$, $\Delta t \approx
10^{-6} - 10^{-5}$ s, and $r = 1$ kpc, they get $h \approx 10^{-24} -
10^{-22}$. In reality, $\Delta t$ could be smaller, thereby enhancing $h$
and $dE/dt$. Note that as the gravitational wave emission is extremely
short lived, net energy lost by the star is negligible compared to
its mass.

In conclusion, transient changes in MI of a pulsar due to phase transition
induced density perturbations give rise to non-vanishing rapidly varying
quadrupole moment and hence quadrupolar gravitational radiation. In
general, the MI tensor in this case gets off diagonal contributions
that cause wobbling of the pulsar and consequent modification of
the peak pulse intensity. This is a special feature of the model of
\cite{Bagchi:2015tna}. Through observation of modulation of peak pulse
intensity and gravitational waves, it is possible to identify the specific
nature of phase transition occurring within the core of the star.

\section{Pulsar as a Weber detector of gravitational waves}
\label{sec:weber_detector}

So far we discussed the effects of internal dynamics of neutron star 
leading to density fluctuations and changes in the nature of
pulses from the pulsar. In this section, and next, we will  discuss the 
effects of an external gravitational wave on the neutron star configuration. 
It is natural to be sceptic, as expected deformations in NS will be 
extremely tiny. However, at the same time we also recall 
the impressive accuracy of pulsar timing observations, better than
1 part in $10^{15}$. Further, as discussed above in Section 
\ref{sec:pulse_modification}, 
possible changes in the pulse profile from induced wobbling of
pulsar may be large,  even if pulse timing changes remain very small.
We will put forward arguments here that pulsars can effectively act as 
{\it remotely stationed} resonant Weber detectors whose GW perturbed 
signals may be observable on earth. Such a possibility is worth
exploring, even if it requires challenging observations. Many  
gravitational wave detectors (like LIGO/Virgo) are being set up 
around the globe, in order to be able to detect gravitational waves with 
good localization of the source in the sky. This will be complemented by 
future  space-based detectors for the search for gravitational 
wave sources with very wide range of wavelengths and strengths.
However, even all these near earth detectors will be limited in
their scope as most of the powerful gravitational wave sources occur 
very far and also, triangulating the location of the sources will also
be limited. Clearly, if one could have a family of detectors placed 
far away in space, so that their signals could be monitored on earth,
that will immensely boost our ability for detecting and identifying
GW sources.

The discussion in this section is taken primarily from
ref.\cite{Das:2018kvy}.  Basic physics proposed here is, to put
it simply, taking the entire neutron star as  a {\it resonant
Weber detector}. While for the conventional Weber detectors
\cite{Weber:1967jye, Weber:1969bz, Aguiar:2010kn, Husa:2009zz}, GW induced
deformations are detected and converted into electrical signals, for
neutron stars (pulsars), deformations induced by external GWs will be
imprinted on the detailed nature of pulses which can be monitored on
earth. As for the conventional Weber detector, resonance will play
a crucial role for pulsar Weber detectors also which will lead to
amplitude enhancement, and more importantly, the {\it ringing effect}
\cite{Husa:2009zz} which will allow folding of a large number of pulses
to tremendously improve the signal to noise ratio.

It is important to make a distinction between the proposal here for using
pulsar itself as a Weber detector, and the conventional technique of
pulsar timing array (PTA) for GW detection \cite{Hobbs:2017oam}. For PTA,
one monitors the pulse arrival times from a network of pulsars, and looks
for gradual changes due to passage of very low frequency gravitational
waves in the intervening region.  By its very nature, PTA technique is
limited to extremely low frequency sources with frequencies of order
$10^{-6}$ Hz or less.  Such GWs are expected to arise from supermassive
black hole mergers or exotic objects such as cosmic strings. In
contrast, for the technique discussed here, a single pulsar acts as a
GW detector. In principle, even a single pulsar Weber detector can give
some information about the GW source direction through changes in its
spin rate and pulse profile.  In the context of the proposal made here,
we note that it was proposed long ago \cite{Dyson:1969zgf} that with
an array of seismometers, the whole earth may be used as gravitational
wave detector for low frequency GWs. Proposals have also been made
that gravitational waves may be detected due to their effects on nearby
stars, in particular on the solar acoustic modes (helioseismology and
astro-seismology) \cite{Lopes:2017xvq,Lopes:2015pca,Lopes:2014dba}. There
is some similarity in spirit of these proposals with the pulsar Weber
detector proposed here. However, to our knowledge, detection of GW
using its effects on pulse modifications of a pulsar \cite{Das:2018kvy},
had not been been discussed earlier.

Consider a monochromatic gravitational wave with wavelength $\lambda$
from a far away GW source, reaching a neutron star. Passing of GW  means
periodic changes in the Riemann curvature tensor $R_{\mu\nu\lambda\rho}$
which will induce deformations of any body in its path. For a neutron
star, one can calculate changes in its quadrupole moment $Q_{ij}$ induced
by a GW in the static limit (essentially, large wavelength limit for
the GW compared to the neutron star size).  It can be written in the
following form \cite{Hinderer:2007mb}
\begin{equation}
Q_{ij} = - \lambda_d E_{ij}~.
\label{eq:qij}
\end{equation}
Here, $E_{ij} = R_{i0j0}$ is the external tidal field. $\lambda_d$ is called
the tidal deformability,
\begin{equation}
\lambda_d = \frac{2}{3} k_2 \frac{R^5}{G}.  
\label{eq:lambda}
\end{equation}
Here $R$ denotes the equilibrium (undisturbed) radius of the neutron star and 
$k_2$ is called the second Love number. There have been theoretical estimates
of this. For polytropic pressure-density relation 
$P = K \rho^{(1 + 1/n)}$, where $K$ is a constant and $n$ is the 
polytropic index, numerical results (for $0.5 \le n \le 1.0$, and 
$0.1 \le (M/R) \le 0.24$)  can be fitted by the formula \cite{Hinderer:2007mb}. 

\begin{equation}
k_2 \simeq \frac{3}{2} \left(-0.41 + \frac{0.56}{n^{0.33}}\right) 
\left(\frac{M}{R}\right)^{-0.003}
\end{equation}

Remarkably, the BNS merger event detected by LIGO/Virgo 
\cite{LIGOScientific:2017vwq} has been used to put a direct observational 
constraint on the value of $k_2$ to lie within the range, 
$k_2 \simeq 0.05 - 0.15$, and we will be using values of $k_2$ within
this allowed range.

For a GW travelling along z direction, the tidal field $E_{ij}$ can be
calculated for the two polarizations (`$+$', and `$\times$' polarizations)
in the transverse traceless (TT) gauge. We consider the `$+$' polarization
and denote the gravitational wave strain amplitude by $h$ for this
polarization. The resulting tidal field amplitude is given by
\cite{Carroll:2004st},
\begin{equation}
E_{xx} = - E_{yy} = \frac{2\pi^2 h c^2}{\lambda^2},
\label{eq:ett}
\end{equation}
We take the neutron star to have a spherical shape and mass $M$. This
tidal field of the gravitational wave will then induce a quadrupole
moment tensor as given above in Eq. \ref{eq:qij}. Taking 
this deformation to be of an ellipsoidal shape, one can estimate the 
resulting change in the moment of inertia tensor of the neutron star.
\begin{equation}
\frac{\Delta I_{xx}}{I} = - \frac{\Delta I_{yy}}{I}
\simeq \frac{k_2}{3} \frac{R^3 c^2}{GM \lambda^2} 20h,
\label{eq:deltami}
\end{equation}
where, $\Delta I_{xx}$ and $\Delta I_{yy}$ are the changes of MI in 
$I_{xx}$ and $I_{yy}$ respectively. 
For sample values, $M = 1.0 M_\odot$ and $R = 10$ km and $\lambda$ for a 
gravitational wave with 1 kHz frequency (for GW from a typical astrophysical 
source, as detected by LIGO/Virgo). 
\begin{equation}
\frac{\Delta I_{xx}}{I} \simeq 10^{-2} h~.
\label{eq:sampledeltami}
\end{equation}

Here we have used a sample value $k_2$ = 0.1 within the allowed
value from BNS merger event \cite{LIGOScientific:2017vwq}. This event
had the GW source about 130 million light years away from earth, and
had the peak strength of the signal $h \simeq 10^{-19}$.
The advantage of the proposed pulsar Weber detector is that it could
be anywhere in space. Most optimistically, one can even imagine that
a pulsar was about 1 light years away from the BNS undergoing merger.
(Such a possibility can be considered as most neutron stars and 
pulsars arise in globular clusters in our galaxy, so one could imagine 
it to be also true for other galaxies, i.e. for this particular BNS event,
though extragalactic pulsar observations are not very frequent).
The value of $h$ at the location of that pulsar will be 
about $h \simeq 10^{-11}$. As the spin rate change for the pulsar is 
directly related to change in its MI, change in the spin rate of the 
pulsar will be
\begin{equation}
\frac{\Delta \nu}{\nu} = \frac{\Delta I}{I} \simeq 10^{-13}~\,,
\end{equation}
where, $\Delta I$ represent the change in the relevant 
component of MI.  
Given the extreme levels of accuracy of measurements of pulsar signals,
this magnitude of spin rate change is well within observations.
It is important to note that for a generic direction of propagation of 
the GW, the MI tensor will develop non-zero off-diagonal components also.
This will induce wobbling of the pulsar leading to the modulation of the 
pulse intensity profile. Exactly this type of modulation was discussed 
above in Section \ref{sec:pulse_modification}. One important 
result from that section is that even for very tiny changes in the spin rate 
due to changes in the diagonal components of MI tensor, induced wobbling (and 
hence  changes in the pulse profile) may be much larger, especially for 
neutron stars with small deformation parameter $\eta$. Thus profile changes 
may become more important as a signal of GW passing through a pulsar. 
However, in this section we will keep focusing on the spin rate change, as 
profile change discussion is much more involved. Discussion of 
Section \ref{sec:pulse_modification} can be 
straightforwardly applied
to the present case of GW induced changes in the MI tensor.

A crucial requirement for the Weber detector is that it needs to work
at resonance. The resulting increase in amplitude is not too large as GW
signals are short pulses, limiting resonant enhancement of the amplitude.
However, at resonance, the solid detector exhibits the so called
{\it ringing effect}. Ringing effect refers to continued vibration of
the detector in the resonant mode for a long time even after complete 
passing of the GW pulse through the detector. This happens because at
resonance energy absorption from the wave is highly efficient. That
energy has to be dissipated in sound and heat, until then the detector
will keep vibrating.  For example, for a GW pulse of duration a few ms,
at resonance, the vibrations of the Weber bar can continue for time
of order 10 min \cite{Aguiar:2010kn,Husa:2009zz}. With the particular
template for the pulse, a large number of pulses, getting repeated during this
ringing, can be folded. This leads to tremendous increase in the signal to
noise ratio. In the same way, for the pulsar Weber detector working at
resonance, one would expect the  pulsar to continue {\it ringing} for a 
long time after the passing of the GW. This should allow folding of many
pulses to separate this {\it ringing} signal. In fact, this is the
standard way in which  a large number of regular pulses from the pulsar 
are folded, leading to impressive accuracy of pulse timings. The difference
is that regular pulses from the pulsar are periodic, and hardly change
in relevant time scales (apart from occasional glitches), so 
pulse folding is standard. The ringing of pulsar Weber detector shows
that even for a short GW pulse, similar folding of large number of
pulses may be possible.

There is a wide range of resonant frequencies for neutron stars with
frequencies of few Hz all the way up to 20 kHz \cite{Kruger:2014pva}.
For specific modes, the resonant frequencies of NS can be in the Range of
100 Hz to 1 kHz,  which is  precisely the range relevant for a typical
BNS merger GW source, also for typical black hole mergers (with masses
within tens of solar masses).

\subsection{The quality factor {\bf Q} for the neutron star matter}
\vskip .1in

Effectiveness of Weber detector crucially depends on the mechanical quality 
factor {\bf Q} of the detector material. Very high values of {\bf Q} will 
lead to better enhancement in the vibration amplitude. More importantly,
the energy dissipated per vibration cycle will be low, prolonging 
the ringing effect, thereby allowing folding of much larger number 
of pulses for better signal to noise ratio. The  quality factor {\bf Q},
for this purpose, is defined as
\begin{equation}
\textrm{\textbf{Q}} \equiv 2 \pi ~\frac{\textrm{energy stored in the system}}
{\textrm{energy dissipated per cycle}}
\end{equation}
Very large value of {\bf Q} is needed to be able to suppress thermal noise in 
order to detect the gravitational wave impulse \cite{Coccia:1996gw, 
Riles:2012yw, Ju:2000va}.

For the neutron star, it has been argued 
that various viscous effects may not be very important even on time scales of 
the orbital decay time  \cite{Lai:1993di}.  Thus it is entirely possible
that the quality factor {\bf Q} for neutron star interior may be very
large, possibly much larger than the material available on earth for
conventional Weber detectors. Thus, one needs to know the {\bf Q} factor
for NS interior. This is a new challenge for QCD calculations, apart
from the well known problem of determining the equation of state, and
transport coefficients like shear viscosity, the quality factor {\bf Q}
for different phases of QCD (at least for the hadronic and the QGP phases)
needs to be calculated. It is illuminating to quote here from the review
article on {\it Detection of gravitational waves} \cite{Ju:2000va}. In the
section on the Antenna materials for the resonant-mass (Weber) detectors,
it is stated,

{\it ``An ideal resonant bar would consist of a piece of nuclear
matter, with high density and a velocity of sound comparable
to the velocity of light! Since this is not available except in
neutron stars, we must find a form of molecular matter
which, to maximize coupling to gravitational waves, combines
high velocity of sound $v_s$, and high density $\rho$. To reduce 
the thermal noise we require a low acoustic loss $Q^{-1}$"}.

The arguments presented in this section propose that, going with the
spirit of above quotation, we realize that neutron star material
can possibly be available for use in a Weber detector, in fact
by using the entire neutron star itself as a pulsar Weber detector.

\section{Re-visiting past gravitational waves via pulsars as 
Weber detectors}
\label{sec:revist}

An interesting outcome of the possibility of using pulsars as
Weber gravitational wave detectors spread out in
the cosmos is that they can be used to revisit past GW events, like
collisions of black holes, neutron stars, supernova explosions etc.,
again and again. These past events are taken to be such that their
signals have already passed through Earth. They might have been detected
by LIGO/Virgo or their signals might have been missed. Supernovae within
our galaxy whose existence have been deduced from astronomical data
collected around the world fall within the latter category.  This novel
idea has been proposed in \cite{Biswal:2019szu}.
\begin{figure}
\centering \includegraphics[width=0.9\linewidth]{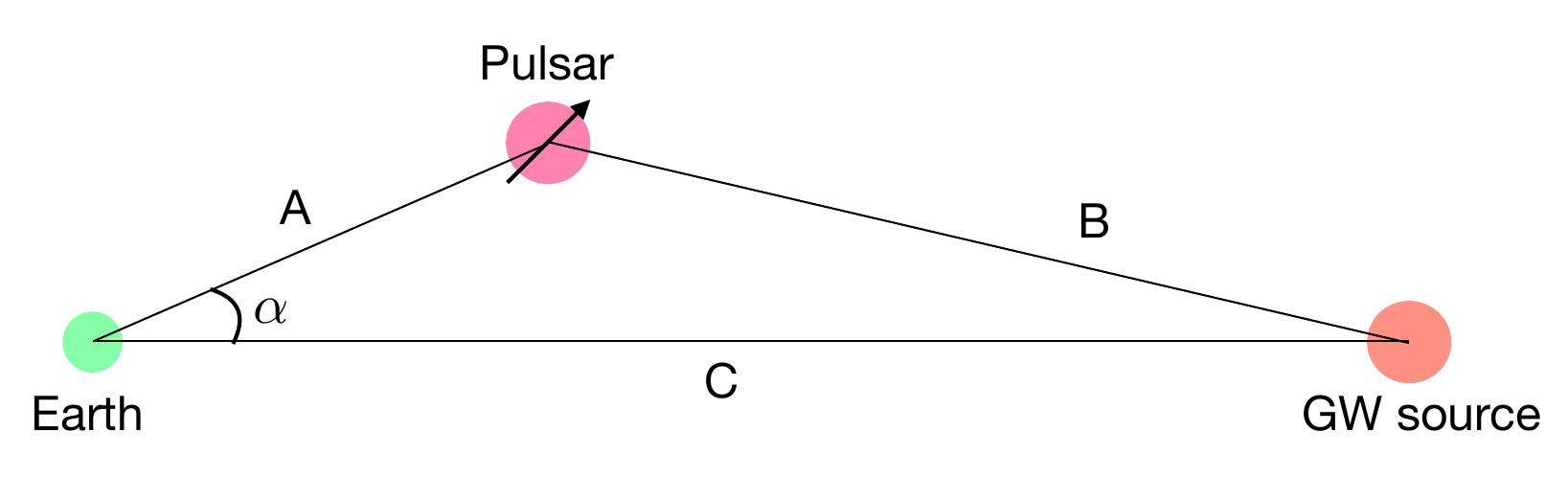} 
\caption{Schematic diagram showing the relative positions of Earth, a distant 
GW source and a pulsar whose change in pulse profile and spin rate we can measure
on Earth. (Fig. taken from \cite{Biswal:2019szu}.)}
\label{fig:trio}
\end{figure}
Figure \ref{fig:trio} schematically represents the situation we are
talking about. Gravitational waves from the source directly travel to
Earth via path C but it also travels to a pulsar via path B, transiently
inducing a quadrupole moment in the pulsar and changing its moment of
inertia and hence its pulse frequency and profile. The modified pulses
reach earth via path A. If $r_{A,B,C}$ are the distances along paths A,
B, C respectively then the path length difference between the direct
path C and the path A+B via a pulsar is,
\begin{align}
\Delta r &= (r_A + r_B) - r_C   \nonumber \\
&= r_A + (r_A^2 + r_C^2 -2 r_A r_C \cos \alpha)^{1/2} - r_C .      
\label{eq:deltar}
\end{align}
Thus, the indirect signal will be detected on Earth a time $t_0$ after the 
arrival of the direct signal, where
\begin{align}
t_0=\frac{\Delta r}{c} .
\end{align}
Angle $\alpha$ between the directions of the GW source and the pulsar is
given by $\alpha = \cos^{-1}(\sin \theta_p \sin \theta_s + \cos \theta_p
\cos \theta_s \cos (\phi_p -\phi_s))$, $\theta_{p,s}$ and $\phi_{p,s}$
being respectively the `Declination' and `Right Ascension' angles for the
pulsar and GW source. Signals from pulsars that lie far away from Earth
(i.e., $r_A$ is relatively large) will arrive on Earth within a reasonably
short interval after its direct detection only if $\alpha$ is small. For
$\alpha$ close to zero, $r_A$ doesn't matter in Eq. \ref{eq:deltar}
and there is almost no time delay between the direct and indirect
signals. (Thus, errors in pulsar distances do not affect the values of
$t_0$ for small $\alpha$.) In \cite{Biswal:2019szu}, the authors have
presented an elaborate list of GW events, pulsars through whose signals
the events may be revisited and the range of computed arrival times of
the affected signals from the pulsars, accommodating for known errors
wherever possible. Their data is available for past events whose earliest
signal arrival dates lie within the next 100y after 1967 and also whose
uncertainty in signal arrival dates is limited to within 100y. A notable
example among many interesting cases is the earliest recorded supernova
event SN185 that may become observable again via pulsars J0900-3144
and J1858-2216 between 2016-2049. A few other interesting cases are of
supernovae SN1885 whose perturbed signal via pulsar B2310+42 is expected
to reach Earth between 2022-2044 and SN1604 whose signal perturbed by
pulsar J1813-1246 should reach between 1971-2052.

Thus, gravitational waves from events that have travelled through pulsars
strewn in the sky can be observed again, not only once but multiple times,
through the observation of perturbed pulses from these pulsars. Nature
of modification of pulsar timing and pulse profile also depends on the
relative directions of GW propagation and pulsar spin. So, even a single
indirect pulsar-detector observation can be used to get information
about source direction. However, an important question that may arise
here is how would one determine whether the observed change in pulse
profile/timing is indeed due to passage of gravitational waves and not
some other phenomena like glitches or phase transition induced density
perturbations in the core of the star. Of course, a careful analysis
using detailed characteristics of the incoming GW signal (its profile
and direction), the internal structure of the pulsar and the direction of
observation on Earth is needed to confirm this. Yet, even without that,
one can keep in mind the simple physical fact that any change in pulse
profile due to passing gravitational waves will be temporary and the spin
rate/pulse profile of the pulsar will be restored to the original value,
unlike cases where the density perturbations arising from 
phase transitions are the source of pulse modification.
Except an important effect that there will in general be a net shift of the 
angular position of the pulse if there was any such transient GW event. 

This is exactly the same effect which was discussed in Section 
\ref{sec:pulse_modification} for any transient effect
of phase transition induced density fluctuations. Thus, a net, residual
shift of the angular position of the pulse could signal a missed phase
transition or a missed GW event. To further distinguish between these
two possibilities, we note that any phase transition will necessarily
also lead to free energy changes which are permanent, thereby leading
to permanent change in spin rate. In contrast, GW event is genuinely
transient, without any permanent change of the NS structure. Thus the
original spin rate will be completely restored. Note
that pulse timing is also changed due to the glitch events. However, the
features of pulsar glitches \cite{Espinoza:2011pq} (i.e., the glitch size,
post-glitch recovery period, etc.) are well-studied and reasonably
understood. One important characteristic feature of glitch events is the
post-glitch recovery phase. Most of the pulsars are observed to recover
their spin frequency monotonically and quasi-exponentially like Vela;
{\it overshoot} is also observed in pulsars like the Crab
\cite{vanEysden:2010ha}. Most importantly, irrespective of the behaviour
of post-glitch relaxation, the long recovery time scale (a few days to
weeks) associated with the pulsar glitches should be able to distinguish
such sources through the pulsar timing.

\section{Conclusions and future directions}
\label{sec:conclusions}

Main focus of this brief review is on certain specific properties
of pulsars, (which are rotating neutron stars), namely the extreme
accuracy of observations of its pulses, and very sensitive dependence
of the detailed nature of pulses, as observed on earth, on the internal
structure of the pulsar. Probes of internal structure of neutron stars
is of paramount importance in astrophysics. This has acquired special
importance in view of the exciting detection of GWs from BNS merger events
which have allowed direct probe of internal structure of the neutron
star matter due to tidal deformation of the neutron stars during last
stages of coalescence. Probing neutron star core structure is of great
importance for understanding of the rich spectrum of various exotic high
baryon density phases of QCD.

This part of the QCD phase diagram has so far eluded experimental
observations in terrestrial experiments (relativistic heavy-ion collision
experiments). Various estimates suggest that extreme baryon densities
needed for several such phases may only become accessible inside neutron
star cores. In fact, even at baryon densities which
are/may become accessible in low energy heavy-ion collision experiments,
the low temperatures available in neutrons stars are simply not possible
in heavy-ion collision experiments. Configuration of a pulsar will
be perturbed during any phase changes occurring in its core, or if
the pulsar is affected by external influences, such as an external
gravitational wave.  This may leave imprints on the pulses coming from
the pulsar as the detailed nature of pulses is extremely sensitive to
any changes occurring in the rotational dynamics of the pulsar, hence
to any changes occurring in the configuration of a pulsar. This makes
pulsar observations as powerful probe of dynamical processes affecting
pulsar structure.

We have discussed how internal changes in pulsar due to various phase
transitions occurring in the pulsar cores can leave detailed imprints on
pulses. Changes in the equation of state directly affects the spin rate
by affecting the diagonal components of the moment of inertia tensor. At
the same time, phase transition induced random density fluctuations
affect entire MI tensor, including its off-diagonal components. This
induces wobbling of the pulsar, leading to modulation of pulses as
observed on earth.  It may thus be possible to infer about the detailed
nature of specific phase transitions occurring in the pulsar core by
detailed analysis of pulsar observations. Random density fluctuations
during phase transitions also induce rapidly changing quadrupole moment
of the neutron star, providing one more possible source of GW emission
from neutron stars.

A particular interesting possibility arises when one considers pulsars
in the presence of external gravitational waves. While acknowledging the
fact that GWs induce extremely tiny configurational changes in bodies,
it has been proposed that extreme accuracy of pulsar observations
may allow detection of such GWs, especially for GW frequencies in the
resonant bands of the neutron star oscillations. This is precisely the
physics of Weber detectors of gravitational waves. Pulsars, thus can
act as remotely stationed resonant Weber detectors of GWs, with the
signal of GW getting transmitted to earth in form of perturbations on
the pulses. This also allows for very exciting possibility of detecting
gravitational wave events of past, that is cases when GWs from distant
sources passed through the earth in past. Same GWs also reach pulsars
affecting its pulses. Those perturbed pulses are detected on earth much
later, allowing us to re-visit past gravitational wave events.

Admittedly, the feasibility of this proposal of using pulsars as Weber
detectors leaves many questions unanswered. Even with the ringing effect,
which will allow for the folding of very large number  of pulses, it is
not very clear  what level of GW strain amplitude will be observable by
analysis of pulsar signals. The extreme accuracy of pulsar signals
is also achieved only by folding huge number of pulses. One thing which
will come to help here is that pulsar will probably be most perfect Weber
detector possible, formed with extreme density QCD matter (recall the
quotation in Section IX.A from ref. \cite{Ju:2000va}). Further, ideal situation
will be to have a BNS system where at least one partner is a pulsar. If one
neutron star emits GW, then the partner pulsar will carry large imprints of
that signal to earth. Pulsar in this  sense becomes permanent probe of any
micro changes happening in the interior of the partner neutron star.

The most important aspect of pulsar being Weber detector will be that the
detector has possibility of being very close to the GW source. It is clear,
that the proposal will be most effective if extra-galactic pulsars can be
monitored with great accuracy. Thus, importance of extra-galactic pulsar
may be viewed in this term also, that, with these {\it pulsar Weber
detectors} being out there, chances of a powerful GW source being close-by
may be significant, allowing us to see the imprints of those GWs
on the signals of that pulsar.

\section*{Acknowledgment}
This article is dedicated to the loving memory of Abira Sarkar. We also
deeply acknowledge her immense support during the writing of this article.

\bibliography{revpulsar} 

\begin{thebibliography}{142}%
\makeatletter
\providecommand \@ifxundefined [1]{%
 \@ifx{#1\undefined}
}%
\providecommand \@ifnum [1]{%
 \ifnum #1\expandafter \@firstoftwo
 \else \expandafter \@secondoftwo
 \fi
}%
\providecommand \@ifx [1]{%
 \ifx #1\expandafter \@firstoftwo
 \else \expandafter \@secondoftwo
 \fi
}%
\providecommand \natexlab [1]{#1}%
\providecommand \enquote  [1]{``#1''}%
\providecommand \bibnamefont  [1]{#1}%
\providecommand \bibfnamefont [1]{#1}%
\providecommand \citenamefont [1]{#1}%
\providecommand \href@noop [0]{\@secondoftwo}%
\providecommand \href [0]{\begingroup \@sanitize@url \@href}%
\providecommand \@href[1]{\@@startlink{#1}\@@href}%
\providecommand \@@href[1]{\endgroup#1\@@endlink}%
\providecommand \@sanitize@url [0]{\catcode `\\12\catcode `\$12\catcode
  `\&12\catcode `\#12\catcode `\^12\catcode `\_12\catcode `\%12\relax}%
\providecommand \@@startlink[1]{}%
\providecommand \@@endlink[0]{}%
\providecommand \url  [0]{\begingroup\@sanitize@url \@url }%
\providecommand \@url [1]{\endgroup\@href {#1}{\urlprefix }}%
\providecommand \urlprefix  [0]{URL }%
\providecommand \Eprint [0]{\href }%
\providecommand \doibase [0]{https://doi.org/}%
\providecommand \selectlanguage [0]{\@gobble}%
\providecommand \bibinfo  [0]{\@secondoftwo}%
\providecommand \bibfield  [0]{\@secondoftwo}%
\providecommand \translation [1]{[#1]}%
\providecommand \BibitemOpen [0]{}%
\providecommand \bibitemStop [0]{}%
\providecommand \bibitemNoStop [0]{.\EOS\space}%
\providecommand \EOS [0]{\spacefactor3000\relax}%
\providecommand \BibitemShut  [1]{\csname bibitem#1\endcsname}%
\let\auto@bib@innerbib\@empty
\bibitem [{\citenamefont {Csernai}(1994)}]{Csernai:1994xw}%
  \BibitemOpen
  \bibfield  {author} {\bibinfo {author} {\bibfnamefont {L.~P.}\ \bibnamefont
  {Csernai}},\ }\href@noop {} {\emph {\bibinfo {title} {{Introduction to
  relativistic heavy ion collisions}}}}\ (\bibinfo {year} {1994})\BibitemShut
  {NoStop}%
\bibitem [{\citenamefont {Wong}(1995)}]{Wong:1995jf}%
  \BibitemOpen
  \bibfield  {author} {\bibinfo {author} {\bibfnamefont {C.~Y.}\ \bibnamefont
  {Wong}},\ }\href@noop {} {\emph {\bibinfo {title} {{Introduction to
  high-energy heavy ion collisions}}}}\ (\bibinfo {year} {1995})\BibitemShut
  {NoStop}%
\bibitem [{\citenamefont {Heinz}\ and\ \citenamefont
  {Snellings}(2013)}]{Heinz:2013th}%
  \BibitemOpen
  \bibfield  {author} {\bibinfo {author} {\bibfnamefont {U.}~\bibnamefont
  {Heinz}}\ and\ \bibinfo {author} {\bibfnamefont {R.}~\bibnamefont
  {Snellings}},\ }\bibfield  {title} {\bibinfo {title} {{Collective flow and
  viscosity in relativistic heavy-ion collisions}},\ }\href
  {https://doi.org/10.1146/annurev-nucl-102212-170540} {\bibfield  {journal}
  {\bibinfo  {journal} {Ann. Rev. Nucl. Part. Sci.}\ }\textbf {\bibinfo
  {volume} {63}},\ \bibinfo {pages} {123} (\bibinfo {year} {2013})},\ \Eprint
  {https://arxiv.org/abs/1301.2826} {arXiv:1301.2826 [nucl-th]} \BibitemShut
  {NoStop}%
\bibitem [{\citenamefont {Rajagopal}\ and\ \citenamefont
  {Wilczek}(2000)}]{Rajagopal:2000wf}%
  \BibitemOpen
  \bibfield  {author} {\bibinfo {author} {\bibfnamefont {K.}~\bibnamefont
  {Rajagopal}}\ and\ \bibinfo {author} {\bibfnamefont {F.}~\bibnamefont
  {Wilczek}},\ }\bibinfo {title} {{The Condensed matter physics of QCD}},\ in\
  \href {https://doi.org/10.1142/9789812810458_0043} {\emph {\bibinfo
  {booktitle} {{At the frontier of particle physics. Handbook of QCD. Vol.
  1-3}}}},\ \bibinfo {editor} {edited by\ \bibinfo {editor} {\bibfnamefont
  {M.}~\bibnamefont {Shifman}}\ and\ \bibinfo {editor} {\bibfnamefont
  {B.}~\bibnamefont {Ioffe}}}\ (\bibinfo {year} {2000})\ pp.\ \bibinfo {pages}
  {2061--2151},\ \Eprint {https://arxiv.org/abs/hep-ph/0011333}
  {arXiv:hep-ph/0011333} \BibitemShut {NoStop}%
\bibitem [{\citenamefont {Alford}\ \emph {et~al.}(2008)\citenamefont {Alford},
  \citenamefont {Schmitt}, \citenamefont {Rajagopal},\ and\ \citenamefont
  {Sch\"afer}}]{Alford:2007xm}%
  \BibitemOpen
  \bibfield  {author} {\bibinfo {author} {\bibfnamefont {M.~G.}\ \bibnamefont
  {Alford}}, \bibinfo {author} {\bibfnamefont {A.}~\bibnamefont {Schmitt}},
  \bibinfo {author} {\bibfnamefont {K.}~\bibnamefont {Rajagopal}},\ and\
  \bibinfo {author} {\bibfnamefont {T.}~\bibnamefont {Sch\"afer}},\ }\bibfield
  {title} {\bibinfo {title} {{Color superconductivity in dense quark matter}},\
  }\href {https://doi.org/10.1103/RevModPhys.80.1455} {\bibfield  {journal}
  {\bibinfo  {journal} {Rev. Mod. Phys.}\ }\textbf {\bibinfo {volume} {80}},\
  \bibinfo {pages} {1455} (\bibinfo {year} {2008})},\ \Eprint
  {https://arxiv.org/abs/0709.4635} {arXiv:0709.4635 [hep-ph]} \BibitemShut
  {NoStop}%
\bibitem [{\citenamefont {Rajagopal}(2002)}]{Rajagopal:2001ngu}%
  \BibitemOpen
  \bibfield  {author} {\bibinfo {author} {\bibfnamefont {K.}~\bibnamefont
  {Rajagopal}},\ }\bibinfo {title} {Color superconductivity},\ in\ \href
  {https://doi.org/10.1007/978-94-010-0267-7_16} {\emph {\bibinfo {booktitle}
  {QCD Perspectives on Hot and Dense Matter}}}\ (\bibinfo  {publisher}
  {Springer Netherlands},\ \bibinfo {address} {Dordrecht},\ \bibinfo {year}
  {2002})\ p.\ \bibinfo {pages} {475}\BibitemShut {NoStop}%
\bibitem [{\citenamefont {Alford}\ \emph
  {et~al.}(2001{\natexlab{a}})\citenamefont {Alford}, \citenamefont {Bowers},\
  and\ \citenamefont {Rajagopal}}]{Alford:2000sx}%
  \BibitemOpen
  \bibfield  {author} {\bibinfo {author} {\bibfnamefont {M.~G.}\ \bibnamefont
  {Alford}}, \bibinfo {author} {\bibfnamefont {J.~A.}\ \bibnamefont {Bowers}},\
  and\ \bibinfo {author} {\bibfnamefont {K.}~\bibnamefont {Rajagopal}},\
  }\bibfield  {title} {\bibinfo {title} {{Color superconductivity in compact
  stars}},\ }\href {https://doi.org/10.1088/0954-3899/27/3/335} {\bibfield
  {journal} {\bibinfo  {journal} {J. Phys. G}\ }\textbf {\bibinfo {volume}
  {27}},\ \bibinfo {pages} {541} (\bibinfo {year} {2001}{\natexlab{a}})},\
  \Eprint {https://arxiv.org/abs/hep-ph/0009357} {arXiv:hep-ph/0009357}
  \BibitemShut {NoStop}%
\bibitem [{\citenamefont {Witten}(1984)}]{Witten:1984rs}%
  \BibitemOpen
  \bibfield  {author} {\bibinfo {author} {\bibfnamefont {E.}~\bibnamefont
  {Witten}},\ }\bibfield  {title} {\bibinfo {title} {{Cosmic Separation of
  Phases}},\ }\href {https://doi.org/10.1103/PhysRevD.30.272} {\bibfield
  {journal} {\bibinfo  {journal} {Phys. Rev. D}\ }\textbf {\bibinfo {volume}
  {30}},\ \bibinfo {pages} {272} (\bibinfo {year} {1984})}\BibitemShut
  {NoStop}%
\bibitem [{\citenamefont {Farhi}\ and\ \citenamefont
  {Jaffe}(1984)}]{Farhi:1984qu}%
  \BibitemOpen
  \bibfield  {author} {\bibinfo {author} {\bibfnamefont {E.}~\bibnamefont
  {Farhi}}\ and\ \bibinfo {author} {\bibfnamefont {R.~L.}\ \bibnamefont
  {Jaffe}},\ }\bibfield  {title} {\bibinfo {title} {{Strange Matter}},\ }\href
  {https://doi.org/10.1103/PhysRevD.30.2379} {\bibfield  {journal} {\bibinfo
  {journal} {Phys. Rev. D}\ }\textbf {\bibinfo {volume} {30}},\ \bibinfo
  {pages} {2379} (\bibinfo {year} {1984})}\BibitemShut {NoStop}%
\bibitem [{\citenamefont {Alcock}\ \emph {et~al.}(1986)\citenamefont {Alcock},
  \citenamefont {Farhi},\ and\ \citenamefont {Olinto}}]{Alcock:1986hz}%
  \BibitemOpen
  \bibfield  {author} {\bibinfo {author} {\bibfnamefont {C.}~\bibnamefont
  {Alcock}}, \bibinfo {author} {\bibfnamefont {E.}~\bibnamefont {Farhi}},\ and\
  \bibinfo {author} {\bibfnamefont {A.}~\bibnamefont {Olinto}},\ }\bibfield
  {title} {\bibinfo {title} {{Strange stars}},\ }\href
  {https://doi.org/10.1086/164679} {\bibfield  {journal} {\bibinfo  {journal}
  {Astrophys. J.}\ }\textbf {\bibinfo {volume} {310}},\ \bibinfo {pages} {261}
  (\bibinfo {year} {1986})}\BibitemShut {NoStop}%
\bibitem [{\citenamefont {Horvath}\ \emph {et~al.}(1992)\citenamefont
  {Horvath}, \citenamefont {Benvenuto},\ and\ \citenamefont
  {Vucetich}}]{Horvath:1992wq}%
  \BibitemOpen
  \bibfield  {author} {\bibinfo {author} {\bibfnamefont {J.~E.}\ \bibnamefont
  {Horvath}}, \bibinfo {author} {\bibfnamefont {O.~G.}\ \bibnamefont
  {Benvenuto}},\ and\ \bibinfo {author} {\bibfnamefont {H.}~\bibnamefont
  {Vucetich}},\ }\bibfield  {title} {\bibinfo {title} {{Nucleation of strange
  matter in dense stellar cores}},\ }\href
  {https://doi.org/10.1103/PhysRevD.45.3865} {\bibfield  {journal} {\bibinfo
  {journal} {Phys. Rev. D}\ }\textbf {\bibinfo {volume} {45}},\ \bibinfo
  {pages} {3865} (\bibinfo {year} {1992})}\BibitemShut {NoStop}%
\bibitem [{\citenamefont {Weber}\ \emph {et~al.}(1994)\citenamefont {Weber},
  \citenamefont {Kettner}, \citenamefont {Weigel},\ and\ \citenamefont
  {Glendenning}}]{Weber:1994yx}%
  \BibitemOpen
  \bibfield  {author} {\bibinfo {author} {\bibfnamefont {F.}~\bibnamefont
  {Weber}}, \bibinfo {author} {\bibfnamefont {C.}~\bibnamefont {Kettner}},
  \bibinfo {author} {\bibfnamefont {M.~K.}\ \bibnamefont {Weigel}},\ and\
  \bibinfo {author} {\bibfnamefont {N.~K.}\ \bibnamefont {Glendenning}},\
  }\bibfield  {title} {\bibinfo {title} {{Strange matter stars}},\ }in\
  \href@noop {} {\emph {\bibinfo {booktitle} {{International Symposium on
  Strangeness and Quark Matter}}}}\ (\bibinfo {year} {1994})\ pp.\ \bibinfo
  {pages} {0308--317}\BibitemShut {NoStop}%
\bibitem [{\citenamefont {Yang}\ and\ \citenamefont
  {Clark}(1971)}]{Yang:1971oux}%
  \BibitemOpen
  \bibfield  {author} {\bibinfo {author} {\bibfnamefont {C.~H.}\ \bibnamefont
  {Yang}}\ and\ \bibinfo {author} {\bibfnamefont {J.~W.}\ \bibnamefont
  {Clark}},\ }\bibfield  {title} {\bibinfo {title} {{Superfluid condensation
  energy of neutron matter}},\ }\href
  {https://doi.org/10.1016/0375-9474(71)91002-5} {\bibfield  {journal}
  {\bibinfo  {journal} {Nucl. Phys. A}\ }\textbf {\bibinfo {volume} {174}},\
  \bibinfo {pages} {49} (\bibinfo {year} {1971})}\BibitemShut {NoStop}%
\bibitem [{\citenamefont {Prakash}\ \emph {et~al.}(2001)\citenamefont
  {Prakash}, \citenamefont {Lattimer}, \citenamefont {Pons}, \citenamefont
  {Steiner},\ and\ \citenamefont {Reddy}}]{Prakash:2000jr}%
  \BibitemOpen
  \bibfield  {author} {\bibinfo {author} {\bibfnamefont {M.}~\bibnamefont
  {Prakash}}, \bibinfo {author} {\bibfnamefont {J.~M.}\ \bibnamefont
  {Lattimer}}, \bibinfo {author} {\bibfnamefont {J.~A.}\ \bibnamefont {Pons}},
  \bibinfo {author} {\bibfnamefont {A.~W.}\ \bibnamefont {Steiner}},\ and\
  \bibinfo {author} {\bibfnamefont {S.}~\bibnamefont {Reddy}},\ }\bibfield
  {title} {\bibinfo {title} {{Evolution of a neutron star from its birth to old
  age}},\ }\href@noop {} {\bibfield  {journal} {\bibinfo  {journal} {Lect.
  Notes Phys.}\ }\textbf {\bibinfo {volume} {578}},\ \bibinfo {pages} {364}
  (\bibinfo {year} {2001})},\ \Eprint {https://arxiv.org/abs/astro-ph/0012136}
  {arXiv:astro-ph/0012136} \BibitemShut {NoStop}%
\bibitem [{\citenamefont {Hulse}\ and\ \citenamefont
  {Taylor}(1975)}]{Hulse:1974eb}%
  \BibitemOpen
  \bibfield  {author} {\bibinfo {author} {\bibfnamefont {R.~A.}\ \bibnamefont
  {Hulse}}\ and\ \bibinfo {author} {\bibfnamefont {J.~H.}\ \bibnamefont
  {Taylor}},\ }\bibfield  {title} {\bibinfo {title} {{Discovery of a pulsar in
  a binary system}},\ }\href {https://doi.org/10.1086/181708} {\bibfield
  {journal} {\bibinfo  {journal} {Astrophys. J. Lett.}\ }\textbf {\bibinfo
  {volume} {195}},\ \bibinfo {pages} {L51} (\bibinfo {year}
  {1975})}\BibitemShut {NoStop}%
\bibitem [{\citenamefont {Taylor}\ and\ \citenamefont
  {Weisberg}(1982)}]{Taylor:1982zz}%
  \BibitemOpen
  \bibfield  {author} {\bibinfo {author} {\bibfnamefont {J.~H.}\ \bibnamefont
  {Taylor}}\ and\ \bibinfo {author} {\bibfnamefont {J.~M.}\ \bibnamefont
  {Weisberg}},\ }\bibfield  {title} {\bibinfo {title} {{A new test of general
  relativity: Gravitational radiation and the binary pulsar PS R 1913+16}},\
  }\href {https://doi.org/10.1086/159690} {\bibfield  {journal} {\bibinfo
  {journal} {Astrophys. J.}\ }\textbf {\bibinfo {volume} {253}},\ \bibinfo
  {pages} {908} (\bibinfo {year} {1982})}\BibitemShut {NoStop}%
\bibitem [{\citenamefont {Abbott}\ \emph {et~al.}(2016)\citenamefont {Abbott}
  \emph {et~al.}}]{LIGOScientific:2016aoc}%
  \BibitemOpen
  \bibfield  {author} {\bibinfo {author} {\bibfnamefont {B.~P.}\ \bibnamefont
  {Abbott}} \emph {et~al.} (\bibinfo {collaboration} {LIGO Scientific,
  Virgo}),\ }\bibfield  {title} {\bibinfo {title} {{Observation of
  Gravitational Waves from a Binary Black Hole Merger}},\ }\href
  {https://doi.org/10.1103/PhysRevLett.116.061102} {\bibfield  {journal}
  {\bibinfo  {journal} {Phys. Rev. Lett.}\ }\textbf {\bibinfo {volume} {116}},\
  \bibinfo {pages} {061102} (\bibinfo {year} {2016})},\ \Eprint
  {https://arxiv.org/abs/1602.03837} {arXiv:1602.03837 [gr-qc]} \BibitemShut
  {NoStop}%
\bibitem [{\citenamefont {Abbott}\ \emph {et~al.}(2017)\citenamefont {Abbott}
  \emph {et~al.}}]{LIGOScientific:2017vwq}%
  \BibitemOpen
  \bibfield  {author} {\bibinfo {author} {\bibfnamefont {B.~P.}\ \bibnamefont
  {Abbott}} \emph {et~al.} (\bibinfo {collaboration} {LIGO Scientific,
  Virgo}),\ }\bibfield  {title} {\bibinfo {title} {{GW170817: Observation of
  Gravitational Waves from a Binary Neutron Star Inspiral}},\ }\href
  {https://doi.org/10.1103/PhysRevLett.119.161101} {\bibfield  {journal}
  {\bibinfo  {journal} {Phys. Rev. Lett.}\ }\textbf {\bibinfo {volume} {119}},\
  \bibinfo {pages} {161101} (\bibinfo {year} {2017})},\ \Eprint
  {https://arxiv.org/abs/1710.05832} {arXiv:1710.05832 [gr-qc]} \BibitemShut
  {NoStop}%
\bibitem [{\citenamefont {Bagchi}\ \emph {et~al.}(2015)\citenamefont {Bagchi},
  \citenamefont {Das}, \citenamefont {Layek},\ and\ \citenamefont
  {Srivastava}}]{Bagchi:2015tna}%
  \BibitemOpen
  \bibfield  {author} {\bibinfo {author} {\bibfnamefont {P.}~\bibnamefont
  {Bagchi}}, \bibinfo {author} {\bibfnamefont {A.}~\bibnamefont {Das}},
  \bibinfo {author} {\bibfnamefont {B.}~\bibnamefont {Layek}},\ and\ \bibinfo
  {author} {\bibfnamefont {A.~M.}\ \bibnamefont {Srivastava}},\ }\bibfield
  {title} {\bibinfo {title} {{Effects of phase transition induced density
  fluctuations on pulsar dynamics}},\ }\href
  {https://doi.org/10.1016/j.physletb.2015.05.055} {\bibfield  {journal}
  {\bibinfo  {journal} {Phys. Lett. B}\ }\textbf {\bibinfo {volume} {747}},\
  \bibinfo {pages} {120} (\bibinfo {year} {2015})},\ \Eprint
  {https://arxiv.org/abs/1506.03287} {arXiv:1506.03287 [astro-ph.HE]}
  \BibitemShut {NoStop}%
\bibitem [{\citenamefont {Srivastava}\ \emph {et~al.}(2017)\citenamefont
  {Srivastava}, \citenamefont {Bagchi}, \citenamefont {Das},\ and\
  \citenamefont {Layek}}]{Srivastava:2017itj}%
  \BibitemOpen
  \bibfield  {author} {\bibinfo {author} {\bibfnamefont {A.~M.}\ \bibnamefont
  {Srivastava}}, \bibinfo {author} {\bibfnamefont {P.}~\bibnamefont {Bagchi}},
  \bibinfo {author} {\bibfnamefont {A.}~\bibnamefont {Das}},\ and\ \bibinfo
  {author} {\bibfnamefont {B.}~\bibnamefont {Layek}},\ }\bibfield  {title}
  {\bibinfo {title} {{High-density QCD phase transitions inside neutron stars:
  Glitches and gravitational waves}},\ }\href
  {https://doi.org/10.1007/s12043-017-1465-1} {\bibfield  {journal} {\bibinfo
  {journal} {Pramana}\ }\textbf {\bibinfo {volume} {89}},\ \bibinfo {pages}
  {68} (\bibinfo {year} {2017})}\BibitemShut {NoStop}%
\bibitem [{\citenamefont {Bagchi}\ \emph {et~al.}(2022)\citenamefont {Bagchi},
  \citenamefont {Layek}, \citenamefont {Sarkar},\ and\ \citenamefont
  {Srivastava}}]{Bagchi:2021etv}%
  \BibitemOpen
  \bibfield  {author} {\bibinfo {author} {\bibfnamefont {P.}~\bibnamefont
  {Bagchi}}, \bibinfo {author} {\bibfnamefont {B.}~\bibnamefont {Layek}},
  \bibinfo {author} {\bibfnamefont {A.}~\bibnamefont {Sarkar}},\ and\ \bibinfo
  {author} {\bibfnamefont {A.~M.}\ \bibnamefont {Srivastava}},\ }\bibfield
  {title} {\bibinfo {title} {{Modulation of pulse profile as a signal for phase
  transitions in a pulsar core}},\ }\href
  {https://doi.org/10.1093/mnras/stac1062} {\bibfield  {journal} {\bibinfo
  {journal} {Mon. Not. Roy. Astron. Soc.}\ }\textbf {\bibinfo {volume} {513}},\
  \bibinfo {pages} {2794} (\bibinfo {year} {2022})},\ \Eprint
  {https://arxiv.org/abs/2111.10805} {arXiv:2111.10805 [astro-ph.HE]}
  \BibitemShut {NoStop}%
\bibitem [{\citenamefont {Das}\ \emph {et~al.}(2019)\citenamefont {Das},
  \citenamefont {Dave}, \citenamefont {Ganguly},\ and\ \citenamefont
  {Srivastava}}]{Das:2018kvy}%
  \BibitemOpen
  \bibfield  {author} {\bibinfo {author} {\bibfnamefont {A.}~\bibnamefont
  {Das}}, \bibinfo {author} {\bibfnamefont {S.~S.}\ \bibnamefont {Dave}},
  \bibinfo {author} {\bibfnamefont {O.}~\bibnamefont {Ganguly}},\ and\ \bibinfo
  {author} {\bibfnamefont {A.~M.}\ \bibnamefont {Srivastava}},\ }\bibfield
  {title} {\bibinfo {title} {{Pulsars as Weber gravitational wave detectors}},\
  }\href {https://doi.org/10.1016/j.physletb.2019.02.031} {\bibfield  {journal}
  {\bibinfo  {journal} {Phys. Lett. B}\ }\textbf {\bibinfo {volume} {791}},\
  \bibinfo {pages} {167} (\bibinfo {year} {2019})},\ \Eprint
  {https://arxiv.org/abs/1804.00453} {arXiv:1804.00453 [astro-ph.HE]}
  \BibitemShut {NoStop}%
\bibitem [{\citenamefont {Biswal}\ \emph {et~al.}(2020)\citenamefont {Biswal},
  \citenamefont {Dave},\ and\ \citenamefont {Srivastava}}]{Biswal:2019szu}%
  \BibitemOpen
  \bibfield  {author} {\bibinfo {author} {\bibfnamefont {M.}~\bibnamefont
  {Biswal}}, \bibinfo {author} {\bibfnamefont {S.~S.}\ \bibnamefont {Dave}},\
  and\ \bibinfo {author} {\bibfnamefont {A.~M.}\ \bibnamefont {Srivastava}},\
  }\bibfield  {title} {\bibinfo {title} {{Re-visiting gravitational wave events
  via pulsars}},\ }\href {https://doi.org/10.1016/j.physletb.2020.135887}
  {\bibfield  {journal} {\bibinfo  {journal} {Phys. Lett. B}\ }\textbf
  {\bibinfo {volume} {811}},\ \bibinfo {pages} {135887} (\bibinfo {year}
  {2020})},\ \Eprint {https://arxiv.org/abs/1909.04476} {arXiv:1909.04476
  [astro-ph.HE]} \BibitemShut {NoStop}%
\bibitem [{\citenamefont {Glendenning}\ \emph {et~al.}(1997)\citenamefont
  {Glendenning}, \citenamefont {Pei},\ and\ \citenamefont
  {Weber}}]{Glendenning:1997fy}%
  \BibitemOpen
  \bibfield  {author} {\bibinfo {author} {\bibfnamefont {N.~K.}\ \bibnamefont
  {Glendenning}}, \bibinfo {author} {\bibfnamefont {S.}~\bibnamefont {Pei}},\
  and\ \bibinfo {author} {\bibfnamefont {F.}~\bibnamefont {Weber}},\ }\bibfield
   {title} {\bibinfo {title} {{Signal of quark deconfinement in the timing
  structure of pulsar spindown}},\ }\href
  {https://doi.org/10.1103/PhysRevLett.79.1603} {\bibfield  {journal} {\bibinfo
   {journal} {Phys. Rev. Lett.}\ }\textbf {\bibinfo {volume} {79}},\ \bibinfo
  {pages} {1603} (\bibinfo {year} {1997})},\ \Eprint
  {https://arxiv.org/abs/astro-ph/9705235} {arXiv:astro-ph/9705235}
  \BibitemShut {NoStop}%
\bibitem [{\citenamefont {Heiselberg}\ and\ \citenamefont
  {Hjorth-Jensen}(1998)}]{Heiselberg:1998vh}%
  \BibitemOpen
  \bibfield  {author} {\bibinfo {author} {\bibfnamefont {H.}~\bibnamefont
  {Heiselberg}}\ and\ \bibinfo {author} {\bibfnamefont {M.}~\bibnamefont
  {Hjorth-Jensen}},\ }\bibfield  {title} {\bibinfo {title} {{Phase transitions
  in rotating neutron stars}},\ }\href
  {https://doi.org/10.1103/PhysRevLett.80.5485} {\bibfield  {journal} {\bibinfo
   {journal} {Phys. Rev. Lett.}\ }\textbf {\bibinfo {volume} {80}},\ \bibinfo
  {pages} {5485} (\bibinfo {year} {1998})},\ \Eprint
  {https://arxiv.org/abs/astro-ph/9801187} {arXiv:astro-ph/9801187}
  \BibitemShut {NoStop}%
\bibitem [{\citenamefont {{Meitner}}\ and\ \citenamefont
  {{Frisch}}(1939)}]{1939Natur.143..471M}%
  \BibitemOpen
  \bibfield  {author} {\bibinfo {author} {\bibfnamefont {L.}~\bibnamefont
  {{Meitner}}}\ and\ \bibinfo {author} {\bibfnamefont {O.~R.}\ \bibnamefont
  {{Frisch}}},\ }\bibfield  {title} {\bibinfo {title} {{Products of the Fission
  of the Uranium Nucleus}},\ }\href {https://doi.org/10.1038/143471a0}
  {\bibfield  {journal} {\bibinfo  {journal} {Nature}\ }\textbf {\bibinfo
  {volume} {143}},\ \bibinfo {pages} {471} (\bibinfo {year}
  {1939})}\BibitemShut {NoStop}%
\bibitem [{\citenamefont {Sime}(1998)}]{Sime:1998}%
  \BibitemOpen
  \bibfield  {author} {\bibinfo {author} {\bibfnamefont {R.~L.}\ \bibnamefont
  {Sime}},\ }\bibfield  {title} {\bibinfo {title} {{Lise Meitner and the
  Discovery of Nuclear Fission}},\ }\href@noop {} {\bibfield  {journal}
  {\bibinfo  {journal} {Scientific American}\ }\textbf {\bibinfo {volume}
  {278}},\ \bibinfo {pages} {80} (\bibinfo {year} {1998})}\BibitemShut
  {NoStop}%
\bibitem [{\citenamefont {Anderson}\ and\ \citenamefont
  {Itoh}(1975)}]{Anderson:1975zze}%
  \BibitemOpen
  \bibfield  {author} {\bibinfo {author} {\bibfnamefont {P.~W.}\ \bibnamefont
  {Anderson}}\ and\ \bibinfo {author} {\bibfnamefont {N.}~\bibnamefont
  {Itoh}},\ }\bibfield  {title} {\bibinfo {title} {{Pulsar glitches and
  restlessness as a hard superfluidity phenomenon}},\ }\href
  {https://doi.org/10.1038/256025a0} {\bibfield  {journal} {\bibinfo  {journal}
  {Nature}\ }\textbf {\bibinfo {volume} {256}},\ \bibinfo {pages} {25}
  (\bibinfo {year} {1975})}\BibitemShut {NoStop}%
\bibitem [{\citenamefont {Nayak}(2020)}]{Nayak:2020bjz}%
  \BibitemOpen
  \bibfield  {author} {\bibinfo {author} {\bibfnamefont {T.~K.}\ \bibnamefont
  {Nayak}},\ }\bibfield  {title} {\bibinfo {title} {{Probing the QCD phase
  structure using event-by-event fluctuations}},\ }\href
  {https://doi.org/10.1088/1742-6596/1602/1/012003} {\bibfield  {journal}
  {\bibinfo  {journal} {J. Phys. Conf. Ser.}\ }\textbf {\bibinfo {volume}
  {1602}},\ \bibinfo {pages} {012003} (\bibinfo {year} {2020})},\ \Eprint
  {https://arxiv.org/abs/2008.04643} {arXiv:2008.04643 [nucl-ex]} \BibitemShut
  {NoStop}%
\bibitem [{\citenamefont {Borsanyi}\ \emph {et~al.}(2020)\citenamefont
  {Borsanyi}, \citenamefont {Fodor}, \citenamefont {Guenther}, \citenamefont
  {Kara}, \citenamefont {Katz}, \citenamefont {Parotto}, \citenamefont
  {Pasztor}, \citenamefont {Ratti},\ and\ \citenamefont
  {Szabo}}]{Borsanyi:2020fev}%
  \BibitemOpen
  \bibfield  {author} {\bibinfo {author} {\bibfnamefont {S.}~\bibnamefont
  {Borsanyi}}, \bibinfo {author} {\bibfnamefont {Z.}~\bibnamefont {Fodor}},
  \bibinfo {author} {\bibfnamefont {J.~N.}\ \bibnamefont {Guenther}}, \bibinfo
  {author} {\bibfnamefont {R.}~\bibnamefont {Kara}}, \bibinfo {author}
  {\bibfnamefont {S.~D.}\ \bibnamefont {Katz}}, \bibinfo {author}
  {\bibfnamefont {P.}~\bibnamefont {Parotto}}, \bibinfo {author} {\bibfnamefont
  {A.}~\bibnamefont {Pasztor}}, \bibinfo {author} {\bibfnamefont
  {C.}~\bibnamefont {Ratti}},\ and\ \bibinfo {author} {\bibfnamefont {K.~K.}\
  \bibnamefont {Szabo}},\ }\bibfield  {title} {\bibinfo {title} {{QCD Crossover
  at Finite Chemical Potential from Lattice Simulations}},\ }\href
  {https://doi.org/10.1103/PhysRevLett.125.052001} {\bibfield  {journal}
  {\bibinfo  {journal} {Phys. Rev. Lett.}\ }\textbf {\bibinfo {volume} {125}},\
  \bibinfo {pages} {052001} (\bibinfo {year} {2020})},\ \Eprint
  {https://arxiv.org/abs/2002.02821} {arXiv:2002.02821 [hep-lat]} \BibitemShut
  {NoStop}%
\bibitem [{\citenamefont {Endrodi}\ \emph {et~al.}(2011)\citenamefont
  {Endrodi}, \citenamefont {Fodor}, \citenamefont {Katz},\ and\ \citenamefont
  {Szabo}}]{Endrodi:2011gv}%
  \BibitemOpen
  \bibfield  {author} {\bibinfo {author} {\bibfnamefont {G.}~\bibnamefont
  {Endrodi}}, \bibinfo {author} {\bibfnamefont {Z.}~\bibnamefont {Fodor}},
  \bibinfo {author} {\bibfnamefont {S.~D.}\ \bibnamefont {Katz}},\ and\
  \bibinfo {author} {\bibfnamefont {K.~K.}\ \bibnamefont {Szabo}},\ }\bibfield
  {title} {\bibinfo {title} {{The QCD phase diagram at nonzero quark
  density}},\ }\href {https://doi.org/10.1007/JHEP04(2011)001} {\bibfield
  {journal} {\bibinfo  {journal} {JHEP}\ }\textbf {\bibinfo {volume} {04}},\
  \bibinfo {pages} {001}},\ \Eprint {https://arxiv.org/abs/1102.1356}
  {arXiv:1102.1356 [hep-lat]} \BibitemShut {NoStop}%
\bibitem [{\citenamefont {Ge}\ \emph {et~al.}(2018)\citenamefont {Ge},
  \citenamefont {Liang},\ and\ \citenamefont {Zhitnitsky}}]{Ge:2017idw}%
  \BibitemOpen
  \bibfield  {author} {\bibinfo {author} {\bibfnamefont {S.}~\bibnamefont
  {Ge}}, \bibinfo {author} {\bibfnamefont {X.}~\bibnamefont {Liang}},\ and\
  \bibinfo {author} {\bibfnamefont {A.}~\bibnamefont {Zhitnitsky}},\ }\bibfield
   {title} {\bibinfo {title} {{Cosmological axion and a quark nugget dark
  matter model}},\ }\href {https://doi.org/10.1103/PhysRevD.97.043008}
  {\bibfield  {journal} {\bibinfo  {journal} {Phys. Rev. D}\ }\textbf {\bibinfo
  {volume} {97}},\ \bibinfo {pages} {043008} (\bibinfo {year} {2018})},\
  \Eprint {https://arxiv.org/abs/1711.06271} {arXiv:1711.06271 [hep-ph]}
  \BibitemShut {NoStop}%
\bibitem [{\citenamefont {Atreya}\ \emph {et~al.}(2014)\citenamefont {Atreya},
  \citenamefont {Sarkar},\ and\ \citenamefont {Srivastava}}]{Atreya:2014sca}%
  \BibitemOpen
  \bibfield  {author} {\bibinfo {author} {\bibfnamefont {A.}~\bibnamefont
  {Atreya}}, \bibinfo {author} {\bibfnamefont {A.}~\bibnamefont {Sarkar}},\
  and\ \bibinfo {author} {\bibfnamefont {A.~M.}\ \bibnamefont {Srivastava}},\
  }\bibfield  {title} {\bibinfo {title} {{Reviving quark nuggets as a candidate
  for dark matter}},\ }\href {https://doi.org/10.1103/PhysRevD.90.045010}
  {\bibfield  {journal} {\bibinfo  {journal} {Phys. Rev. D}\ }\textbf {\bibinfo
  {volume} {90}},\ \bibinfo {pages} {045010} (\bibinfo {year} {2014})},\
  \Eprint {https://arxiv.org/abs/1405.6492} {arXiv:1405.6492 [hep-ph]}
  \BibitemShut {NoStop}%
\bibitem [{\citenamefont {Philipsen}(2021)}]{Philipsen:2021qji}%
  \BibitemOpen
  \bibfield  {author} {\bibinfo {author} {\bibfnamefont {O.}~\bibnamefont
  {Philipsen}},\ }\bibfield  {title} {\bibinfo {title} {{Lattice Constraints on
  the QCD Chiral Phase Transition at Finite Temperature and Baryon Density}},\
  }\href {https://doi.org/10.3390/sym13112079} {\bibfield  {journal} {\bibinfo
  {journal} {Symmetry}\ }\textbf {\bibinfo {volume} {13}},\ \bibinfo {pages}
  {2079} (\bibinfo {year} {2021})},\ \Eprint {https://arxiv.org/abs/2111.03590}
  {arXiv:2111.03590 [hep-lat]} \BibitemShut {NoStop}%
\bibitem [{\citenamefont {McLerran}\ and\ \citenamefont
  {Pisarski}(2007)}]{McLerran:2007qj}%
  \BibitemOpen
  \bibfield  {author} {\bibinfo {author} {\bibfnamefont {L.}~\bibnamefont
  {McLerran}}\ and\ \bibinfo {author} {\bibfnamefont {R.~D.}\ \bibnamefont
  {Pisarski}},\ }\bibfield  {title} {\bibinfo {title} {{Phases of cold, dense
  quarks at large N(c)}},\ }\href
  {https://doi.org/10.1016/j.nuclphysa.2007.08.013} {\bibfield  {journal}
  {\bibinfo  {journal} {Nucl. Phys. A}\ }\textbf {\bibinfo {volume} {796}},\
  \bibinfo {pages} {83} (\bibinfo {year} {2007})},\ \Eprint
  {https://arxiv.org/abs/0706.2191} {arXiv:0706.2191 [hep-ph]} \BibitemShut
  {NoStop}%
\bibitem [{\citenamefont {McLerran}(2020)}]{McLerran:2020rnw}%
  \BibitemOpen
  \bibfield  {author} {\bibinfo {author} {\bibfnamefont {L.}~\bibnamefont
  {McLerran}},\ }\bibfield  {title} {\bibinfo {title} {{A Pedagogical
  Discussion of Quarkyonic Matter and Its Implication for Neutron Stars}},\
  }\href {https://doi.org/10.5506/APhysPolB.51.1067} {\bibfield  {journal}
  {\bibinfo  {journal} {Acta Phys. Polon. B}\ }\textbf {\bibinfo {volume}
  {51}},\ \bibinfo {pages} {1067} (\bibinfo {year} {2020})}\BibitemShut
  {NoStop}%
\bibitem [{\citenamefont {Kojo}\ \emph {et~al.}(2012)\citenamefont {Kojo},
  \citenamefont {Hidaka}, \citenamefont {Fukushima}, \citenamefont {McLerran},\
  and\ \citenamefont {Pisarski}}]{Kojo:2011cn}%
  \BibitemOpen
  \bibfield  {author} {\bibinfo {author} {\bibfnamefont {T.}~\bibnamefont
  {Kojo}}, \bibinfo {author} {\bibfnamefont {Y.}~\bibnamefont {Hidaka}},
  \bibinfo {author} {\bibfnamefont {K.}~\bibnamefont {Fukushima}}, \bibinfo
  {author} {\bibfnamefont {L.~D.}\ \bibnamefont {McLerran}},\ and\ \bibinfo
  {author} {\bibfnamefont {R.~D.}\ \bibnamefont {Pisarski}},\ }\bibfield
  {title} {\bibinfo {title} {{Interweaving Chiral Spirals}},\ }\href
  {https://doi.org/10.1016/j.nuclphysa.2011.11.007} {\bibfield  {journal}
  {\bibinfo  {journal} {Nucl. Phys. A}\ }\textbf {\bibinfo {volume} {875}},\
  \bibinfo {pages} {94} (\bibinfo {year} {2012})},\ \Eprint
  {https://arxiv.org/abs/1107.2124} {arXiv:1107.2124 [hep-ph]} \BibitemShut
  {NoStop}%
\bibitem [{\citenamefont {Larkin}\ and\ \citenamefont
  {Ovchinnikov}(1964)}]{Larkin:1964wok}%
  \BibitemOpen
  \bibfield  {author} {\bibinfo {author} {\bibfnamefont {A.~I.}\ \bibnamefont
  {Larkin}}\ and\ \bibinfo {author} {\bibfnamefont {Y.~N.}\ \bibnamefont
  {Ovchinnikov}},\ }\bibfield  {title} {\bibinfo {title} {{Nonuniform state of
  superconductors}},\ }\href@noop {} {\bibfield  {journal} {\bibinfo  {journal}
  {Zh. Eksp. Teor. Fiz.}\ }\textbf {\bibinfo {volume} {47}},\ \bibinfo {pages}
  {1136} (\bibinfo {year} {1964})}\BibitemShut {NoStop}%
\bibitem [{\citenamefont {Fulde}\ and\ \citenamefont
  {Ferrell}(1964)}]{Fulde:1964zz}%
  \BibitemOpen
  \bibfield  {author} {\bibinfo {author} {\bibfnamefont {P.}~\bibnamefont
  {Fulde}}\ and\ \bibinfo {author} {\bibfnamefont {R.~A.}\ \bibnamefont
  {Ferrell}},\ }\bibfield  {title} {\bibinfo {title} {{Superconductivity in a
  Strong Spin-Exchange Field}},\ }\href
  {https://doi.org/10.1103/PhysRev.135.A550} {\bibfield  {journal} {\bibinfo
  {journal} {Phys. Rev.}\ }\textbf {\bibinfo {volume} {135}},\ \bibinfo {pages}
  {A550} (\bibinfo {year} {1964})}\BibitemShut {NoStop}%
\bibitem [{\citenamefont {Alford}\ \emph
  {et~al.}(2001{\natexlab{b}})\citenamefont {Alford}, \citenamefont {Bowers},\
  and\ \citenamefont {Rajagopal}}]{Alford:2000ze}%
  \BibitemOpen
  \bibfield  {author} {\bibinfo {author} {\bibfnamefont {M.~G.}\ \bibnamefont
  {Alford}}, \bibinfo {author} {\bibfnamefont {J.~A.}\ \bibnamefont {Bowers}},\
  and\ \bibinfo {author} {\bibfnamefont {K.}~\bibnamefont {Rajagopal}},\
  }\bibfield  {title} {\bibinfo {title} {{Crystalline color
  superconductivity}},\ }\href {https://doi.org/10.1103/PhysRevD.63.074016}
  {\bibfield  {journal} {\bibinfo  {journal} {Phys. Rev. D}\ }\textbf {\bibinfo
  {volume} {63}},\ \bibinfo {pages} {074016} (\bibinfo {year}
  {2001}{\natexlab{b}})},\ \Eprint {https://arxiv.org/abs/hep-ph/0008208}
  {arXiv:hep-ph/0008208} \BibitemShut {NoStop}%
\bibitem [{\citenamefont {Hosaka}\ and\ \citenamefont
  {Toki}(2001)}]{Hosaka:2001ux}%
  \BibitemOpen
  \bibfield  {author} {\bibinfo {author} {\bibfnamefont {A.}~\bibnamefont
  {Hosaka}}\ and\ \bibinfo {author} {\bibfnamefont {H.}~\bibnamefont {Toki}},\
  }\href@noop {} {\emph {\bibinfo {title} {{Quarks, baryons and chiral
  symmetry}}}}\ (\bibinfo {year} {2001})\BibitemShut {NoStop}%
\bibitem [{\citenamefont {Baade}\ and\ \citenamefont
  {Zwicky}(1934)}]{Baade:1934onh}%
  \BibitemOpen
  \bibfield  {author} {\bibinfo {author} {\bibfnamefont {W.}~\bibnamefont
  {Baade}}\ and\ \bibinfo {author} {\bibfnamefont {F.}~\bibnamefont {Zwicky}},\
  }\bibfield  {title} {\bibinfo {title} {Cosmic rays from super-novae},\ }\href
  {https://doi.org/10.1073/pnas.20.5.259} {\bibfield  {journal} {\bibinfo
  {journal} {Proceedings of the National Academy of Sciences}\ }\textbf
  {\bibinfo {volume} {20}},\ \bibinfo {pages} {259 } (\bibinfo {year}
  {1934})}\BibitemShut {NoStop}%
\bibitem [{\citenamefont {Hewish}\ \emph {et~al.}(1968)\citenamefont {Hewish},
  \citenamefont {Bell}, \citenamefont {Pilkington}, \citenamefont {Scott},\
  and\ \citenamefont {Collins}}]{Hewish:1968bj}%
  \BibitemOpen
  \bibfield  {author} {\bibinfo {author} {\bibfnamefont {A.}~\bibnamefont
  {Hewish}}, \bibinfo {author} {\bibfnamefont {S.~J.}\ \bibnamefont {Bell}},
  \bibinfo {author} {\bibfnamefont {J.~D.~H.}\ \bibnamefont {Pilkington}},
  \bibinfo {author} {\bibfnamefont {P.~F.}\ \bibnamefont {Scott}},\ and\
  \bibinfo {author} {\bibfnamefont {R.~A.}\ \bibnamefont {Collins}},\
  }\bibfield  {title} {\bibinfo {title} {{Observation of a rapidly pulsating
  radio source}},\ }\href {https://doi.org/10.1038/217709a0} {\bibfield
  {journal} {\bibinfo  {journal} {Nature}\ }\textbf {\bibinfo {volume} {217}},\
  \bibinfo {pages} {709} (\bibinfo {year} {1968})}\BibitemShut {NoStop}%
\bibitem [{\citenamefont {{Lorimer}}\ and\ \citenamefont
  {{Kramer}}(2004)}]{2004hpa..book.....L}%
  \BibitemOpen
  \bibfield  {author} {\bibinfo {author} {\bibfnamefont {D.~R.}\ \bibnamefont
  {{Lorimer}}}\ and\ \bibinfo {author} {\bibfnamefont {M.}~\bibnamefont
  {{Kramer}}},\ }\href@noop {} {\emph {\bibinfo {title} {{Handbook of Pulsar
  Astronomy}}}},\ Vol.~\bibinfo {volume} {4}\ (\bibinfo {year}
  {2004})\BibitemShut {NoStop}%
\bibitem [{\citenamefont {{Andersson}}(2019)}]{2019gwa..book.....A}%
  \BibitemOpen
  \bibfield  {author} {\bibinfo {author} {\bibfnamefont {N.}~\bibnamefont
  {{Andersson}}},\ }\href
  {https://doi.org/10.1093/oso/9780198568032.001.0001/oso-9780198568032} {\emph
  {\bibinfo {title} {{Gravitational-Wave Astronomy: Exploring the Dark Side of
  the Universe}}}}\ (\bibinfo {year} {2019})\BibitemShut {NoStop}%
\bibitem [{\citenamefont {Tolman}(1939)}]{Tolman:1939jz}%
  \BibitemOpen
  \bibfield  {author} {\bibinfo {author} {\bibfnamefont {R.~C.}\ \bibnamefont
  {Tolman}},\ }\bibfield  {title} {\bibinfo {title} {{Static solutions of
  Einstein's field equations for spheres of fluid}},\ }\href
  {https://doi.org/10.1103/PhysRev.55.364} {\bibfield  {journal} {\bibinfo
  {journal} {Phys. Rev.}\ }\textbf {\bibinfo {volume} {55}},\ \bibinfo {pages}
  {364} (\bibinfo {year} {1939})}\BibitemShut {NoStop}%
\bibitem [{\citenamefont {Oppenheimer}\ and\ \citenamefont
  {Volkoff}(1939)}]{Oppenheimer:1939ne}%
  \BibitemOpen
  \bibfield  {author} {\bibinfo {author} {\bibfnamefont {J.~R.}\ \bibnamefont
  {Oppenheimer}}\ and\ \bibinfo {author} {\bibfnamefont {G.~M.}\ \bibnamefont
  {Volkoff}},\ }\bibfield  {title} {\bibinfo {title} {{On massive neutron
  cores}},\ }\href {https://doi.org/10.1103/PhysRev.55.374} {\bibfield
  {journal} {\bibinfo  {journal} {Phys. Rev.}\ }\textbf {\bibinfo {volume}
  {55}},\ \bibinfo {pages} {374} (\bibinfo {year} {1939})}\BibitemShut
  {NoStop}%
\bibitem [{\citenamefont {Chamel}\ and\ \citenamefont
  {Haensel}(2008)}]{Chamel:2008ca}%
  \BibitemOpen
  \bibfield  {author} {\bibinfo {author} {\bibfnamefont {N.}~\bibnamefont
  {Chamel}}\ and\ \bibinfo {author} {\bibfnamefont {P.}~\bibnamefont
  {Haensel}},\ }\bibfield  {title} {\bibinfo {title} {{Physics of Neutron Star
  Crusts}},\ }\href {https://doi.org/10.12942/lrr-2008-10} {\bibfield
  {journal} {\bibinfo  {journal} {Living Rev. Rel.}\ }\textbf {\bibinfo
  {volume} {11}},\ \bibinfo {pages} {10} (\bibinfo {year} {2008})},\ \Eprint
  {https://arxiv.org/abs/0812.3955} {arXiv:0812.3955 [astro-ph]} \BibitemShut
  {NoStop}%
\bibitem [{\citenamefont {Link}\ \emph {et~al.}(1999)\citenamefont {Link},
  \citenamefont {Epstein},\ and\ \citenamefont {Lattimer}}]{Link:1999ca}%
  \BibitemOpen
  \bibfield  {author} {\bibinfo {author} {\bibfnamefont {B.}~\bibnamefont
  {Link}}, \bibinfo {author} {\bibfnamefont {R.~I.}\ \bibnamefont {Epstein}},\
  and\ \bibinfo {author} {\bibfnamefont {J.~M.}\ \bibnamefont {Lattimer}},\
  }\bibfield  {title} {\bibinfo {title} {{Pulsar constraints on neutron star
  structure and equation of state}},\ }\href
  {https://doi.org/10.1103/PhysRevLett.83.3362} {\bibfield  {journal} {\bibinfo
   {journal} {Phys. Rev. Lett.}\ }\textbf {\bibinfo {volume} {83}},\ \bibinfo
  {pages} {3362} (\bibinfo {year} {1999})},\ \Eprint
  {https://arxiv.org/abs/astro-ph/9909146} {arXiv:astro-ph/9909146}
  \BibitemShut {NoStop}%
\bibitem [{\citenamefont {Lattimer}\ and\ \citenamefont
  {Prakash}(2001)}]{Lattimer:2000nx}%
  \BibitemOpen
  \bibfield  {author} {\bibinfo {author} {\bibfnamefont {J.~M.}\ \bibnamefont
  {Lattimer}}\ and\ \bibinfo {author} {\bibfnamefont {M.}~\bibnamefont
  {Prakash}},\ }\bibfield  {title} {\bibinfo {title} {{Neutron star structure
  and the equation of state}},\ }\href {https://doi.org/10.1086/319702}
  {\bibfield  {journal} {\bibinfo  {journal} {Astrophys. J.}\ }\textbf
  {\bibinfo {volume} {550}},\ \bibinfo {pages} {426} (\bibinfo {year}
  {2001})},\ \Eprint {https://arxiv.org/abs/astro-ph/0002232}
  {arXiv:astro-ph/0002232} \BibitemShut {NoStop}%
\bibitem [{\citenamefont {\"Ozel}\ and\ \citenamefont
  {Freire}(2016)}]{Ozel:2016oaf}%
  \BibitemOpen
  \bibfield  {author} {\bibinfo {author} {\bibfnamefont {F.}~\bibnamefont
  {\"Ozel}}\ and\ \bibinfo {author} {\bibfnamefont {P.}~\bibnamefont
  {Freire}},\ }\bibfield  {title} {\bibinfo {title} {{Masses, Radii, and the
  Equation of State of Neutron Stars}},\ }\href
  {https://doi.org/10.1146/annurev-astro-081915-023322} {\bibfield  {journal}
  {\bibinfo  {journal} {Ann. Rev. Astron. Astrophys.}\ }\textbf {\bibinfo
  {volume} {54}},\ \bibinfo {pages} {401} (\bibinfo {year} {2016})},\ \Eprint
  {https://arxiv.org/abs/1603.02698} {arXiv:1603.02698 [astro-ph.HE]}
  \BibitemShut {NoStop}%
\bibitem [{\citenamefont {Madsen}(2000)}]{Madsen:1999ci}%
  \BibitemOpen
  \bibfield  {author} {\bibinfo {author} {\bibfnamefont {J.}~\bibnamefont
  {Madsen}},\ }\bibfield  {title} {\bibinfo {title} {{Probing strange stars and
  color superconductivity by r-mode instabilities in millisecond pulsars}},\
  }\href {https://doi.org/10.1103/PhysRevLett.85.10} {\bibfield  {journal}
  {\bibinfo  {journal} {Phys. Rev. Lett.}\ }\textbf {\bibinfo {volume} {85}},\
  \bibinfo {pages} {10} (\bibinfo {year} {2000})},\ \Eprint
  {https://arxiv.org/abs/astro-ph/9912418} {arXiv:astro-ph/9912418}
  \BibitemShut {NoStop}%
\bibitem [{\citenamefont {Jha}\ \emph {et~al.}(2008)\citenamefont {Jha},
  \citenamefont {Mishra},\ and\ \citenamefont {Sreekanth}}]{Jha:2007ej}%
  \BibitemOpen
  \bibfield  {author} {\bibinfo {author} {\bibfnamefont {T.~K.}\ \bibnamefont
  {Jha}}, \bibinfo {author} {\bibfnamefont {H.}~\bibnamefont {Mishra}},\ and\
  \bibinfo {author} {\bibfnamefont {V.}~\bibnamefont {Sreekanth}},\ }\bibfield
  {title} {\bibinfo {title} {{On attributes of a Rotating Neutron star with a
  Hyperon core}},\ }\href {https://doi.org/10.1103/PhysRevC.77.045801}
  {\bibfield  {journal} {\bibinfo  {journal} {Phys. Rev. C}\ }\textbf {\bibinfo
  {volume} {77}},\ \bibinfo {pages} {045801} (\bibinfo {year} {2008})},\
  \Eprint {https://arxiv.org/abs/0710.5392} {arXiv:0710.5392 [nucl-th]}
  \BibitemShut {NoStop}%
\bibitem [{\citenamefont {Kunjipurayil}\ \emph {et~al.}(2022)\citenamefont
  {Kunjipurayil}, \citenamefont {Zhao}, \citenamefont {Kumar}, \citenamefont
  {Agrawal},\ and\ \citenamefont {Prakash}}]{Kunjipurayil:2022zah}%
  \BibitemOpen
  \bibfield  {author} {\bibinfo {author} {\bibfnamefont {A.}~\bibnamefont
  {Kunjipurayil}}, \bibinfo {author} {\bibfnamefont {T.}~\bibnamefont {Zhao}},
  \bibinfo {author} {\bibfnamefont {B.}~\bibnamefont {Kumar}}, \bibinfo
  {author} {\bibfnamefont {B.~K.}\ \bibnamefont {Agrawal}},\ and\ \bibinfo
  {author} {\bibfnamefont {M.}~\bibnamefont {Prakash}},\ }\bibfield  {title}
  {\bibinfo {title} {{Impact of the equation of state on f- and p- mode
  oscillations of neutron stars}},\ }\href
  {https://doi.org/10.1103/PhysRevD.106.063005} {\bibfield  {journal} {\bibinfo
   {journal} {Phys. Rev. D}\ }\textbf {\bibinfo {volume} {106}},\ \bibinfo
  {pages} {063005} (\bibinfo {year} {2022})},\ \Eprint
  {https://arxiv.org/abs/2205.02081} {arXiv:2205.02081 [nucl-th]} \BibitemShut
  {NoStop}%
\bibitem [{\citenamefont {Cromartie}\ \emph {et~al.}(2019)\citenamefont
  {Cromartie} \emph {et~al.}}]{NANOGrav:2019jur}%
  \BibitemOpen
  \bibfield  {author} {\bibinfo {author} {\bibfnamefont {H.~T.}\ \bibnamefont
  {Cromartie}} \emph {et~al.} (\bibinfo {collaboration} {NANOGrav}),\
  }\bibfield  {title} {\bibinfo {title} {{Relativistic Shapiro delay
  measurements of an extremely massive millisecond pulsar}},\ }\href
  {https://doi.org/10.1038/s41550-019-0880-2} {\bibfield  {journal} {\bibinfo
  {journal} {Nature Astron.}\ }\textbf {\bibinfo {volume} {4}},\ \bibinfo
  {pages} {72} (\bibinfo {year} {2019})},\ \Eprint
  {https://arxiv.org/abs/1904.06759} {arXiv:1904.06759 [astro-ph.HE]}
  \BibitemShut {NoStop}%
\bibitem [{\citenamefont {Koehn}\ \emph {et~al.}(2024)\citenamefont {Koehn}
  \emph {et~al.}}]{Koehn:2024set}%
  \BibitemOpen
  \bibfield  {author} {\bibinfo {author} {\bibfnamefont {H.}~\bibnamefont
  {Koehn}} \emph {et~al.},\ }\bibfield  {title} {\bibinfo {title} {{An overview
  of existing and new nuclear and astrophysical constraints on the equation of
  state of neutron-rich dense matter}},\ }\href@noop {} {\  (\bibinfo {year}
  {2024})},\ \Eprint {https://arxiv.org/abs/2402.04172} {arXiv:2402.04172
  [astro-ph.HE]} \BibitemShut {NoStop}%
\bibitem [{\citenamefont {Hinderer}\ \emph {et~al.}(2010)\citenamefont
  {Hinderer}, \citenamefont {Lackey}, \citenamefont {Lang},\ and\ \citenamefont
  {Read}}]{Hinderer:2009ca}%
  \BibitemOpen
  \bibfield  {author} {\bibinfo {author} {\bibfnamefont {T.}~\bibnamefont
  {Hinderer}}, \bibinfo {author} {\bibfnamefont {B.~D.}\ \bibnamefont
  {Lackey}}, \bibinfo {author} {\bibfnamefont {R.~N.}\ \bibnamefont {Lang}},\
  and\ \bibinfo {author} {\bibfnamefont {J.~S.}\ \bibnamefont {Read}},\
  }\bibfield  {title} {\bibinfo {title} {{Tidal deformability of neutron stars
  with realistic equations of state and their gravitational wave signatures in
  binary inspiral}},\ }\href {https://doi.org/10.1103/PhysRevD.81.123016}
  {\bibfield  {journal} {\bibinfo  {journal} {Phys. Rev. D}\ }\textbf {\bibinfo
  {volume} {81}},\ \bibinfo {pages} {123016} (\bibinfo {year} {2010})},\
  \Eprint {https://arxiv.org/abs/0911.3535} {arXiv:0911.3535 [astro-ph.HE]}
  \BibitemShut {NoStop}%
\bibitem [{\citenamefont {Bauswein}\ \emph {et~al.}(2019)\citenamefont
  {Bauswein}, \citenamefont {Bastian}, \citenamefont {Blaschke}, \citenamefont
  {Chatziioannou}, \citenamefont {Clark}, \citenamefont {Fischer},\ and\
  \citenamefont {Oertel}}]{Bauswein:2018bma}%
  \BibitemOpen
  \bibfield  {author} {\bibinfo {author} {\bibfnamefont {A.}~\bibnamefont
  {Bauswein}}, \bibinfo {author} {\bibfnamefont {N.-U.~F.}\ \bibnamefont
  {Bastian}}, \bibinfo {author} {\bibfnamefont {D.~B.}\ \bibnamefont
  {Blaschke}}, \bibinfo {author} {\bibfnamefont {K.}~\bibnamefont
  {Chatziioannou}}, \bibinfo {author} {\bibfnamefont {J.~A.}\ \bibnamefont
  {Clark}}, \bibinfo {author} {\bibfnamefont {T.}~\bibnamefont {Fischer}},\
  and\ \bibinfo {author} {\bibfnamefont {M.}~\bibnamefont {Oertel}},\
  }\bibfield  {title} {\bibinfo {title} {{Identifying a first-order phase
  transition in neutron star mergers through gravitational waves}},\ }\href
  {https://doi.org/10.1103/PhysRevLett.122.061102} {\bibfield  {journal}
  {\bibinfo  {journal} {Phys. Rev. Lett.}\ }\textbf {\bibinfo {volume} {122}},\
  \bibinfo {pages} {061102} (\bibinfo {year} {2019})},\ \Eprint
  {https://arxiv.org/abs/1809.01116} {arXiv:1809.01116 [astro-ph.HE]}
  \BibitemShut {NoStop}%
\bibitem [{\citenamefont {Most}\ \emph {et~al.}(2019)\citenamefont {Most},
  \citenamefont {Papenfort}, \citenamefont {Dexheimer}, \citenamefont
  {Hanauske}, \citenamefont {Schramm}, \citenamefont {St\"ocker},\ and\
  \citenamefont {Rezzolla}}]{Most:2018eaw}%
  \BibitemOpen
  \bibfield  {author} {\bibinfo {author} {\bibfnamefont {E.~R.}\ \bibnamefont
  {Most}}, \bibinfo {author} {\bibfnamefont {L.~J.}\ \bibnamefont {Papenfort}},
  \bibinfo {author} {\bibfnamefont {V.}~\bibnamefont {Dexheimer}}, \bibinfo
  {author} {\bibfnamefont {M.}~\bibnamefont {Hanauske}}, \bibinfo {author}
  {\bibfnamefont {S.}~\bibnamefont {Schramm}}, \bibinfo {author} {\bibfnamefont
  {H.}~\bibnamefont {St\"ocker}},\ and\ \bibinfo {author} {\bibfnamefont
  {L.}~\bibnamefont {Rezzolla}},\ }\bibfield  {title} {\bibinfo {title}
  {{Signatures of quark-hadron phase transitions in general-relativistic
  neutron-star mergers}},\ }\href
  {https://doi.org/10.1103/PhysRevLett.122.061101} {\bibfield  {journal}
  {\bibinfo  {journal} {Phys. Rev. Lett.}\ }\textbf {\bibinfo {volume} {122}},\
  \bibinfo {pages} {061101} (\bibinfo {year} {2019})},\ \Eprint
  {https://arxiv.org/abs/1807.03684} {arXiv:1807.03684 [astro-ph.HE]}
  \BibitemShut {NoStop}%
\bibitem [{\citenamefont {Manchester}(2017)}]{Manchester:2017ykr}%
  \BibitemOpen
  \bibfield  {author} {\bibinfo {author} {\bibfnamefont {R.~N.}\ \bibnamefont
  {Manchester}},\ }\bibfield  {title} {\bibinfo {title} {{Pulsar timing and its
  applications}},\ }\href {https://doi.org/10.1088/1742-6596/932/1/012002}
  {\bibfield  {journal} {\bibinfo  {journal} {J. Phys. Conf. Ser.}\ }\textbf
  {\bibinfo {volume} {932}},\ \bibinfo {pages} {012002} (\bibinfo {year}
  {2017})},\ \Eprint {https://arxiv.org/abs/1801.04318} {arXiv:1801.04318
  [astro-ph.HE]} \BibitemShut {NoStop}%
\bibitem [{\citenamefont {Weisberg}\ \emph {et~al.}(1981)\citenamefont
  {Weisberg}, \citenamefont {Taylor},\ and\ \citenamefont
  {Fowler}}]{Weisberg:1981mt}%
  \BibitemOpen
  \bibfield  {author} {\bibinfo {author} {\bibfnamefont {J.~M.}\ \bibnamefont
  {Weisberg}}, \bibinfo {author} {\bibfnamefont {J.~H.}\ \bibnamefont
  {Taylor}},\ and\ \bibinfo {author} {\bibfnamefont {L.~A.}\ \bibnamefont
  {Fowler}},\ }\bibfield  {title} {\bibinfo {title} {{Gravitational waves from
  an orbiting pulsar}},\ }\href
  {https://doi.org/10.1038/scientificamerican1081-74} {\bibfield  {journal}
  {\bibinfo  {journal} {Sci. Am.}\ }\textbf {\bibinfo {volume} {245}},\
  \bibinfo {pages} {66} (\bibinfo {year} {1981})}\BibitemShut {NoStop}%
\bibitem [{\citenamefont {Taylor}\ and\ \citenamefont
  {Weisberg}(1989)}]{Taylor:1989sw}%
  \BibitemOpen
  \bibfield  {author} {\bibinfo {author} {\bibfnamefont {J.~H.}\ \bibnamefont
  {Taylor}}\ and\ \bibinfo {author} {\bibfnamefont {J.~M.}\ \bibnamefont
  {Weisberg}},\ }\bibfield  {title} {\bibinfo {title} {{Further experimental
  tests of relativistic gravity using the binary pulsar PSR 1913+16}},\ }\href
  {https://doi.org/10.1086/167917} {\bibfield  {journal} {\bibinfo  {journal}
  {Astrophys. J.}\ }\textbf {\bibinfo {volume} {345}},\ \bibinfo {pages} {434}
  (\bibinfo {year} {1989})}\BibitemShut {NoStop}%
\bibitem [{\citenamefont {Weisberg}\ \emph {et~al.}(2010)\citenamefont
  {Weisberg}, \citenamefont {Nice},\ and\ \citenamefont
  {Taylor}}]{Weisberg:2010zz}%
  \BibitemOpen
  \bibfield  {author} {\bibinfo {author} {\bibfnamefont {J.~M.}\ \bibnamefont
  {Weisberg}}, \bibinfo {author} {\bibfnamefont {D.~J.}\ \bibnamefont {Nice}},\
  and\ \bibinfo {author} {\bibfnamefont {J.~H.}\ \bibnamefont {Taylor}},\
  }\bibfield  {title} {\bibinfo {title} {{Timing Measurements of the
  Relativistic Binary Pulsar PSR B1913+16}},\ }\href
  {https://doi.org/10.1088/0004-637X/722/2/1030} {\bibfield  {journal}
  {\bibinfo  {journal} {Astrophys. J.}\ }\textbf {\bibinfo {volume} {722}},\
  \bibinfo {pages} {1030} (\bibinfo {year} {2010})},\ \Eprint
  {https://arxiv.org/abs/1011.0718} {arXiv:1011.0718 [astro-ph.GA]}
  \BibitemShut {NoStop}%
\bibitem [{\citenamefont {Bildsten}(1998)}]{Bildsten:1998ey}%
  \BibitemOpen
  \bibfield  {author} {\bibinfo {author} {\bibfnamefont {L.}~\bibnamefont
  {Bildsten}},\ }\bibfield  {title} {\bibinfo {title} {{Gravitational radiation
  and rotation of accreting neutron stars}},\ }\href
  {https://doi.org/10.1086/311440} {\bibfield  {journal} {\bibinfo  {journal}
  {Astrophys. J. Lett.}\ }\textbf {\bibinfo {volume} {501}},\ \bibinfo {pages}
  {L89} (\bibinfo {year} {1998})},\ \Eprint
  {https://arxiv.org/abs/astro-ph/9804325} {arXiv:astro-ph/9804325}
  \BibitemShut {NoStop}%
\bibitem [{\citenamefont {Jones}(2002)}]{Jones:2001ui}%
  \BibitemOpen
  \bibfield  {author} {\bibinfo {author} {\bibfnamefont {D.~I.}\ \bibnamefont
  {Jones}},\ }\bibfield  {title} {\bibinfo {title} {{Gravitational waves from
  rotating neutron stars}},\ }\href
  {https://doi.org/10.1088/0264-9381/19/7/304} {\bibfield  {journal} {\bibinfo
  {journal} {Class. Quant. Grav.}\ }\textbf {\bibinfo {volume} {19}},\ \bibinfo
  {pages} {1255} (\bibinfo {year} {2002})},\ \Eprint
  {https://arxiv.org/abs/gr-qc/0111007} {arXiv:gr-qc/0111007} \BibitemShut
  {NoStop}%
\bibitem [{\citenamefont {Layek}\ and\ \citenamefont
  {Yadav}(2020{\natexlab{a}})}]{Layek:2019ede}%
  \BibitemOpen
  \bibfield  {author} {\bibinfo {author} {\bibfnamefont {B.}~\bibnamefont
  {Layek}}\ and\ \bibinfo {author} {\bibfnamefont {P.}~\bibnamefont {Yadav}},\
  }\bibfield  {title} {\bibinfo {title} {{Bursts of Gravitational Waves due to
  Crustquake from Pulsars}},\ }\href
  {https://doi.org/10.1007/s12036-020-09631-0} {\bibfield  {journal} {\bibinfo
  {journal} {J. Astrophys. Astron.}\ }\textbf {\bibinfo {volume} {41}},\
  \bibinfo {pages} {14} (\bibinfo {year} {2020}{\natexlab{a}})},\ \Eprint
  {https://arxiv.org/abs/1904.04570} {arXiv:1904.04570 [astro-ph.HE]}
  \BibitemShut {NoStop}%
\bibitem [{\citenamefont {Abadie}\ \emph {et~al.}(2011)\citenamefont {Abadie}
  \emph {et~al.}}]{LIGOScientific:2010epv}%
  \BibitemOpen
  \bibfield  {author} {\bibinfo {author} {\bibfnamefont {J.}~\bibnamefont
  {Abadie}} \emph {et~al.} (\bibinfo {collaboration} {LIGO Scientific}),\
  }\bibfield  {title} {\bibinfo {title} {{A search for gravitational waves
  associated with the August 2006 timing glitch of the Vela pulsar}},\ }\href
  {https://doi.org/10.1103/PhysRevD.83.042001} {\bibfield  {journal} {\bibinfo
  {journal} {Phys. Rev. D}\ }\textbf {\bibinfo {volume} {83}},\ \bibinfo
  {pages} {042001} (\bibinfo {year} {2011})},\ \Eprint
  {https://arxiv.org/abs/1011.1357} {arXiv:1011.1357 [gr-qc]} \BibitemShut
  {NoStop}%
\bibitem [{\citenamefont {Manchester}(2013)}]{Manchester:2013ndt}%
  \BibitemOpen
  \bibfield  {author} {\bibinfo {author} {\bibfnamefont {R.~N.}\ \bibnamefont
  {Manchester}},\ }\bibfield  {title} {\bibinfo {title} {{The International
  Pulsar Timing Array}},\ }\href
  {https://doi.org/10.1088/0264-9381/30/22/224010} {\bibfield  {journal}
  {\bibinfo  {journal} {Class. Quant. Grav.}\ }\textbf {\bibinfo {volume}
  {30}},\ \bibinfo {pages} {224010} (\bibinfo {year} {2013})},\ \Eprint
  {https://arxiv.org/abs/1309.7392} {arXiv:1309.7392 [astro-ph.IM]}
  \BibitemShut {NoStop}%
\bibitem [{\citenamefont {{Radhakrishnan}}\ and\ \citenamefont
  {{Manchester}}(1969)}]{1969Natur.222..228R}%
  \BibitemOpen
  \bibfield  {author} {\bibinfo {author} {\bibfnamefont {V.}~\bibnamefont
  {{Radhakrishnan}}}\ and\ \bibinfo {author} {\bibfnamefont {R.~N.}\
  \bibnamefont {{Manchester}}},\ }\bibfield  {title} {\bibinfo {title}
  {{Detection of a Change of State in the Pulsar PSR 0833-45}},\ }\href
  {https://doi.org/10.1038/222228a0} {\bibfield  {journal} {\bibinfo  {journal}
  {Nature}\ }\textbf {\bibinfo {volume} {222}},\ \bibinfo {pages} {228}
  (\bibinfo {year} {1969})}\BibitemShut {NoStop}%
\bibitem [{\citenamefont {Espinoza}\ \emph {et~al.}(2011)\citenamefont
  {Espinoza}, \citenamefont {Lyne}, \citenamefont {Stappers},\ and\
  \citenamefont {Kramer}}]{Espinoza:2011pq}%
  \BibitemOpen
  \bibfield  {author} {\bibinfo {author} {\bibfnamefont {C.~M.}\ \bibnamefont
  {Espinoza}}, \bibinfo {author} {\bibfnamefont {A.~G.}\ \bibnamefont {Lyne}},
  \bibinfo {author} {\bibfnamefont {B.~W.}\ \bibnamefont {Stappers}},\ and\
  \bibinfo {author} {\bibfnamefont {M.}~\bibnamefont {Kramer}},\ }\bibfield
  {title} {\bibinfo {title} {{A study of 315 glitches in the rotation of 102
  pulsars}},\ }\href {https://doi.org/10.1111/j.1365-2966.2011.18503.x}
  {\bibfield  {journal} {\bibinfo  {journal} {Mon. Not. Roy. Astron. Soc.}\
  }\textbf {\bibinfo {volume} {414}},\ \bibinfo {pages} {1679} (\bibinfo {year}
  {2011})},\ \Eprint {https://arxiv.org/abs/1102.1743} {arXiv:1102.1743
  [astro-ph.HE]} \BibitemShut {NoStop}%
\bibitem [{\citenamefont {{Ruderman}}(1969)}]{1969Natur.223..597R}%
  \BibitemOpen
  \bibfield  {author} {\bibinfo {author} {\bibfnamefont {M.}~\bibnamefont
  {{Ruderman}}},\ }\bibfield  {title} {\bibinfo {title} {{Neutron Starquakes
  and Pulsar Periods}},\ }\href {https://doi.org/10.1038/223597b0} {\bibfield
  {journal} {\bibinfo  {journal} {Nature}\ }\textbf {\bibinfo {volume} {223}},\
  \bibinfo {pages} {597} (\bibinfo {year} {1969})}\BibitemShut {NoStop}%
\bibitem [{\citenamefont {{Baym}}\ and\ \citenamefont
  {{Pines}}(1971)}]{1971AnPhy..66..816B}%
  \BibitemOpen
  \bibfield  {author} {\bibinfo {author} {\bibfnamefont {G.}~\bibnamefont
  {{Baym}}}\ and\ \bibinfo {author} {\bibfnamefont {D.}~\bibnamefont
  {{Pines}}},\ }\bibfield  {title} {\bibinfo {title} {{Neutron starquakes and
  pulsar speedup}},\ }\href {https://doi.org/10.1016/0003-4916(71)90084-4}
  {\bibfield  {journal} {\bibinfo  {journal} {Annals of Physics}\ }\textbf
  {\bibinfo {volume} {66}},\ \bibinfo {pages} {816} (\bibinfo {year}
  {1971})}\BibitemShut {NoStop}%
\bibitem [{\citenamefont {Migdal}(1959)}]{1959NucPh..13..655M}%
  \BibitemOpen
  \bibfield  {author} {\bibinfo {author} {\bibfnamefont {A.~B.}\ \bibnamefont
  {Migdal}},\ }\bibfield  {title} {\bibinfo {title} {Superfluidity and the
  moments of inertia of nuclei},\ }\href
  {https://doi.org/https://doi.org/10.1016/0029-5582(59)90264-0} {\bibfield
  {journal} {\bibinfo  {journal} {Nuclear Physics}\ }\textbf {\bibinfo {volume}
  {13}},\ \bibinfo {pages} {655} (\bibinfo {year} {1959})}\BibitemShut
  {NoStop}%
\bibitem [{\citenamefont {Haskell}\ and\ \citenamefont
  {Melatos}(2015)}]{Haskell:2015jra}%
  \BibitemOpen
  \bibfield  {author} {\bibinfo {author} {\bibfnamefont {B.}~\bibnamefont
  {Haskell}}\ and\ \bibinfo {author} {\bibfnamefont {A.}~\bibnamefont
  {Melatos}},\ }\bibfield  {title} {\bibinfo {title} {{Models of Pulsar
  Glitches}},\ }\href {https://doi.org/10.1142/S0218271815300086} {\bibfield
  {journal} {\bibinfo  {journal} {Int. J. Mod. Phys. D}\ }\textbf {\bibinfo
  {volume} {24}},\ \bibinfo {pages} {1530008} (\bibinfo {year} {2015})},\
  \Eprint {https://arxiv.org/abs/1502.07062} {arXiv:1502.07062 [astro-ph.SR]}
  \BibitemShut {NoStop}%
\bibitem [{\citenamefont {Layek}\ and\ \citenamefont
  {Yadav}(2020{\natexlab{b}})}]{Layek:2020ocz}%
  \BibitemOpen
  \bibfield  {author} {\bibinfo {author} {\bibfnamefont {B.}~\bibnamefont
  {Layek}}\ and\ \bibinfo {author} {\bibfnamefont {P.}~\bibnamefont {Yadav}},\
  }\bibfield  {title} {\bibinfo {title} {{Vortex unpinning due to crustquake
  initiated neutron excitation and pulsar glitches}},\ }\href
  {https://doi.org/10.1093/mnras/staa2880} {\bibfield  {journal} {\bibinfo
  {journal} {Mon. Not. Roy. Astron. Soc.}\ }\textbf {\bibinfo {volume} {499}},\
  \bibinfo {pages} {455} (\bibinfo {year} {2020}{\natexlab{b}})},\ \Eprint
  {https://arxiv.org/abs/2009.08085} {arXiv:2009.08085 [astro-ph.HE]}
  \BibitemShut {NoStop}%
\bibitem [{\citenamefont {Layek}\ \emph {et~al.}(2023)\citenamefont {Layek},
  \citenamefont {Venkata},\ and\ \citenamefont {Yadav}}]{Layek:2022kja}%
  \BibitemOpen
  \bibfield  {author} {\bibinfo {author} {\bibfnamefont {B.}~\bibnamefont
  {Layek}}, \bibinfo {author} {\bibfnamefont {D.~G.}\ \bibnamefont {Venkata}},\
  and\ \bibinfo {author} {\bibfnamefont {P.}~\bibnamefont {Yadav}},\ }\bibfield
   {title} {\bibinfo {title} {{Glitches due to quasineutron-vortex scattering
  in the superfluid inner crust of a pulsar}},\ }\href
  {https://doi.org/10.1103/PhysRevD.107.023004} {\bibfield  {journal} {\bibinfo
   {journal} {Phys. Rev. D}\ }\textbf {\bibinfo {volume} {107}},\ \bibinfo
  {pages} {023004} (\bibinfo {year} {2023})},\ \Eprint
  {https://arxiv.org/abs/2207.07834} {arXiv:2207.07834 [astro-ph.HE]}
  \BibitemShut {NoStop}%
\bibitem [{\citenamefont {{Baym}}\ \emph {et~al.}(1969)\citenamefont {{Baym}},
  \citenamefont {{Pethick}},\ and\ \citenamefont
  {{Pines}}}]{1969Natur.224..673B}%
  \BibitemOpen
  \bibfield  {author} {\bibinfo {author} {\bibfnamefont {G.}~\bibnamefont
  {{Baym}}}, \bibinfo {author} {\bibfnamefont {C.}~\bibnamefont {{Pethick}}},\
  and\ \bibinfo {author} {\bibfnamefont {D.}~\bibnamefont {{Pines}}},\
  }\bibfield  {title} {\bibinfo {title} {{Superfluidity in Neutron Stars}},\
  }\href {https://doi.org/10.1038/224673a0} {\bibfield  {journal} {\bibinfo
  {journal} {Nature}\ }\textbf {\bibinfo {volume} {224}},\ \bibinfo {pages}
  {673} (\bibinfo {year} {1969})}\BibitemShut {NoStop}%
\bibitem [{\citenamefont {Archibald}\ \emph {et~al.}(2013)\citenamefont
  {Archibald}, \citenamefont {Kaspi}, \citenamefont {Ng}, \citenamefont
  {Gourgouliatos}, \citenamefont {Tsang}, \citenamefont {Scholz}, \citenamefont
  {Beardmore}, \citenamefont {Gehrels},\ and\ \citenamefont
  {Kennea}}]{Archibald:2013kla}%
  \BibitemOpen
  \bibfield  {author} {\bibinfo {author} {\bibfnamefont {R.~F.}\ \bibnamefont
  {Archibald}}, \bibinfo {author} {\bibfnamefont {V.~M.}\ \bibnamefont
  {Kaspi}}, \bibinfo {author} {\bibfnamefont {C.~Y.}\ \bibnamefont {Ng}},
  \bibinfo {author} {\bibfnamefont {K.~N.}\ \bibnamefont {Gourgouliatos}},
  \bibinfo {author} {\bibfnamefont {D.}~\bibnamefont {Tsang}}, \bibinfo
  {author} {\bibfnamefont {P.}~\bibnamefont {Scholz}}, \bibinfo {author}
  {\bibfnamefont {A.~P.}\ \bibnamefont {Beardmore}}, \bibinfo {author}
  {\bibfnamefont {N.}~\bibnamefont {Gehrels}},\ and\ \bibinfo {author}
  {\bibfnamefont {J.~A.}\ \bibnamefont {Kennea}},\ }\bibfield  {title}
  {\bibinfo {title} {{An Anti-Glitch in a Magnetar}},\ }\href
  {https://doi.org/10.1038/nature12159} {\bibfield  {journal} {\bibinfo
  {journal} {Nature}\ }\textbf {\bibinfo {volume} {497}},\ \bibinfo {pages}
  {591} (\bibinfo {year} {2013})},\ \Eprint {https://arxiv.org/abs/1305.6894}
  {arXiv:1305.6894 [astro-ph.HE]} \BibitemShut {NoStop}%
\bibitem [{\citenamefont {Campana}\ \emph {et~al.}(2012)\citenamefont
  {Campana}, \citenamefont {Salvaterra}, \citenamefont {Melandri},
  \citenamefont {Vergani}, \citenamefont {Covino}, \citenamefont {D'Avanzo},
  \citenamefont {Fugazza}, \citenamefont {Ghisellini}, \citenamefont
  {Sbarufatti},\ and\ \citenamefont {Tagliaferri}}]{Campana:2011ht}%
  \BibitemOpen
  \bibfield  {author} {\bibinfo {author} {\bibfnamefont {S.}~\bibnamefont
  {Campana}}, \bibinfo {author} {\bibfnamefont {R.}~\bibnamefont {Salvaterra}},
  \bibinfo {author} {\bibfnamefont {A.}~\bibnamefont {Melandri}}, \bibinfo
  {author} {\bibfnamefont {S.~D.}\ \bibnamefont {Vergani}}, \bibinfo {author}
  {\bibfnamefont {S.}~\bibnamefont {Covino}}, \bibinfo {author} {\bibfnamefont
  {P.}~\bibnamefont {D'Avanzo}}, \bibinfo {author} {\bibfnamefont
  {D.}~\bibnamefont {Fugazza}}, \bibinfo {author} {\bibfnamefont
  {G.}~\bibnamefont {Ghisellini}}, \bibinfo {author} {\bibfnamefont
  {B.}~\bibnamefont {Sbarufatti}},\ and\ \bibinfo {author} {\bibfnamefont
  {G.}~\bibnamefont {Tagliaferri}},\ }\bibfield  {title} {\bibinfo {title}
  {{The X-ray absorbing column density of a complete sample of bright Swift
  Gamma-Ray Bursts}},\ }\href
  {https://doi.org/10.1111/j.1365-2966.2012.20428.x} {\bibfield  {journal}
  {\bibinfo  {journal} {Mon. Not. Roy. Astron. Soc.}\ }\textbf {\bibinfo
  {volume} {421}},\ \bibinfo {pages} {1697} (\bibinfo {year} {2012})},\ \Eprint
  {https://arxiv.org/abs/1112.5111} {arXiv:1112.5111 [astro-ph.HE]}
  \BibitemShut {NoStop}%
\bibitem [{\citenamefont {Ray}\ \emph {et~al.}(2019)\citenamefont {Ray} \emph
  {et~al.}}]{Ray:2018vuy}%
  \BibitemOpen
  \bibfield  {author} {\bibinfo {author} {\bibfnamefont {P.~S.}\ \bibnamefont
  {Ray}} \emph {et~al.},\ }\bibfield  {title} {\bibinfo {title} {{Anti-glitches
  in the Ultraluminous Accreting Pulsar NGC 300 ULX-1 Observed with $NICER$}},\
  }\href {https://doi.org/10.3847/1538-4357/ab24d8} {\bibfield  {journal}
  {\bibinfo  {journal} {Astrophys. J.}\ }\textbf {\bibinfo {volume} {879}},\
  \bibinfo {pages} {130} (\bibinfo {year} {2019})},\ \Eprint
  {https://arxiv.org/abs/1811.09218} {arXiv:1811.09218 [astro-ph.HE]}
  \BibitemShut {NoStop}%
\bibitem [{\citenamefont {Migdal}(1972)}]{Migdal:1971cu}%
  \BibitemOpen
  \bibfield  {author} {\bibinfo {author} {\bibfnamefont {A.~B.}\ \bibnamefont
  {Migdal}},\ }\bibfield  {title} {\bibinfo {title} {Stability of vacuum and
  limiting fields},\ }\href@noop {} {\bibfield  {journal} {\bibinfo  {journal}
  {Sov. Phys. JETP}\ }\textbf {\bibinfo {volume} {34}},\ \bibinfo {pages}
  {1184} (\bibinfo {year} {1972})}\BibitemShut {NoStop}%
\bibitem [{\citenamefont {Sawyer}(1972)}]{Sawyer:1972cq}%
  \BibitemOpen
  \bibfield  {author} {\bibinfo {author} {\bibfnamefont {R.~F.}\ \bibnamefont
  {Sawyer}},\ }\bibfield  {title} {\bibinfo {title} {{Condensed pi- phase in
  neutron star matter}},\ }\href {https://doi.org/10.1103/PhysRevLett.29.382}
  {\bibfield  {journal} {\bibinfo  {journal} {Phys. Rev. Lett.}\ }\textbf
  {\bibinfo {volume} {29}},\ \bibinfo {pages} {382} (\bibinfo {year}
  {1972})}\BibitemShut {NoStop}%
\bibitem [{\citenamefont {Ohnishi}\ \emph {et~al.}(2009)\citenamefont
  {Ohnishi}, \citenamefont {Jido}, \citenamefont {Sekihara},\ and\
  \citenamefont {Tsubakihara}}]{Ohnishi:2008ng}%
  \BibitemOpen
  \bibfield  {author} {\bibinfo {author} {\bibfnamefont {A.}~\bibnamefont
  {Ohnishi}}, \bibinfo {author} {\bibfnamefont {D.}~\bibnamefont {Jido}},
  \bibinfo {author} {\bibfnamefont {T.}~\bibnamefont {Sekihara}},\ and\
  \bibinfo {author} {\bibfnamefont {K.}~\bibnamefont {Tsubakihara}},\
  }\bibfield  {title} {\bibinfo {title} {{Possibility of s-wave pion
  condensates in neutron stars revisited}},\ }\href
  {https://doi.org/10.1103/PhysRevC.80.038202} {\bibfield  {journal} {\bibinfo
  {journal} {Phys. Rev. C}\ }\textbf {\bibinfo {volume} {80}},\ \bibinfo
  {pages} {038202} (\bibinfo {year} {2009})},\ \Eprint
  {https://arxiv.org/abs/0810.3531} {arXiv:0810.3531 [nucl-th]} \BibitemShut
  {NoStop}%
\bibitem [{\citenamefont {Akmal}\ \emph {et~al.}(1998)\citenamefont {Akmal},
  \citenamefont {Pandharipande},\ and\ \citenamefont
  {Ravenhall}}]{Akmal:1998cf}%
  \BibitemOpen
  \bibfield  {author} {\bibinfo {author} {\bibfnamefont {A.}~\bibnamefont
  {Akmal}}, \bibinfo {author} {\bibfnamefont {V.~R.}\ \bibnamefont
  {Pandharipande}},\ and\ \bibinfo {author} {\bibfnamefont {D.~G.}\
  \bibnamefont {Ravenhall}},\ }\bibfield  {title} {\bibinfo {title} {{The
  Equation of state of nucleon matter and neutron star structure}},\ }\href
  {https://doi.org/10.1103/PhysRevC.58.1804} {\bibfield  {journal} {\bibinfo
  {journal} {Phys. Rev. C}\ }\textbf {\bibinfo {volume} {58}},\ \bibinfo
  {pages} {1804} (\bibinfo {year} {1998})},\ \Eprint
  {https://arxiv.org/abs/nucl-th/9804027} {arXiv:nucl-th/9804027} \BibitemShut
  {NoStop}%
\bibitem [{\citenamefont {Li}\ \emph {et~al.}(1997)\citenamefont {Li},
  \citenamefont {Lee},\ and\ \citenamefont {Brown}}]{Li:1997zb}%
  \BibitemOpen
  \bibfield  {author} {\bibinfo {author} {\bibfnamefont {G.-Q.}\ \bibnamefont
  {Li}}, \bibinfo {author} {\bibfnamefont {C.~H.}\ \bibnamefont {Lee}},\ and\
  \bibinfo {author} {\bibfnamefont {G.~E.}\ \bibnamefont {Brown}},\ }\bibfield
  {title} {\bibinfo {title} {{Kaons in dense matter, kaon production in heavy
  ion collisions, and kaon condensation in neutron stars}},\ }\href
  {https://doi.org/10.1016/S0375-9474(97)00489-2} {\bibfield  {journal}
  {\bibinfo  {journal} {Nucl. Phys. A}\ }\textbf {\bibinfo {volume} {625}},\
  \bibinfo {pages} {372} (\bibinfo {year} {1997})},\ \Eprint
  {https://arxiv.org/abs/nucl-th/9706057} {arXiv:nucl-th/9706057} \BibitemShut
  {NoStop}%
\bibitem [{\citenamefont {Kaplan}\ and\ \citenamefont
  {Nelson}(1988)}]{Kaplan:1987sc}%
  \BibitemOpen
  \bibfield  {author} {\bibinfo {author} {\bibfnamefont {D.~B.}\ \bibnamefont
  {Kaplan}}\ and\ \bibinfo {author} {\bibfnamefont {A.~E.}\ \bibnamefont
  {Nelson}},\ }\bibfield  {title} {\bibinfo {title} {{Kaon Condensation in
  Dense Matter}},\ }\href {https://doi.org/10.1016/0375-9474(88)90442-3}
  {\bibfield  {journal} {\bibinfo  {journal} {Nucl. Phys. A}\ }\textbf
  {\bibinfo {volume} {479}},\ \bibinfo {pages} {273c} (\bibinfo {year}
  {1988})}\BibitemShut {NoStop}%
\bibitem [{\citenamefont {Glendenning}(1985)}]{Glendenning:1984jr}%
  \BibitemOpen
  \bibfield  {author} {\bibinfo {author} {\bibfnamefont {N.~K.}\ \bibnamefont
  {Glendenning}},\ }\bibfield  {title} {\bibinfo {title} {{Neutron Stars Are
  Giant Hypernuclei?}},\ }\href {https://doi.org/10.1086/163253} {\bibfield
  {journal} {\bibinfo  {journal} {Astrophys. J.}\ }\textbf {\bibinfo {volume}
  {293}},\ \bibinfo {pages} {470} (\bibinfo {year} {1985})}\BibitemShut
  {NoStop}%
\bibitem [{\citenamefont {Baym}\ \emph {et~al.}(2018)\citenamefont {Baym},
  \citenamefont {Hatsuda}, \citenamefont {Kojo}, \citenamefont {Powell},
  \citenamefont {Song},\ and\ \citenamefont {Takatsuka}}]{Baym:2017whm}%
  \BibitemOpen
  \bibfield  {author} {\bibinfo {author} {\bibfnamefont {G.}~\bibnamefont
  {Baym}}, \bibinfo {author} {\bibfnamefont {T.}~\bibnamefont {Hatsuda}},
  \bibinfo {author} {\bibfnamefont {T.}~\bibnamefont {Kojo}}, \bibinfo {author}
  {\bibfnamefont {P.~D.}\ \bibnamefont {Powell}}, \bibinfo {author}
  {\bibfnamefont {Y.}~\bibnamefont {Song}},\ and\ \bibinfo {author}
  {\bibfnamefont {T.}~\bibnamefont {Takatsuka}},\ }\bibfield  {title} {\bibinfo
  {title} {{From hadrons to quarks in neutron stars: a review}},\ }\href
  {https://doi.org/10.1088/1361-6633/aaae14} {\bibfield  {journal} {\bibinfo
  {journal} {Rept. Prog. Phys.}\ }\textbf {\bibinfo {volume} {81}},\ \bibinfo
  {pages} {056902} (\bibinfo {year} {2018})},\ \Eprint
  {https://arxiv.org/abs/1707.04966} {arXiv:1707.04966 [astro-ph.HE]}
  \BibitemShut {NoStop}%
\bibitem [{\citenamefont {Weber}\ \emph {et~al.}(2010)\citenamefont {Weber},
  \citenamefont {Hamil}, \citenamefont {Mimura},\ and\ \citenamefont
  {Negreiros}}]{Weber:2010zza}%
  \BibitemOpen
  \bibfield  {author} {\bibinfo {author} {\bibfnamefont {F.}~\bibnamefont
  {Weber}}, \bibinfo {author} {\bibfnamefont {O.}~\bibnamefont {Hamil}},
  \bibinfo {author} {\bibfnamefont {K.}~\bibnamefont {Mimura}},\ and\ \bibinfo
  {author} {\bibfnamefont {R.}~\bibnamefont {Negreiros}},\ }\bibfield  {title}
  {\bibinfo {title} {{From crust to core: A brief review of quark matter in
  neutron stars}},\ }\href {https://doi.org/10.1142/S0218271810017329}
  {\bibfield  {journal} {\bibinfo  {journal} {Int. J. Mod. Phys. D}\ }\textbf
  {\bibinfo {volume} {19}},\ \bibinfo {pages} {1427} (\bibinfo {year}
  {2010})}\BibitemShut {NoStop}%
\bibitem [{\citenamefont {Yamamoto}\ \emph {et~al.}(2022)\citenamefont
  {Yamamoto}, \citenamefont {Yasutake},\ and\ \citenamefont
  {Rijken}}]{Yamamoto:2021htv}%
  \BibitemOpen
  \bibfield  {author} {\bibinfo {author} {\bibfnamefont {Y.}~\bibnamefont
  {Yamamoto}}, \bibinfo {author} {\bibfnamefont {N.}~\bibnamefont {Yasutake}},\
  and\ \bibinfo {author} {\bibfnamefont {T.~A.}\ \bibnamefont {Rijken}},\
  }\bibfield  {title} {\bibinfo {title} {{Quark-quark interaction and quark
  matter in neutron stars}},\ }\href
  {https://doi.org/10.1103/PhysRevC.105.015804} {\bibfield  {journal} {\bibinfo
   {journal} {Phys. Rev. C}\ }\textbf {\bibinfo {volume} {105}},\ \bibinfo
  {pages} {015804} (\bibinfo {year} {2022})},\ \Eprint
  {https://arxiv.org/abs/2112.12931} {arXiv:2112.12931 [nucl-th]} \BibitemShut
  {NoStop}%
\bibitem [{\citenamefont {Annala}\ \emph {et~al.}(2020)\citenamefont {Annala},
  \citenamefont {Gorda}, \citenamefont {Kurkela}, \citenamefont {N\"attil\"a},\
  and\ \citenamefont {Vuorinen}}]{Annala:2019puf}%
  \BibitemOpen
  \bibfield  {author} {\bibinfo {author} {\bibfnamefont {E.}~\bibnamefont
  {Annala}}, \bibinfo {author} {\bibfnamefont {T.}~\bibnamefont {Gorda}},
  \bibinfo {author} {\bibfnamefont {A.}~\bibnamefont {Kurkela}}, \bibinfo
  {author} {\bibfnamefont {J.}~\bibnamefont {N\"attil\"a}},\ and\ \bibinfo
  {author} {\bibfnamefont {A.}~\bibnamefont {Vuorinen}},\ }\bibfield  {title}
  {\bibinfo {title} {{Evidence for quark-matter cores in massive neutron
  stars}},\ }\href {https://doi.org/10.1038/s41567-020-0914-9} {\bibfield
  {journal} {\bibinfo  {journal} {Nature Phys.}\ }\textbf {\bibinfo {volume}
  {16}},\ \bibinfo {pages} {907} (\bibinfo {year} {2020})},\ \Eprint
  {https://arxiv.org/abs/1903.09121} {arXiv:1903.09121 [astro-ph.HE]}
  \BibitemShut {NoStop}%
\bibitem [{\citenamefont {Alford}\ \emph {et~al.}(2019)\citenamefont {Alford},
  \citenamefont {Han},\ and\ \citenamefont {Schwenzer}}]{Alford:2019oge}%
  \BibitemOpen
  \bibfield  {author} {\bibinfo {author} {\bibfnamefont {M.~G.}\ \bibnamefont
  {Alford}}, \bibinfo {author} {\bibfnamefont {S.}~\bibnamefont {Han}},\ and\
  \bibinfo {author} {\bibfnamefont {K.}~\bibnamefont {Schwenzer}},\ }\bibfield
  {title} {\bibinfo {title} {{Signatures for quark matter from multi-messenger
  observations}},\ }\href {https://doi.org/10.1088/1361-6471/ab337a} {\bibfield
   {journal} {\bibinfo  {journal} {J. Phys. G}\ }\textbf {\bibinfo {volume}
  {46}},\ \bibinfo {pages} {114001} (\bibinfo {year} {2019})},\ \Eprint
  {https://arxiv.org/abs/1904.05471} {arXiv:1904.05471 [nucl-th]} \BibitemShut
  {NoStop}%
\bibitem [{\citenamefont {Kuroda}\ \emph {et~al.}(2022)\citenamefont {Kuroda},
  \citenamefont {Fischer}, \citenamefont {Takiwaki},\ and\ \citenamefont
  {Kotake}}]{Kuroda:2021eiv}%
  \BibitemOpen
  \bibfield  {author} {\bibinfo {author} {\bibfnamefont {T.}~\bibnamefont
  {Kuroda}}, \bibinfo {author} {\bibfnamefont {T.}~\bibnamefont {Fischer}},
  \bibinfo {author} {\bibfnamefont {T.}~\bibnamefont {Takiwaki}},\ and\
  \bibinfo {author} {\bibfnamefont {K.}~\bibnamefont {Kotake}},\ }\bibfield
  {title} {\bibinfo {title} {{Core-collapse Supernova Simulations and the
  Formation of Neutron Stars, Hybrid Stars, and Black Holes}},\ }\href
  {https://doi.org/10.3847/1538-4357/ac31a8} {\bibfield  {journal} {\bibinfo
  {journal} {Astrophys. J.}\ }\textbf {\bibinfo {volume} {924}},\ \bibinfo
  {pages} {38} (\bibinfo {year} {2022})},\ \Eprint
  {https://arxiv.org/abs/2109.01508} {arXiv:2109.01508 [astro-ph.HE]}
  \BibitemShut {NoStop}%
\bibitem [{\citenamefont {Stark}\ and\ \citenamefont
  {Piran}(1985)}]{Stark:1985da}%
  \BibitemOpen
  \bibfield  {author} {\bibinfo {author} {\bibfnamefont {R.~F.}\ \bibnamefont
  {Stark}}\ and\ \bibinfo {author} {\bibfnamefont {T.}~\bibnamefont {Piran}},\
  }\bibfield  {title} {\bibinfo {title} {{Gravitational Wave Emission from
  Rotating Gravitational Collapse}},\ }\href
  {https://doi.org/10.1103/PhysRevLett.55.891} {\bibfield  {journal} {\bibinfo
  {journal} {Phys. Rev. Lett.}\ }\textbf {\bibinfo {volume} {55}},\ \bibinfo
  {pages} {891} (\bibinfo {year} {1985})},\ \bibinfo {note} {[Erratum:
  Phys.Rev.Lett. 56, 97 (1986)]}\BibitemShut {NoStop}%
\bibitem [{\citenamefont {Giacomazzo}\ and\ \citenamefont
  {Perna}(2012)}]{Giacomazzo:2012bw}%
  \BibitemOpen
  \bibfield  {author} {\bibinfo {author} {\bibfnamefont {B.}~\bibnamefont
  {Giacomazzo}}\ and\ \bibinfo {author} {\bibfnamefont {R.}~\bibnamefont
  {Perna}},\ }\bibfield  {title} {\bibinfo {title} {{General Relativistic
  Simulations of Accretion Induced Collapse of Neutron Stars to Black Holes}},\
  }\href {https://doi.org/10.1088/2041-8205/758/1/L8} {\bibfield  {journal}
  {\bibinfo  {journal} {Astrophys. J. Lett.}\ }\textbf {\bibinfo {volume}
  {758}},\ \bibinfo {pages} {L8} (\bibinfo {year} {2012})},\ \Eprint
  {https://arxiv.org/abs/1209.0783} {arXiv:1209.0783 [astro-ph.HE]}
  \BibitemShut {NoStop}%
\bibitem [{\citenamefont {Giacomazzo}\ \emph {et~al.}(2011)\citenamefont
  {Giacomazzo}, \citenamefont {Rezzolla},\ and\ \citenamefont
  {Stergioulas}}]{Giacomazzo:2011cv}%
  \BibitemOpen
  \bibfield  {author} {\bibinfo {author} {\bibfnamefont {B.}~\bibnamefont
  {Giacomazzo}}, \bibinfo {author} {\bibfnamefont {L.}~\bibnamefont
  {Rezzolla}},\ and\ \bibinfo {author} {\bibfnamefont {N.}~\bibnamefont
  {Stergioulas}},\ }\bibfield  {title} {\bibinfo {title} {{Collapse of
  differentially rotating neutron stars and cosmic censorship}},\ }\href
  {https://doi.org/10.1103/PhysRevD.84.024022} {\bibfield  {journal} {\bibinfo
  {journal} {Phys. Rev. D}\ }\textbf {\bibinfo {volume} {84}},\ \bibinfo
  {pages} {024022} (\bibinfo {year} {2011})},\ \Eprint
  {https://arxiv.org/abs/1105.0122} {arXiv:1105.0122 [gr-qc]} \BibitemShut
  {NoStop}%
\bibitem [{\citenamefont {Richardson}(1972)}]{Richardson:1972xn}%
  \BibitemOpen
  \bibfield  {author} {\bibinfo {author} {\bibfnamefont {R.~W.}\ \bibnamefont
  {Richardson}},\ }\bibfield  {title} {\bibinfo {title} {{Ginzburg-landau
  theory of anisotropic superfluid neutron-star matter}},\ }\href
  {https://doi.org/10.1103/PhysRevD.5.1883} {\bibfield  {journal} {\bibinfo
  {journal} {Phys. Rev. D}\ }\textbf {\bibinfo {volume} {5}},\ \bibinfo {pages}
  {1883} (\bibinfo {year} {1972})}\BibitemShut {NoStop}%
\bibitem [{\citenamefont {Kibble}(1976)}]{Kibble:1976sj}%
  \BibitemOpen
  \bibfield  {author} {\bibinfo {author} {\bibfnamefont {T.~W.~B.}\
  \bibnamefont {Kibble}},\ }\bibfield  {title} {\bibinfo {title} {{Topology of
  Cosmic Domains and Strings}},\ }\href
  {https://doi.org/10.1088/0305-4470/9/8/029} {\bibfield  {journal} {\bibinfo
  {journal} {J. Phys. A}\ }\textbf {\bibinfo {volume} {9}},\ \bibinfo {pages}
  {1387} (\bibinfo {year} {1976})}\BibitemShut {NoStop}%
\bibitem [{\citenamefont {Kibble}(1980)}]{Kibble:1980mv}%
  \BibitemOpen
  \bibfield  {author} {\bibinfo {author} {\bibfnamefont {T.~W.~B.}\
  \bibnamefont {Kibble}},\ }\bibfield  {title} {\bibinfo {title} {{Some
  Implications of a Cosmological Phase Transition}},\ }\href
  {https://doi.org/10.1016/0370-1573(80)90091-5} {\bibfield  {journal}
  {\bibinfo  {journal} {Phys. Rept.}\ }\textbf {\bibinfo {volume} {67}},\
  \bibinfo {pages} {183} (\bibinfo {year} {1980})}\BibitemShut {NoStop}%
\bibitem [{\citenamefont {Gupta}\ \emph {et~al.}(2010)\citenamefont {Gupta},
  \citenamefont {Mohapatra}, \citenamefont {Srivastava},\ and\ \citenamefont
  {Tiwari}}]{Gupta:2010pp}%
  \BibitemOpen
  \bibfield  {author} {\bibinfo {author} {\bibfnamefont {U.~S.}\ \bibnamefont
  {Gupta}}, \bibinfo {author} {\bibfnamefont {R.~K.}\ \bibnamefont
  {Mohapatra}}, \bibinfo {author} {\bibfnamefont {A.~M.}\ \bibnamefont
  {Srivastava}},\ and\ \bibinfo {author} {\bibfnamefont {V.~K.}\ \bibnamefont
  {Tiwari}},\ }\bibfield  {title} {\bibinfo {title} {{Simulation of Z(3) walls
  and string production via bubble nucleation in a quark-hadron transition}},\
  }\href {https://doi.org/10.1103/PhysRevD.82.074020} {\bibfield  {journal}
  {\bibinfo  {journal} {Phys. Rev. D}\ }\textbf {\bibinfo {volume} {82}},\
  \bibinfo {pages} {074020} (\bibinfo {year} {2010})},\ \Eprint
  {https://arxiv.org/abs/1007.5001} {arXiv:1007.5001 [hep-ph]} \BibitemShut
  {NoStop}%
\bibitem [{\citenamefont {Vachaspati}\ and\ \citenamefont
  {Vilenkin}(1984)}]{Vachaspati:1984dz}%
  \BibitemOpen
  \bibfield  {author} {\bibinfo {author} {\bibfnamefont {T.}~\bibnamefont
  {Vachaspati}}\ and\ \bibinfo {author} {\bibfnamefont {A.}~\bibnamefont
  {Vilenkin}},\ }\bibfield  {title} {\bibinfo {title} {{Formation and Evolution
  of Cosmic Strings}},\ }\href {https://doi.org/10.1103/PhysRevD.30.2036}
  {\bibfield  {journal} {\bibinfo  {journal} {Phys. Rev. D}\ }\textbf {\bibinfo
  {volume} {30}},\ \bibinfo {pages} {2036} (\bibinfo {year}
  {1984})}\BibitemShut {NoStop}%
\bibitem [{\citenamefont {McLerran}\ and\ \citenamefont
  {Svetitsky}(1981)}]{PhysRevD.24.450}%
  \BibitemOpen
  \bibfield  {author} {\bibinfo {author} {\bibfnamefont {L.~D.}\ \bibnamefont
  {McLerran}}\ and\ \bibinfo {author} {\bibfnamefont {B.}~\bibnamefont
  {Svetitsky}},\ }\bibfield  {title} {\bibinfo {title} {Quark liberation at
  high temperature: A monte carlo study of $su(2)$ gauge theory},\ }\href
  {https://doi.org/10.1103/PhysRevD.24.450} {\bibfield  {journal} {\bibinfo
  {journal} {Phys. Rev. D}\ }\textbf {\bibinfo {volume} {24}},\ \bibinfo
  {pages} {450} (\bibinfo {year} {1981})}\BibitemShut {NoStop}%
\bibitem [{\citenamefont {Svetitsky}(1986)}]{Svetitsky:1985ye}%
  \BibitemOpen
  \bibfield  {author} {\bibinfo {author} {\bibfnamefont {B.}~\bibnamefont
  {Svetitsky}},\ }\bibfield  {title} {\bibinfo {title} {{Symmetry Aspects of
  Finite Temperature Confinement Transitions}},\ }\href
  {https://doi.org/10.1016/0370-1573(86)90014-1} {\bibfield  {journal}
  {\bibinfo  {journal} {Phys. Rept.}\ }\textbf {\bibinfo {volume} {132}},\
  \bibinfo {pages} {1} (\bibinfo {year} {1986})}\BibitemShut {NoStop}%
\bibitem [{\citenamefont {Bhattacharya}\ \emph {et~al.}(1992)\citenamefont
  {Bhattacharya}, \citenamefont {Gocksch}, \citenamefont {Korthals~Altes},\
  and\ \citenamefont {Pisarski}}]{Bhattacharya:1992qb}%
  \BibitemOpen
  \bibfield  {author} {\bibinfo {author} {\bibfnamefont {T.}~\bibnamefont
  {Bhattacharya}}, \bibinfo {author} {\bibfnamefont {A.}~\bibnamefont
  {Gocksch}}, \bibinfo {author} {\bibfnamefont {C.}~\bibnamefont
  {Korthals~Altes}},\ and\ \bibinfo {author} {\bibfnamefont {R.~D.}\
  \bibnamefont {Pisarski}},\ }\bibfield  {title} {\bibinfo {title} {{Z(N)
  interface tension in a hot SU(N) gauge theory}},\ }\href
  {https://doi.org/10.1016/0550-3213(92)90086-Q} {\bibfield  {journal}
  {\bibinfo  {journal} {Nucl. Phys. B}\ }\textbf {\bibinfo {volume} {383}},\
  \bibinfo {pages} {497} (\bibinfo {year} {1992})},\ \Eprint
  {https://arxiv.org/abs/hep-ph/9205231} {arXiv:hep-ph/9205231} \BibitemShut
  {NoStop}%
\bibitem [{\citenamefont {Layek}\ \emph {et~al.}(2005)\citenamefont {Layek},
  \citenamefont {Mishra},\ and\ \citenamefont {Srivastava}}]{Layek:2005fn}%
  \BibitemOpen
  \bibfield  {author} {\bibinfo {author} {\bibfnamefont {B.}~\bibnamefont
  {Layek}}, \bibinfo {author} {\bibfnamefont {A.~P.}\ \bibnamefont {Mishra}},\
  and\ \bibinfo {author} {\bibfnamefont {A.~M.}\ \bibnamefont {Srivastava}},\
  }\bibfield  {title} {\bibinfo {title} {{Strings with a confining core in a
  quark-gluon plasma}},\ }\href {https://doi.org/10.1103/PhysRevD.71.074015}
  {\bibfield  {journal} {\bibinfo  {journal} {Phys. Rev. D}\ }\textbf {\bibinfo
  {volume} {71}},\ \bibinfo {pages} {074015} (\bibinfo {year} {2005})},\
  \Eprint {https://arxiv.org/abs/hep-ph/0502250} {arXiv:hep-ph/0502250}
  \BibitemShut {NoStop}%
\bibitem [{\citenamefont {Gupta}\ \emph {et~al.}(2012)\citenamefont {Gupta},
  \citenamefont {Mohapatra}, \citenamefont {Srivastava},\ and\ \citenamefont
  {Tiwari}}]{Gupta:2011ag}%
  \BibitemOpen
  \bibfield  {author} {\bibinfo {author} {\bibfnamefont {U.~S.}\ \bibnamefont
  {Gupta}}, \bibinfo {author} {\bibfnamefont {R.~K.}\ \bibnamefont
  {Mohapatra}}, \bibinfo {author} {\bibfnamefont {A.~M.}\ \bibnamefont
  {Srivastava}},\ and\ \bibinfo {author} {\bibfnamefont {V.~K.}\ \bibnamefont
  {Tiwari}},\ }\bibfield  {title} {\bibinfo {title} {{Effects of Quarks on the
  Formation and Evolution of $Z(3)$ Walls and Strings in Relativistic Heavy-Ion
  Collisions}},\ }\href {https://doi.org/10.1103/PhysRevD.86.125016} {\bibfield
   {journal} {\bibinfo  {journal} {Phys. Rev. D}\ }\textbf {\bibinfo {volume}
  {86}},\ \bibinfo {pages} {125016} (\bibinfo {year} {2012})},\ \Eprint
  {https://arxiv.org/abs/1111.5402} {arXiv:1111.5402 [hep-ph]} \BibitemShut
  {NoStop}%
\bibitem [{\citenamefont {Mohapatra}\ and\ \citenamefont
  {Srivastava}(2013)}]{Mohapatra:2012ck}%
  \BibitemOpen
  \bibfield  {author} {\bibinfo {author} {\bibfnamefont {R.~K.}\ \bibnamefont
  {Mohapatra}}\ and\ \bibinfo {author} {\bibfnamefont {A.~M.}\ \bibnamefont
  {Srivastava}},\ }\bibfield  {title} {\bibinfo {title} {{Domain growth and
  fluctuations during quenched transition to quark-gluon plasma in relativistic
  heavy-ion collisions}},\ }\href {https://doi.org/10.1103/PhysRevC.88.044901}
  {\bibfield  {journal} {\bibinfo  {journal} {Phys. Rev. C}\ }\textbf {\bibinfo
  {volume} {88}},\ \bibinfo {pages} {044901} (\bibinfo {year} {2013})},\
  \Eprint {https://arxiv.org/abs/1210.4718} {arXiv:1210.4718 [hep-ph]}
  \BibitemShut {NoStop}%
\bibitem [{\citenamefont {Alford}\ \emph {et~al.}(1998)\citenamefont {Alford},
  \citenamefont {Rajagopal},\ and\ \citenamefont {Wilczek}}]{Alford:1997zt}%
  \BibitemOpen
  \bibfield  {author} {\bibinfo {author} {\bibfnamefont {M.~G.}\ \bibnamefont
  {Alford}}, \bibinfo {author} {\bibfnamefont {K.}~\bibnamefont {Rajagopal}},\
  and\ \bibinfo {author} {\bibfnamefont {F.}~\bibnamefont {Wilczek}},\
  }\bibfield  {title} {\bibinfo {title} {{QCD at finite baryon density: Nucleon
  droplets and color superconductivity}},\ }\href
  {https://doi.org/10.1016/S0370-2693(98)00051-3} {\bibfield  {journal}
  {\bibinfo  {journal} {Phys. Lett. B}\ }\textbf {\bibinfo {volume} {422}},\
  \bibinfo {pages} {247} (\bibinfo {year} {1998})},\ \Eprint
  {https://arxiv.org/abs/hep-ph/9711395} {arXiv:hep-ph/9711395} \BibitemShut
  {NoStop}%
\bibitem [{\citenamefont {Haensel}\ \emph {et~al.}(2007)\citenamefont
  {Haensel}, \citenamefont {Potekhin},\ and\ \citenamefont
  {Yakovlev}}]{Haensel:2007yy}%
  \BibitemOpen
  \bibfield  {author} {\bibinfo {author} {\bibfnamefont {P.}~\bibnamefont
  {Haensel}}, \bibinfo {author} {\bibfnamefont {A.~Y.}\ \bibnamefont
  {Potekhin}},\ and\ \bibinfo {author} {\bibfnamefont {D.~G.}\ \bibnamefont
  {Yakovlev}},\ }\href {https://doi.org/10.1007/978-0-387-47301-7} {\emph
  {\bibinfo {title} {{Neutron stars 1: Equation of state and structure}}}},\
  Vol.\ \bibinfo {volume} {326}\ (\bibinfo  {publisher} {Springer},\ \bibinfo
  {address} {New York, USA},\ \bibinfo {year} {2007})\BibitemShut {NoStop}%
\bibitem [{\citenamefont {Horowitz}\ and\ \citenamefont
  {Kadau}(2009)}]{Horowitz:2009ya}%
  \BibitemOpen
  \bibfield  {author} {\bibinfo {author} {\bibfnamefont {C.~J.}\ \bibnamefont
  {Horowitz}}\ and\ \bibinfo {author} {\bibfnamefont {K.}~\bibnamefont
  {Kadau}},\ }\bibfield  {title} {\bibinfo {title} {{The Breaking Strain of
  Neutron Star Crust and Gravitational Waves}},\ }\href
  {https://doi.org/10.1103/PhysRevLett.102.191102} {\bibfield  {journal}
  {\bibinfo  {journal} {Phys. Rev. Lett.}\ }\textbf {\bibinfo {volume} {102}},\
  \bibinfo {pages} {191102} (\bibinfo {year} {2009})},\ \Eprint
  {https://arxiv.org/abs/0904.1986} {arXiv:0904.1986 [astro-ph.SR]}
  \BibitemShut {NoStop}%
\bibitem [{\citenamefont {Baiko}\ and\ \citenamefont
  {Chugunov}(2018)}]{Baiko:2018jax}%
  \BibitemOpen
  \bibfield  {author} {\bibinfo {author} {\bibfnamefont {D.~A.}\ \bibnamefont
  {Baiko}}\ and\ \bibinfo {author} {\bibfnamefont {A.~I.}\ \bibnamefont
  {Chugunov}},\ }\bibfield  {title} {\bibinfo {title} {{Breaking properties of
  neutron star crust}},\ }\href {https://doi.org/10.1093/mnras/sty2259}
  {\bibfield  {journal} {\bibinfo  {journal} {Mon. Not. Roy. Astron. Soc.}\
  }\textbf {\bibinfo {volume} {480}},\ \bibinfo {pages} {5511} (\bibinfo {year}
  {2018})},\ \Eprint {https://arxiv.org/abs/1808.06415} {arXiv:1808.06415
  [astro-ph.HE]} \BibitemShut {NoStop}%
\bibitem [{\citenamefont {Makishima}\ \emph {et~al.}(2014)\citenamefont
  {Makishima}, \citenamefont {Enoto}, \citenamefont {Hiraga}, \citenamefont
  {Nakano}, \citenamefont {Nakazawa}, \citenamefont {Sakurai}, \citenamefont
  {Sasano},\ and\ \citenamefont {Murakami}}]{Makishima:2014dua}%
  \BibitemOpen
  \bibfield  {author} {\bibinfo {author} {\bibfnamefont {K.}~\bibnamefont
  {Makishima}}, \bibinfo {author} {\bibfnamefont {T.}~\bibnamefont {Enoto}},
  \bibinfo {author} {\bibfnamefont {J.~S.}\ \bibnamefont {Hiraga}}, \bibinfo
  {author} {\bibfnamefont {T.}~\bibnamefont {Nakano}}, \bibinfo {author}
  {\bibfnamefont {K.}~\bibnamefont {Nakazawa}}, \bibinfo {author}
  {\bibfnamefont {S.}~\bibnamefont {Sakurai}}, \bibinfo {author} {\bibfnamefont
  {M.}~\bibnamefont {Sasano}},\ and\ \bibinfo {author} {\bibfnamefont
  {H.}~\bibnamefont {Murakami}},\ }\bibfield  {title} {\bibinfo {title}
  {{Possible Evidence for Free Precession of a Strongly Magnetized Neutron Star
  in the Magnetar 4U 0142+61}},\ }\href
  {https://doi.org/10.1103/PhysRevLett.112.171102} {\bibfield  {journal}
  {\bibinfo  {journal} {Phys. Rev. Lett.}\ }\textbf {\bibinfo {volume} {112}},\
  \bibinfo {pages} {171102} (\bibinfo {year} {2014})},\ \Eprint
  {https://arxiv.org/abs/1404.3705} {arXiv:1404.3705 [astro-ph.HE]}
  \BibitemShut {NoStop}%
\bibitem [{\citenamefont {Abbott}\ \emph {et~al.}(2020)\citenamefont {Abbott}
  \emph {et~al.}}]{LIGOScientific:2020gml}%
  \BibitemOpen
  \bibfield  {author} {\bibinfo {author} {\bibfnamefont {R.}~\bibnamefont
  {Abbott}} \emph {et~al.} (\bibinfo {collaboration} {LIGO Scientific,
  Virgo}),\ }\bibfield  {title} {\bibinfo {title} {{Gravitational-wave
  Constraints on the Equatorial Ellipticity of Millisecond Pulsars}},\ }\href
  {https://doi.org/10.3847/2041-8213/abb655} {\bibfield  {journal} {\bibinfo
  {journal} {Astrophys. J. Lett.}\ }\textbf {\bibinfo {volume} {902}},\
  \bibinfo {pages} {L21} (\bibinfo {year} {2020})},\ \Eprint
  {https://arxiv.org/abs/2007.14251} {arXiv:2007.14251 [astro-ph.HE]}
  \BibitemShut {NoStop}%
\bibitem [{\citenamefont {Aasi}\ \emph {et~al.}(2014)\citenamefont {Aasi} \emph
  {et~al.}}]{LIGOScientific:2013rhu}%
  \BibitemOpen
  \bibfield  {author} {\bibinfo {author} {\bibfnamefont {J.}~\bibnamefont
  {Aasi}} \emph {et~al.} (\bibinfo {collaboration} {LIGO Scientific}),\
  }\bibfield  {title} {\bibinfo {title} {{Gravitational waves from known
  pulsars: results from the initial detector era}},\ }\href
  {https://doi.org/10.1088/0004-637X/785/2/119} {\bibfield  {journal} {\bibinfo
   {journal} {Astrophys. J.}\ }\textbf {\bibinfo {volume} {785}},\ \bibinfo
  {pages} {119} (\bibinfo {year} {2014})},\ \Eprint
  {https://arxiv.org/abs/1309.4027} {arXiv:1309.4027 [astro-ph.HE]}
  \BibitemShut {NoStop}%
\bibitem [{\citenamefont {Kleppner}\ and\ \citenamefont
  {Kolenkow}(2014)}]{Kleppner:2014int}%
  \BibitemOpen
  \bibfield  {author} {\bibinfo {author} {\bibfnamefont {D.}~\bibnamefont
  {Kleppner}}\ and\ \bibinfo {author} {\bibfnamefont {R.}~\bibnamefont
  {Kolenkow}},\ }\href {https://books.google.co.in/books?id=Se7CAQAAQBAJ}
  {\emph {\bibinfo {title} {An Introduction to Mechanics}}}\ (\bibinfo
  {publisher} {Cambridge University Press},\ \bibinfo {year}
  {2014})\BibitemShut {NoStop}%
\bibitem [{\citenamefont {Goldstein}(2002)}]{Goldstein:2002cla}%
  \BibitemOpen
  \bibfield  {author} {\bibinfo {author} {\bibfnamefont {H.}~\bibnamefont
  {Goldstein}},\ }\href {https://books.google.co.in/books?id=Spy6xHWFJIEC}
  {\emph {\bibinfo {title} {Classical Mechanics}}}\ (\bibinfo  {publisher}
  {Pearson Education},\ \bibinfo {year} {2002})\BibitemShut {NoStop}%
\bibitem [{\citenamefont {{Colgate}}\ and\ \citenamefont
  {{Petschek}}(1981)}]{1981ApJ...248..771C}%
  \BibitemOpen
  \bibfield  {author} {\bibinfo {author} {\bibfnamefont {S.~A.}\ \bibnamefont
  {{Colgate}}}\ and\ \bibinfo {author} {\bibfnamefont {A.~G.}\ \bibnamefont
  {{Petschek}}},\ }\bibfield  {title} {\bibinfo {title} {{Gamma ray bursts and
  neutron star accretion of a solid body}},\ }\href
  {https://doi.org/10.1086/159201} {\bibfield  {journal} {\bibinfo  {journal}
  {Astrophys. J.}\ }\textbf {\bibinfo {volume} {248}},\ \bibinfo {pages} {771}
  (\bibinfo {year} {1981})}\BibitemShut {NoStop}%
\bibitem [{\citenamefont {Bagchi}\ \emph {et~al.}(2024)\citenamefont {Bagchi},
  \citenamefont {Layek}, \citenamefont {Sarkar},\ and\ \citenamefont
  {Srivastava}}]{asteroid}%
  \BibitemOpen
  \bibfield  {author} {\bibinfo {author} {\bibfnamefont {P.}~\bibnamefont
  {Bagchi}}, \bibinfo {author} {\bibfnamefont {B.}~\bibnamefont {Layek}},
  \bibinfo {author} {\bibfnamefont {A.}~\bibnamefont {Sarkar}},\ and\ \bibinfo
  {author} {\bibfnamefont {A.~M.}\ \bibnamefont {Srivastava}},\ }\href@noop {}
  {\emph {\bibinfo {title} {Manuscript in preparation}}}\ (\bibinfo {year}
  {2024})\BibitemShut {NoStop}%
\bibitem [{\citenamefont {{Thorne}}\ and\ \citenamefont
  {{Campolattaro}}(1967)}]{1967ApJ...149..591T}%
  \BibitemOpen
  \bibfield  {author} {\bibinfo {author} {\bibfnamefont {K.~S.}\ \bibnamefont
  {{Thorne}}}\ and\ \bibinfo {author} {\bibfnamefont {A.}~\bibnamefont
  {{Campolattaro}}},\ }\bibfield  {title} {\bibinfo {title} {{Non-Radial
  Pulsation of General-Relativistic Stellar Models. I. Analytic Analysis for L
  $\ge$ 2}},\ }\href {https://doi.org/10.1086/149288} {\bibfield  {journal}
  {\bibinfo  {journal} {Astrophys. J.}\ ,\ \bibinfo {pages} {591}} (\bibinfo
  {year} {1967})}\BibitemShut {NoStop}%
\bibitem [{\citenamefont {{Price}}\ and\ \citenamefont
  {{Thorne}}(1969)}]{1969ApJ...155..163P}%
  \BibitemOpen
  \bibfield  {author} {\bibinfo {author} {\bibfnamefont {R.}~\bibnamefont
  {{Price}}}\ and\ \bibinfo {author} {\bibfnamefont {K.~S.}\ \bibnamefont
  {{Thorne}}},\ }\bibfield  {title} {\bibinfo {title} {{Non-Radial Pulsation of
  General-Relativistic Stellar Models. II. Properties of the Gravitational
  Waves}},\ }\href {https://doi.org/10.1086/149857} {\bibfield  {journal}
  {\bibinfo  {journal} {Astrophys. J.}\ }\textbf {\bibinfo {volume} {155}},\
  \bibinfo {pages} {163} (\bibinfo {year} {1969})}\BibitemShut {NoStop}%
\bibitem [{\citenamefont {{Thorne}}(1969)}]{1969ApJ...158....1T}%
  \BibitemOpen
  \bibfield  {author} {\bibinfo {author} {\bibfnamefont {K.~S.}\ \bibnamefont
  {{Thorne}}},\ }\bibfield  {title} {\bibinfo {title} {{Nonradial Pulsation of
  General-Relativistic Stellar Models. III. Analytic and Numerical Results for
  Neutron Stars}},\ }\href {https://doi.org/10.1086/150168} {\bibfield
  {journal} {\bibinfo  {journal} {Astrophys. J.}\ }\textbf {\bibinfo {volume}
  {158}},\ \bibinfo {pages} {1} (\bibinfo {year} {1969})}\BibitemShut {NoStop}%
\bibitem [{\citenamefont {Thorne}(1969)}]{Thorne:1969rba}%
  \BibitemOpen
  \bibfield  {author} {\bibinfo {author} {\bibfnamefont {K.~S.}\ \bibnamefont
  {Thorne}},\ }\bibfield  {title} {\bibinfo {title} {{Nonradial Pulsation of
  General-Relativistic Stellar Models. IV. The Weakfield Limit}},\ }\href
  {https://doi.org/10.1086/150259} {\bibfield  {journal} {\bibinfo  {journal}
  {Astrophys. J.}\ }\textbf {\bibinfo {volume} {158}},\ \bibinfo {pages} {997}
  (\bibinfo {year} {1969})}\BibitemShut {NoStop}%
\bibitem [{\citenamefont {Zimmermann}\ and\ \citenamefont
  {Szedenits}(1979)}]{Zimmermann:1979ip}%
  \BibitemOpen
  \bibfield  {author} {\bibinfo {author} {\bibfnamefont {M.}~\bibnamefont
  {Zimmermann}}\ and\ \bibinfo {author} {\bibfnamefont {E.}~\bibnamefont
  {Szedenits}},\ }\bibfield  {title} {\bibinfo {title} {Gravitational waves
  from rotating and precessing rigid bodies: Simple models and applications to
  pulsars},\ }\href {https://doi.org/10.1103/PhysRevD.20.351} {\bibfield
  {journal} {\bibinfo  {journal} {Phys. Rev. D}\ }\textbf {\bibinfo {volume}
  {20}},\ \bibinfo {pages} {351} (\bibinfo {year} {1979})}\BibitemShut
  {NoStop}%
\bibitem [{\citenamefont {Lasky}(2015)}]{Lasky:2015uia}%
  \BibitemOpen
  \bibfield  {author} {\bibinfo {author} {\bibfnamefont {P.~D.}\ \bibnamefont
  {Lasky}},\ }\bibfield  {title} {\bibinfo {title} {{Gravitational Waves from
  Neutron Stars: A Review}},\ }\href {https://doi.org/10.1017/pasa.2015.35}
  {\bibfield  {journal} {\bibinfo  {journal} {Publ. Astron. Soc. Austral.}\
  }\textbf {\bibinfo {volume} {32}},\ \bibinfo {pages} {e034} (\bibinfo {year}
  {2015})},\ \Eprint {https://arxiv.org/abs/1508.06643} {arXiv:1508.06643
  [astro-ph.HE]} \BibitemShut {NoStop}%
\bibitem [{\citenamefont {Keer}\ and\ \citenamefont
  {Jones}(2015)}]{Keer:2014uva}%
  \BibitemOpen
  \bibfield  {author} {\bibinfo {author} {\bibfnamefont {L.}~\bibnamefont
  {Keer}}\ and\ \bibinfo {author} {\bibfnamefont {D.~I.}\ \bibnamefont
  {Jones}},\ }\bibfield  {title} {\bibinfo {title} {{Developing a model for
  neutron star oscillations following starquakes}},\ }\href
  {https://doi.org/10.1093/mnras/stu2123} {\bibfield  {journal} {\bibinfo
  {journal} {Mon. Not. Roy. Astron. Soc.}\ }\textbf {\bibinfo {volume} {446}},\
  \bibinfo {pages} {865} (\bibinfo {year} {2015})},\ \Eprint
  {https://arxiv.org/abs/1408.1249} {arXiv:1408.1249 [astro-ph.SR]}
  \BibitemShut {NoStop}%
\bibitem [{\citenamefont {Riles}(2013)}]{Riles:2012yw}%
  \BibitemOpen
  \bibfield  {author} {\bibinfo {author} {\bibfnamefont {K.}~\bibnamefont
  {Riles}},\ }\bibfield  {title} {\bibinfo {title} {{Gravitational Waves:
  Sources, Detectors and Searches}},\ }\href
  {https://doi.org/10.1016/j.ppnp.2012.08.001} {\bibfield  {journal} {\bibinfo
  {journal} {Prog. Part. Nucl. Phys.}\ }\textbf {\bibinfo {volume} {68}},\
  \bibinfo {pages} {1} (\bibinfo {year} {2013})},\ \Eprint
  {https://arxiv.org/abs/1209.0667} {arXiv:1209.0667 [hep-ex]} \BibitemShut
  {NoStop}%
\bibitem [{\citenamefont {Weber}(1967)}]{Weber:1967jye}%
  \BibitemOpen
  \bibfield  {author} {\bibinfo {author} {\bibfnamefont {J.}~\bibnamefont
  {Weber}},\ }\bibfield  {title} {\bibinfo {title} {{Gravitational
  Radiation}},\ }\href {https://doi.org/10.1103/PhysRevLett.18.498} {\bibfield
  {journal} {\bibinfo  {journal} {Phys. Rev. Lett.}\ }\textbf {\bibinfo
  {volume} {18}},\ \bibinfo {pages} {498} (\bibinfo {year} {1967})}\BibitemShut
  {NoStop}%
\bibitem [{\citenamefont {Weber}(1969)}]{Weber:1969bz}%
  \BibitemOpen
  \bibfield  {author} {\bibinfo {author} {\bibfnamefont {J.}~\bibnamefont
  {Weber}},\ }\bibfield  {title} {\bibinfo {title} {{Evidence for discovery of
  gravitational radiation}},\ }\href
  {https://doi.org/10.1103/PhysRevLett.22.1320} {\bibfield  {journal} {\bibinfo
   {journal} {Phys. Rev. Lett.}\ }\textbf {\bibinfo {volume} {22}},\ \bibinfo
  {pages} {1320} (\bibinfo {year} {1969})}\BibitemShut {NoStop}%
\bibitem [{\citenamefont {Aguiar}(2011)}]{Aguiar:2010kn}%
  \BibitemOpen
  \bibfield  {author} {\bibinfo {author} {\bibfnamefont {O.~D.}\ \bibnamefont
  {Aguiar}},\ }\bibfield  {title} {\bibinfo {title} {{The Past, Present and
  Future of the Resonant-Mass Gravitational Wave Detectors}},\ }\href
  {https://doi.org/10.1088/1674-4527/11/1/001} {\bibfield  {journal} {\bibinfo
  {journal} {Res. Astron. Astrophys.}\ }\textbf {\bibinfo {volume} {11}},\
  \bibinfo {pages} {1} (\bibinfo {year} {2011})},\ \Eprint
  {https://arxiv.org/abs/1009.1138} {arXiv:1009.1138 [astro-ph.IM]}
  \BibitemShut {NoStop}%
\bibitem [{\citenamefont {Husa}(2009)}]{Husa:2009zz}%
  \BibitemOpen
  \bibfield  {author} {\bibinfo {author} {\bibfnamefont {S.}~\bibnamefont
  {Husa}},\ }\bibfield  {title} {\bibinfo {title} {{Michele Maggiore:
  Gravitational waves. Volume 1: Theory and experiments}},\ }\href
  {https://doi.org/10.1007/s10714-009-0762-5} {\bibfield  {journal} {\bibinfo
  {journal} {Gen. Rel. Grav.}\ }\textbf {\bibinfo {volume} {41}},\ \bibinfo
  {pages} {1667} (\bibinfo {year} {2009})}\BibitemShut {NoStop}%
\bibitem [{\citenamefont {Hobbs}\ and\ \citenamefont
  {Dai}(2017)}]{Hobbs:2017oam}%
  \BibitemOpen
  \bibfield  {author} {\bibinfo {author} {\bibfnamefont {G.}~\bibnamefont
  {Hobbs}}\ and\ \bibinfo {author} {\bibfnamefont {S.}~\bibnamefont {Dai}},\
  }\bibfield  {title} {\bibinfo {title} {{Gravitational wave research using
  pulsar timing arrays}},\ }\href {https://doi.org/10.1093/nsr/nwx126}
  {\bibfield  {journal} {\bibinfo  {journal} {Natl. Sci. Rev.}\ }\textbf
  {\bibinfo {volume} {4}},\ \bibinfo {pages} {707} (\bibinfo {year} {2017})},\
  \Eprint {https://arxiv.org/abs/1707.01615} {arXiv:1707.01615 [astro-ph.IM]}
  \BibitemShut {NoStop}%
\bibitem [{\citenamefont {Dyson}(1969)}]{Dyson:1969zgf}%
  \BibitemOpen
  \bibfield  {author} {\bibinfo {author} {\bibfnamefont {F.~J.}\ \bibnamefont
  {Dyson}},\ }\bibfield  {title} {\bibinfo {title} {{Seismic response of the
  earth to a gravitational wave in the 1-Hz band}},\ }\href@noop {} {\bibfield
  {journal} {\bibinfo  {journal} {Astrophys. J.}\ }\textbf {\bibinfo {volume}
  {156}},\ \bibinfo {pages} {529} (\bibinfo {year} {1969})}\BibitemShut
  {NoStop}%
\bibitem [{\citenamefont {Lopes}\ and\ \citenamefont
  {Silk}(2017)}]{Lopes:2017xvq}%
  \BibitemOpen
  \bibfield  {author} {\bibinfo {author} {\bibfnamefont {I.}~\bibnamefont
  {Lopes}}\ and\ \bibinfo {author} {\bibfnamefont {J.}~\bibnamefont {Silk}},\
  }\bibfield  {title} {\bibinfo {title} {{Gravitational Waves from Stellar
  Black Hole Binaries and the Impact on Nearby Sun-like Stars}},\ }\href
  {https://doi.org/10.3847/1538-4357/aa7758} {\bibfield  {journal} {\bibinfo
  {journal} {Astrophys. J.}\ }\textbf {\bibinfo {volume} {844}},\ \bibinfo
  {pages} {39} (\bibinfo {year} {2017})},\ \Eprint
  {https://arxiv.org/abs/1707.06249} {arXiv:1707.06249 [astro-ph.GA]}
  \BibitemShut {NoStop}%
\bibitem [{\citenamefont {Lopes}\ and\ \citenamefont
  {Silk}(2015)}]{Lopes:2015pca}%
  \BibitemOpen
  \bibfield  {author} {\bibinfo {author} {\bibfnamefont {I.}~\bibnamefont
  {Lopes}}\ and\ \bibinfo {author} {\bibfnamefont {J.}~\bibnamefont {Silk}},\
  }\bibfield  {title} {\bibinfo {title} {{Nearby Stars as Gravitational Wave
  Detectors}},\ }\href {https://doi.org/10.1088/0004-637X/807/2/135} {\bibfield
   {journal} {\bibinfo  {journal} {Astrophys. J.}\ }\textbf {\bibinfo {volume}
  {807}},\ \bibinfo {pages} {135} (\bibinfo {year} {2015})},\ \Eprint
  {https://arxiv.org/abs/1507.03212} {arXiv:1507.03212 [astro-ph.SR]}
  \BibitemShut {NoStop}%
\bibitem [{\citenamefont {Lopes}\ and\ \citenamefont
  {Silk}(2014)}]{Lopes:2014dba}%
  \BibitemOpen
  \bibfield  {author} {\bibinfo {author} {\bibfnamefont {I.}~\bibnamefont
  {Lopes}}\ and\ \bibinfo {author} {\bibfnamefont {J.}~\bibnamefont {Silk}},\
  }\bibfield  {title} {\bibinfo {title} {{Helioseismology and Asteroseismology:
  Looking for Gravitational Waves in acoustic oscillations}},\ }\href
  {https://doi.org/10.1088/0004-637X/794/1/32} {\bibfield  {journal} {\bibinfo
  {journal} {Astrophys. J.}\ }\textbf {\bibinfo {volume} {794}},\ \bibinfo
  {pages} {32} (\bibinfo {year} {2014})},\ \Eprint
  {https://arxiv.org/abs/1405.0292} {arXiv:1405.0292 [astro-ph.CO]}
  \BibitemShut {NoStop}%
\bibitem [{\citenamefont {Hinderer}(2008)}]{Hinderer:2007mb}%
  \BibitemOpen
  \bibfield  {author} {\bibinfo {author} {\bibfnamefont {T.}~\bibnamefont
  {Hinderer}},\ }\bibfield  {title} {\bibinfo {title} {{Tidal Love numbers of
  neutron stars}},\ }\href {https://doi.org/10.1086/533487} {\bibfield
  {journal} {\bibinfo  {journal} {Astrophys. J.}\ }\textbf {\bibinfo {volume}
  {677}},\ \bibinfo {pages} {1216} (\bibinfo {year} {2008})},\ \Eprint
  {https://arxiv.org/abs/0711.2420} {arXiv:0711.2420 [astro-ph]} \BibitemShut
  {NoStop}%
\bibitem [{\citenamefont {Carroll}(2019)}]{Carroll:2004st}%
  \BibitemOpen
  \bibfield  {author} {\bibinfo {author} {\bibfnamefont {S.~M.}\ \bibnamefont
  {Carroll}},\ }\href@noop {} {\emph {\bibinfo {title} {{Spacetime and
  Geometry}}}}\ (\bibinfo  {publisher} {Cambridge University Press},\ \bibinfo
  {year} {2019})\BibitemShut {NoStop}%
\bibitem [{\citenamefont {Kr\"uger}\ \emph {et~al.}(2015)\citenamefont
  {Kr\"uger}, \citenamefont {Ho},\ and\ \citenamefont
  {Andersson}}]{Kruger:2014pva}%
  \BibitemOpen
  \bibfield  {author} {\bibinfo {author} {\bibfnamefont {C.~J.}\ \bibnamefont
  {Kr\"uger}}, \bibinfo {author} {\bibfnamefont {W.~C.~G.}\ \bibnamefont
  {Ho}},\ and\ \bibinfo {author} {\bibfnamefont {N.}~\bibnamefont
  {Andersson}},\ }\bibfield  {title} {\bibinfo {title} {{Seismology of
  adolescent neutron stars: Accounting for thermal effects and crust
  elasticity}},\ }\href {https://doi.org/10.1103/PhysRevD.92.063009} {\bibfield
   {journal} {\bibinfo  {journal} {Phys. Rev. D}\ }\textbf {\bibinfo {volume}
  {92}},\ \bibinfo {pages} {063009} (\bibinfo {year} {2015})},\ \Eprint
  {https://arxiv.org/abs/1402.5656} {arXiv:1402.5656 [gr-qc]} \BibitemShut
  {NoStop}%
\bibitem [{\citenamefont {Coccia}\ \emph {et~al.}(1996)\citenamefont {Coccia},
  \citenamefont {Fafone}, \citenamefont {Frossati}, \citenamefont {ter Haar},\
  and\ \citenamefont {Meisel}}]{Coccia:1996gw}%
  \BibitemOpen
  \bibfield  {author} {\bibinfo {author} {\bibfnamefont {E.}~\bibnamefont
  {Coccia}}, \bibinfo {author} {\bibfnamefont {V.}~\bibnamefont {Fafone}},
  \bibinfo {author} {\bibfnamefont {G.}~\bibnamefont {Frossati}}, \bibinfo
  {author} {\bibfnamefont {E.}~\bibnamefont {ter Haar}},\ and\ \bibinfo
  {author} {\bibfnamefont {M.~W.}\ \bibnamefont {Meisel}},\ }\bibfield  {title}
  {\bibinfo {title} {{Eigenfrequencies and quality factors of vibration of
  aluminium alloy spherical resonators}},\ }\href
  {https://doi.org/10.1016/0375-9601(96)00452-5} {\bibfield  {journal}
  {\bibinfo  {journal} {Phys. Lett. A}\ }\textbf {\bibinfo {volume} {219}},\
  \bibinfo {pages} {263} (\bibinfo {year} {1996})}\BibitemShut {NoStop}%
\bibitem [{\citenamefont {Ju}\ \emph {et~al.}(2000)\citenamefont {Ju},
  \citenamefont {Blair},\ and\ \citenamefont {Zhao}}]{Ju:2000va}%
  \BibitemOpen
  \bibfield  {author} {\bibinfo {author} {\bibfnamefont {L.}~\bibnamefont
  {Ju}}, \bibinfo {author} {\bibfnamefont {D.~G.}\ \bibnamefont {Blair}},\ and\
  \bibinfo {author} {\bibfnamefont {C.}~\bibnamefont {Zhao}},\ }\bibfield
  {title} {\bibinfo {title} {{Detection of gravitational waves}},\ }\href
  {https://doi.org/10.1088/0034-4885/63/9/201} {\bibfield  {journal} {\bibinfo
  {journal} {Rept. Prog. Phys.}\ }\textbf {\bibinfo {volume} {63}},\ \bibinfo
  {pages} {1317} (\bibinfo {year} {2000})}\BibitemShut {NoStop}%
\bibitem [{\citenamefont {Lai}(1994)}]{Lai:1993di}%
  \BibitemOpen
  \bibfield  {author} {\bibinfo {author} {\bibfnamefont {D.}~\bibnamefont
  {Lai}},\ }\bibfield  {title} {\bibinfo {title} {{Resonant oscillations and
  tidal heating in coalescing binary neutron stars}},\ }\href
  {https://doi.org/10.1093/mnras/270.3.611} {\bibfield  {journal} {\bibinfo
  {journal} {Mon. Not. Roy. Astron. Soc.}\ }\textbf {\bibinfo {volume} {270}},\
  \bibinfo {pages} {611} (\bibinfo {year} {1994})},\ \Eprint
  {https://arxiv.org/abs/astro-ph/9404062} {arXiv:astro-ph/9404062}
  \BibitemShut {NoStop}%
\bibitem [{\citenamefont {van Eysden}\ and\ \citenamefont
  {Melatos}(2010)}]{vanEysden:2010ha}%
  \BibitemOpen
  \bibfield  {author} {\bibinfo {author} {\bibfnamefont {C.~A.}\ \bibnamefont
  {van Eysden}}\ and\ \bibinfo {author} {\bibfnamefont {A.}~\bibnamefont
  {Melatos}},\ }\bibfield  {title} {\bibinfo {title} {{Pulsar glitch recovery
  and the superfluidity coefficients of bulk nuclear matter}},\ }\href
  {https://doi.org/10.1111/j.1365-2966.2010.17387.x} {\bibfield  {journal}
  {\bibinfo  {journal} {Mon. Not. Roy. Astron. Soc.}\ }\textbf {\bibinfo
  {volume} {409}},\ \bibinfo {pages} {1253} (\bibinfo {year} {2010})},\ \Eprint
  {https://arxiv.org/abs/1007.4360} {arXiv:1007.4360 [astro-ph.SR]}
  \BibitemShut {NoStop}%
\end{thebibliography}%
\end{document}